# Revealing the internal luminescence quantum efficiency of perovskite films via accurate quantification of photon recycling

Paul Fassl[1,2,5*], Vincent Lami[5], Felix J. Berger[4], Lukas M. Falk[5], Jana Zaumseil[4], Bryce S. Richards[1,2], Ian A. Howard[1,2], Yana Vaynzof[3,5], Ulrich W. Paetzold[1,2*]

[1]Institute of Microstructure Technology (IMT), Karlsruhe Institute of Technology, Eggenstein-Leopoldshafen, Germany

[2]Light Technology Institute (LTI), Karlsruhe Institute of Technology, Karlsruhe, Germany

[3]Integrated Center for Applied Physics and Photonic Materials (IAPP) and Centre for Advancing Electronics Dresden (cfaed), Dresden University of Technology, Dresden, Germany

[4]Institute for Physical Chemistry (PCI) and Centre for Advanced Materials (CAM), Heidelberg University, Heidelberg, Germany

[5]Kirchhoff-Institute for Physics (KIP) and Centre for Advanced Materials (CAM), Heidelberg University, Heidelberg, Germany

[*]e-mail: paul.fassl@kit.edu; ulrich.paetzold@kit.edu

**The internal luminescence quantum efficiency ($\boldsymbol{Q_{\mathrm{i}}^{\mathrm{lum}}}$) provides an excellent assessment of the optoelectronic quality of semiconductors. To determine $\boldsymbol{Q_{\mathrm{i}}^{\mathrm{lum}}}$ of perovskite films from the experimentally accessible external luminescence quantum efficiency ($\boldsymbol{Q_{\mathrm{e}}^{\mathrm{lum}}}$) it is essential to account for photon recycling, and this requires knowledge of the photon escape probability ($\boldsymbol{\overline{p}_{\mathrm{e}}}$). Here, we establish an analysis procedure based on a curve fitting model that accurately determines $\boldsymbol{\overline{p}_{\mathrm{e}}}$ of perovskite films from photoluminescence (PL) spectra measured with a confocal microscope and an integrating sphere setup. We show that scattering-induced outcoupling of initially-trapped PL explains commonly observed red-shifted and broadened PL spectral shapes and leads to $\boldsymbol{\overline{p}_{\mathrm{e}}}$ being more than 10% higher in absolute terms compared to earlier assumptions. Applying our model to $CH_3NH_3PbI_3$ films with exceptionally high $\boldsymbol{Q_{\mathrm{e}}^{\mathrm{lum}}}$ up to 47.4% sets a real benchmark for $\boldsymbol{Q_{\mathrm{i}}^{\mathrm{lum}}}$ at 78.0 ± 0.5%, revealing there is beyond a factor of two more scope for reducing non-radiative recombination than previously thought.**

## Introduction

The power conversion efficiency (PCE) of perovskite solar cells (PSCs) has reached a certified 25.5%.[1] Further advance of the PCE towards the fundamental efficiency limit of ~30.5% (for a bandgap of 1.6 eV)[2] requires maximizing the external luminescence quantum efficiency ($Q_{\mathrm{e}}^{\mathrm{lum}}$) by minimizing non-radiative recombination[3] in order to bring the open-circuit voltage ($V_{\mathrm{OC}}$) close to its radiative limit ($V_{\mathrm{OC,rad}}$):[2,4,5]

$$V_{\mathrm{OC}} = V_{\mathrm{OC,rad}} + \frac{kT}{q} \ln Q_{\mathrm{e}}^{\mathrm{lum}} \quad (1)$$

Adapting established assumptions for the photon escape probability ($\bar{p}_{\mathrm{e}}$) in perfectly flat GaAs films,[5–7] polycrystalline perovskite films are associated with a low $\bar{p}_{\mathrm{e}}$[2,8,9] owing to their rather low surface roughness and relatively high refractive index ($n$ ~ 2.6).[10] Therefore, most of the isotropically emitted photoluminescence (PL) is considered trapped and reabsorbed in the perovskite due to the overlap of the emission and absorption spectra.[2,11] As previously shown for GaAs,[5–7] to maximize photon extraction and therefore $Q_{\mathrm{e}}^{\mathrm{lum}}$ in this situation, perovskites need to demonstrate efficient photon recycling (PR),[8,11,12] which requires a high internal luminescence quantum efficiency ($Q_{\mathrm{i}}^{\mathrm{lum}}$).[2,8,13–16] Considering multiple reabsorption and reemission events, $Q_{\mathrm{e}}^{\mathrm{lum}}$ can be expressed as (Figure S1)[7,13]

$$Q_{\mathrm{e}}^{\mathrm{lum}} = \frac{\bar{p}_{\mathrm{e}} \cdot Q_{\mathrm{i}}^{\mathrm{lum}}}{1 - \bar{p}_{\mathrm{r}} \cdot Q_{\mathrm{i}}^{\mathrm{lum}}} \quad (2),$$

where $\bar{p}_{\mathrm{r}} = 1 - \bar{p}_{\mathrm{e}} - \bar{p}_{\mathrm{a}}$ is the probability for photon reabsorption in the perovskite and $\bar{p}_{\mathrm{a}}$ is the probability for parasitic absorption within the charge transport layers or electrodes.[13] For a bare perovskite film on glass, $Q_{\mathrm{i}}^{\mathrm{lum}}$ can be determined from the experimentally accessible $Q_{\mathrm{e}}^{\mathrm{lum}}$ by rearranging Equation (2) and making the assumption of no parasitic absorption:[8,9]

$$Q_\mathrm{i}^\mathrm{lum} = \frac{Q_\mathrm{e}^\mathrm{lum}}{\bar{p}_\mathrm{e} + \bar{p}_\mathrm{r} \cdot Q_\mathrm{e}^\mathrm{lum}} \xrightarrow{\bar{p}_\mathrm{a}=0} \frac{Q_\mathrm{e}^\mathrm{lum}}{\bar{p}_\mathrm{e} + (1-\bar{p}_\mathrm{e}) \cdot Q_\mathrm{e}^\mathrm{lum}} \quad (3)$$

Conceptual studies demonstrated that PR could contribute to enhance the $V_\mathrm{OC}$ of a PSC ($\Delta V_\mathrm{OC}^\mathrm{PR}$) by up to ~80 mV in the radiative limit and consequently the PCE by ~2% absolute.[2,14–16] However, enhancements due to PR are only significant for very high $Q_\mathrm{i}^\mathrm{lum}$ and $\bar{p}_\mathrm{r}$ as derived by Abebe *et al*. (see Figure S2):[14]

$$\Delta V_\mathrm{OC}^\mathrm{PR} = \frac{kT}{q} \cdot \ln \frac{1}{1 - \bar{p}_\mathrm{r} \cdot Q_\mathrm{i}^\mathrm{lum}} \quad (4)$$

Therefore, accurate knowledge of $\bar{p}_\mathrm{e}$ is paramount to determine $Q_\mathrm{i}^\mathrm{lum}$ and thus the scope for minimizing non-radiative recombination in perovskite films as well as to estimate the potential $\Delta V_\mathrm{OC}^\mathrm{PR}$ of PSCs.

Here, we reveal that it is key to account for PL that is initially trapped in guided modes, but coupled out before reabsorption upon scattering after some length of travel within the perovskite film.[17–20] Scattered PL exhibits a red-shifted spectrum given the much stronger reabsorption of high-energy photons.[11,18,20–22] Extending recent findings,[18,20] we show that the measured fractions of directly emitted and scattered PL critically depend on the collection method and film properties. Thereby, the large variety of reported PL spectral shapes (asymmetric/red-shifted/broadened; examples in Figure S12 and Table S1) for perovskite films with direct bandgap behavior[23,24] can be explained by simply considering light propagation and scattering. Importantly, previous reports neglected such scattering with approximations for $\bar{p}_\mathrm{e}$ ranging from ~5-13%;[2,8,9,11,14–16,25–27] these were derived based on simple geometric optics or more rigorous considerations developed for perfectly flat films (see Note S3). Employing such values for $\bar{p}_\mathrm{e}$ in Equation (3), extraordinary high $Q_\mathrm{i}^\mathrm{lum}$ ~90% (for experimentally measured $Q_\mathrm{e}^\mathrm{lum}$ ~40-50%; Figure S1) have recently been reported for perovskite films.[25–27]

In response to this problem, we establish a novel curve fitting model that reveals the real value of $Q_i^{lum}$ by accurately quantifying the effective escape probability $\bar{p}_e = \bar{p}_{e_d} + \bar{p}_{e_s}$, where $\bar{p}_{e_d}$ and $\bar{p}_{e_s}$ are the probabilities for direct emission and scattering-induced outcoupling, respectively. Our model only requires the following inputs: 1) the internal PL spectrum $I(\lambda)$ (determined by analyzing confocal PL measurements); 2) the total externally emitted PL spectrum $E_{tot}^{exp}(\lambda)$ (measured with an integrating sphere (IS) setup); and 3) the corresponding $Q_e^{lum}$. We validate our model with Monte Carlo simulations and apply it to high-quality polycrystalline $CH_3NH_3PbI_3$ (MAPI) films which exhibit exceptionally high $Q_e^{lum}$ up to 47.4% without any additional passivation layer. We show that $\bar{p}_e$ of these films is more than 10% higher in absolute terms compared to earlier assumptions and set a real benchmark for $Q_i^{lum}$ at $78.0 \pm 0.5\%$. This means that there is at least a factor of two more scope for reducing non-radiative recombination in perovskite films than previously thought; there remains more room for material optimization. Our curve fitting model, provided as a user-friendly open-source Matlab app, can be easily employed by researchers to accurately determine $Q_i^{lum}$. We also make use of the scattering properties of perovskite films and propose a new procedure to analyze parasitic absorption in a layer stack. Finally, we discuss the implications of our results on the effect that PR has on enhancing the PCE of state-of-the-art PSCs and underline the importance of further reducing interfacial recombination and parasitic absorption losses on the way towards the radiative efficiency limit.

## Results and discussion

### PL spectrum dependence on collection method

It has been recently revealed that the PL spectrum of a perovskite film measured at a lateral distance r $>$ 3 μm away from a small illumination spot ($<$ 1 μm) already exhibits asymmetry and a small red-shift (Figure S3).[18,28] Bercegol *et al.* modelled the observed spectral shapes using a linear combination of two PL spectra:[18] the high-energy part was assigned to PR,[11] that is, reemission of waveguided PL that is reabsorbed outside the illuminated region. This spectrum is equal to that of the direct emission at $r = 0$ μm due to phonon-assisted upconversion;[11,29–31] the low-energy part was assigned to outcoupling of initially-trapped PL (that is, scattered PL) that becomes increasingly red-shifted with lateral distance given the much stronger reabsorption of high-energy photons.[11,18,20–22] In addition, Cho *et al*. recently demonstrated that scattering is one of the main optical loss mechanism competing with photon amplification in a perovskite waveguide.[20] Previously, asymmetric and/or red-shifted PL spectra have been regularly observed in case of perovskite single crystals and, in many cases, have been correctly assigned to reabsorption of high-energy photons as the PL travels through the bulk of the crystal before exiting the material.[22,31–34] However, in many other cases, such observations (especially for perovskite films) have either been neglected or misinterpreted by trying to explain them with complex physical mechanisms such as the Rashba effect[35–37] as recently summarized by Schötz *et al*.[22]

We exemplary highlight the large variety of reported PL spectra for MAPI films fabricated by similar perovskite recipes in Note S1, Figure S12 and Table S1. In the following, we demonstrate that scattered PL indeed leads to the observation of dissimilar PL spectral shapes for different collection methods and film properties. We compare the PL spectra for MAPI films with thickness of 80/160/260 nm and corresponding surface roughness ($R_q$) of ~10.7/19/44 nm (Figure S4) measured by (1) a confocal PL setup, (2) a standard IS setup, and (3) a modified IS setup to observe total front emission (Figure 1A). Film fabrication is based on a lead acetate trihydrate precursor (see Experimental Procedures) which yields high-quality pin-hole free MAPI films with high $Q_e^{lum}$.[25,38–42] The conclusions drawn in the following are general for spin-coated perovskite films with various compositions as shown in Figure S5.

(1) The confocal setup used in this work detects only PL emitted within a limited field of view (FOV) (~3 μm radius) around a small illumination spot (~2 μm) (see Experimental Procedures and Figure S6). Direct emission dominates the measured PL spectral shape on these length scales with no impact of scattered PL.[18,28] The PL spectrum of an 80 nm-thick MAPI film exhibits a peak at ~768.5 nm (full-width-half-maximum (FWHM) ~39.5 nm) (Figure 1B). It proves to be identical at various spots of different samples, revealing the homogeneity and reproducibility of our films (Figure S7).[38,39] In comparison, the spectrum of a 260 nm-thick MAPI film exhibits a slightly red-shifted peak at ~771 nm (FWHM ~39.3 nm), which is attributed to more high-energy photons being reabsorbed with increasing film thickness according to the Lambert-Beer law and in line with previous reports.[21,43,44] Considering the fundamental interdependence of absorptance and emissivity,[43,45,46] in Note S3 we present a comprehensive procedure detailing how to extract the absorption coefficient $\alpha(\lambda)$ from the confocal spectra (Figures S8 and S10) and provide a comparison with literature (Figure S13). $\alpha(\lambda)$ is used to determine the internal PL spectrum $I(\lambda)$ considering the generalized Planck's law[43] (Figures S9 and S10; Equations S1 and S2) and both spectra are used as inputs of our curve fitting model. To confirm the quality of the MAPI films, we determine the Urbach energy (that is, the slope of the exponential decrease in absorption below the bandgap) of MAPI films with different thickness from $\alpha(\lambda, T)$ extracted at various irradiation intensities (that is, temperatures) and find $E_{\mathrm{U}} = 13.2 \pm 0.1$ meV at room temperature (more details in Figure S11), which is one of the lowest reported values for perovskite films to date.[3,21,40,45]

(2) The standard IS setup detects all PL externally emitted from a film ($E_{\mathrm{tot}}^{\mathrm{exp}}(\lambda)$), encompassing PL from the entire front and rear surfaces as well as the edges of the glass substrate; $E_{\mathrm{tot}}^{\mathrm{exp}}(\lambda)$ is fitted by our curve fitting model. The PL spectra of exactly the same MAPI samples as measured with setup (1) exhibit an asymmetric and broadened shape (FWHM of ~55.6 nm and ~43.3 nm, respectively) with a significant red-shift that is stronger for the thicker film (peak positions of ~780 nm and ~789 nm, respectively) (Figure 1B). These characteristics result from the measured signal being a superposition of directly emitted and red-shifted scattered PL as in detail discussed in the next section and consistent with recent experimental findings.[18,20–22,28,47]

(3) A modified IS setup is used to analyze the contribution of scattered PL to the total front emission. We compare the PL spectra of a 160 nm-thick MAPI film measured with setups (1) - (3) in Figure 1C. Note that the peaks are normalized to their intensity at $\lambda = 730$ nm as this representation highlights alterations in the peak position and shape due to thickness and scattering, which appear for $\lambda \gtrsim 735$ nm in case of MAPI (see Figure S5 for other compositions). The peak position (~775.8 nm) and FWHM (~50.1 nm) measured with setup (3) are in between the values measured with setup (1) (peak at ~770.3 nm, FWHM of ~39.2 nm) and setup (2) (peak at ~786.2 nm, FWHM of ~50.7 nm). Firstly, this illustrative experiment reveals the significant contribution of scattered PL to the total front emission, which is therefore also detected by typically employed standard front collection PL setups.[18,20–22,28,47] Secondly, the reduced red-shift for setup (3) as compared to setup (2) shows that a considerable amount of red-shifted PL is emitted from the substrate edges.

In Note S2 we provide a more in-depth analysis of the absolute photon flux (Figure S14) and PL spectral shapes (Figure S15) observed with various setups for the 260 nm-thick MAPI film with the highest $Q_e^{lum}$ measured in this work. The results indicate that edge emission, which is a well-studied phenomenon in organic LEDs[48,49] yet has not been reported to play a large role for perovskite films before, in fact accounts for ~50% of the total external PL emission and exhibits a strongly red-shifted spectrum (Figure S14a). In Figures S14b and S15 we compare our results to that for MAPI films with similar thickness reported by Braly *et al.*[27] who employed an absolute intensity confocal PL setup with a large FOV (~2.2 mm) in order to collect all photons emitted from the front.[50] The shape of their PL spectrum is red-shifted and broadened compared to the one measured with our confocal setup and very similar to that observed with the setups where edge emission is suppressed (Figure S15), indicating a comparable fraction of scattered PL to the measured signal.

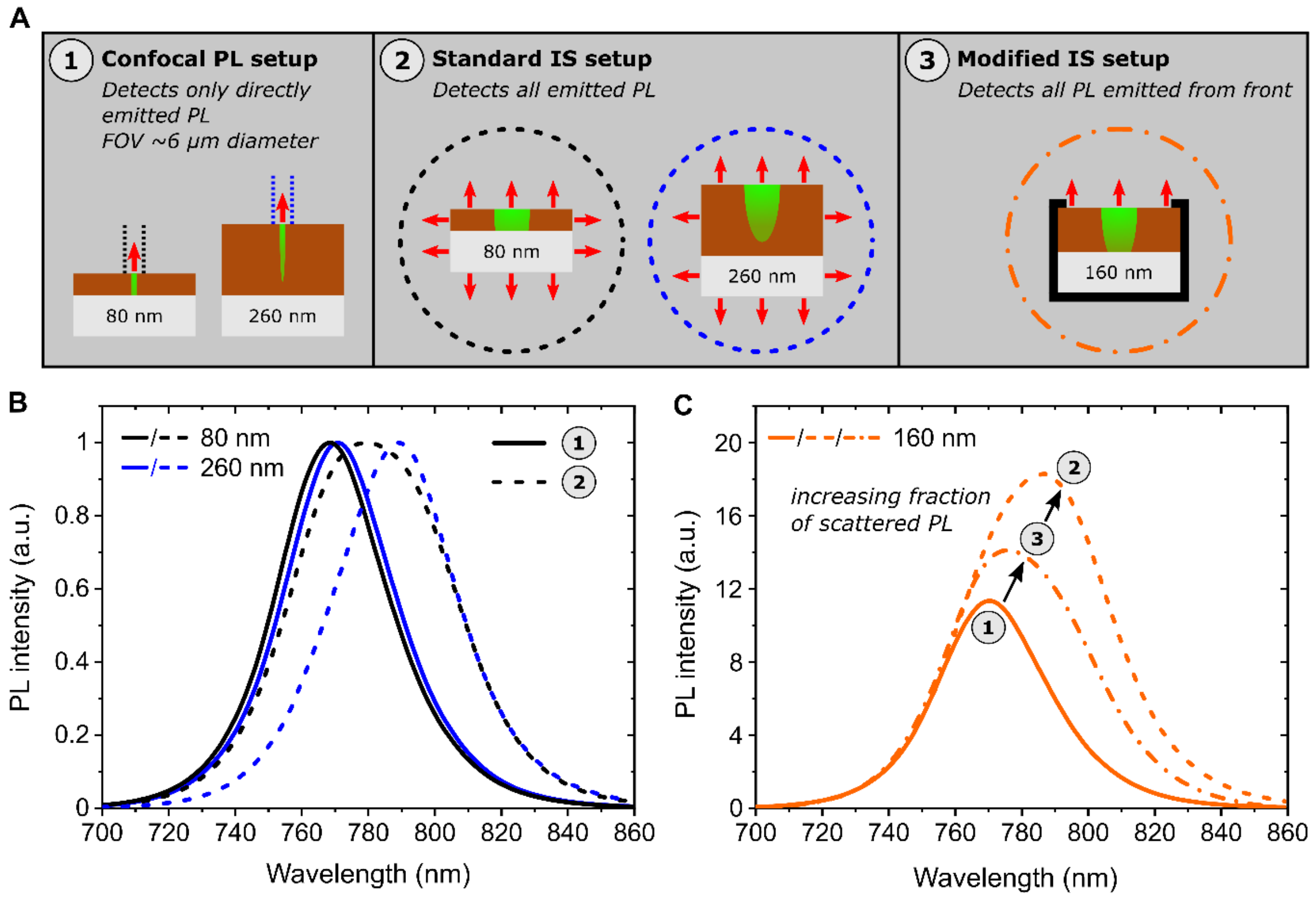

**Figure 1. Photoluminescence spectrum dependence on collection method. (A)** Schematic of the PL setups employed to study the PL spectra of MAPI films with various thicknesses on glass: (1) the confocal PL setup detects only PL emitted within a small field of view (FOV) (~6 µm diameter) around the illumination spot (~2 µm) (see Experimental Procedures; Figure S6); (2) the standard integrating sphere (IS) setup detects all emitted PL; (3) the modified IS setup detects the total front emission by covering the edges and back side of the substrate. The excitation volume using a pulsed 530 nm laser for setup (1) and 525 nm CW laser for setups (2) and (3) is roughly visualized. **(B)** Normalized PL spectra of an 80 nm- and a 260 nm-thick MAPI film measured with setup (1) (solid lines) and setup (2) (dotted lines). **(C)** PL spectra of a 160 nm-thick MAPI film measured with setups (1) - (3) normalized to the value at $\lambda$ = 730 nm. The arrows indicate the increasing fraction of red-shifted scattered PL to the total externally emitted PL.

**Photoluminescence emission from perovskite films**

Compared to solar cells based on the epitaxial grown mature semiconductor GaAs which has a near-atomically flat surface and an extremely low density of defects (~$10^7$-$10^8$ cm$^{-3}$),[51] perovskite films show heterogeneous material properties on various length scales,[52,53] among them: variations in crystal orientation and lattice strain;[41,47] inhomogeneous composition,[39] electronic structure[54] and defect density (~$10^{13}$-$10^{18}$ cm$^{-3}$);[38,41] and photo-induced changes in refractive index.[55] In addition, polycrystalline perovskite films employed in state-of-the-art PSCs are commonly fabricated by spin-coating and exhibit numerous grain boundaries (grain size ~100-2000 nm) with a low yet non-negligible surface roughness (~10-40 nm).[26,27,38–41,54,56–58] Considering these aspects, it can be expected that some initially-trapped PL couples out upon scattering. Surface scattering effects have been previously discussed for perovskite waveguides,[18,20,59,60] luminescent solar concentrators[61] and perovskite LEDs (PeLEDs)[30] and are a well-known phenomenon in high-index-contrast microphotonic devices.[17,19] While it is difficult to ascertain the relative contributions of the above discussed properties to the total magnitude of scattering, we believe that surface roughness plays the major role.[17,19,20]

In the following, we introduce the fundamental underlying optical processes in a perovskite film, leading to the previously discussed asymmetric, broadened and red-shifted PL spectra. Note that for simplicity we assume a perfect back-reflector here and discuss the more general case of a film on glass in Note S4 and Figure S16. Figure 2A schematically summarizes the various individual events after (0) local initial photoexcitation, namely: (1) initial isotropic PL emission; (A) PL emitted within the escape cone ($\theta \lessapprox \theta_c$); (2) reabsorption and isotropic reemission, that is, PR; ($B_i$) PL emitted within the escape cone following event (2); and ($C_i$) PL emitted at angles $\theta \gtrapprox \theta_c$ (initially trapped by total internal reflections) coupling out upon scattering. Note that PL can also be scattered into the bottom escape cone or such that it remains trapped within the film. The corresponding alterations of the external PL spectrum $E(\lambda)$ with respect to $I(\lambda)$ are shown in Figure 2B and 2C together with $\alpha(\lambda)$ (see Figure S10): PL emitted within the escape cone (events (A) and ($B_i$)) exhibits a slightly red-shifted spectrum considering the finite distance $d_{\mathrm{ec}}$ from the emission center to the film surface, following the Lambert-Beer law:[11,18,22,32,44]

$$E_d(\lambda, d_{\mathrm{ec}}, \theta \lessapprox \theta_{\mathrm{c}}) = I(\lambda) \cdot e^{-\alpha(\lambda)\frac{d_{\mathrm{ec}}}{\cos\theta}} \quad (5)$$

Scattered PL (events ($C_i$)) exhibits a progressively increasing red-shift with increasing number $i$ of total internal reflections prior to a scattering event, since initially-trapped PL propagates a total distance $z_i$ within the film before coupling out.[18,20,21] The value of $z_i$ depends on the emission angle, the vertical depth $d_{\mathrm{ec}}$ of the emission center and the film thickness $d$:

$$E_s(\lambda, z_i, \theta_{\mathrm{c}} \lessapprox \theta < 90°) = I(\lambda) \cdot e^{-\alpha(\lambda)\cdot z_i} = I(\lambda) \cdot e^{-\alpha(\lambda)\cdot\left(\frac{d_{\mathrm{ec}}}{\cos\theta} + 2(i-1)\cdot\frac{d}{\cos\theta}\right)} \quad (6)$$

It is important to stress that the discussed scattering is not to be mistaken with Lambertian light scattering like in textured silicon solar cells[5,62] and assumed for modelling light trapping schemes in PSCs.[2,14–16,63] Instead, it describes a small scattering probability every time a photon impinges on a surface, which is strongly suppressed in near-atomically flat III-V semiconductors.[19,20,51] Specifically, since low-energy photons ($\lambda \gtrapprox 790\ \mathrm{nm}$) are barely reabsorbed in the perovskite given the exponential drop in absorption, they can travel long distances within the film (Figures 2C, S10 and S17) and thus experience numerous total internal reflections before a scattering event that results in outcoupling.[20,30] The magnitude of scattering for a given propagation distance $z_i$ is therefore governed by the film thickness (compare Equation (6)) and surface roughness.[17,19,20,61] To conclude, while most high-energy photons are reabsorbed within a short distance from the initial point of emission, even a small magnitude of scattering is sufficient to couple out the low-energy photons before possible reabsorption.

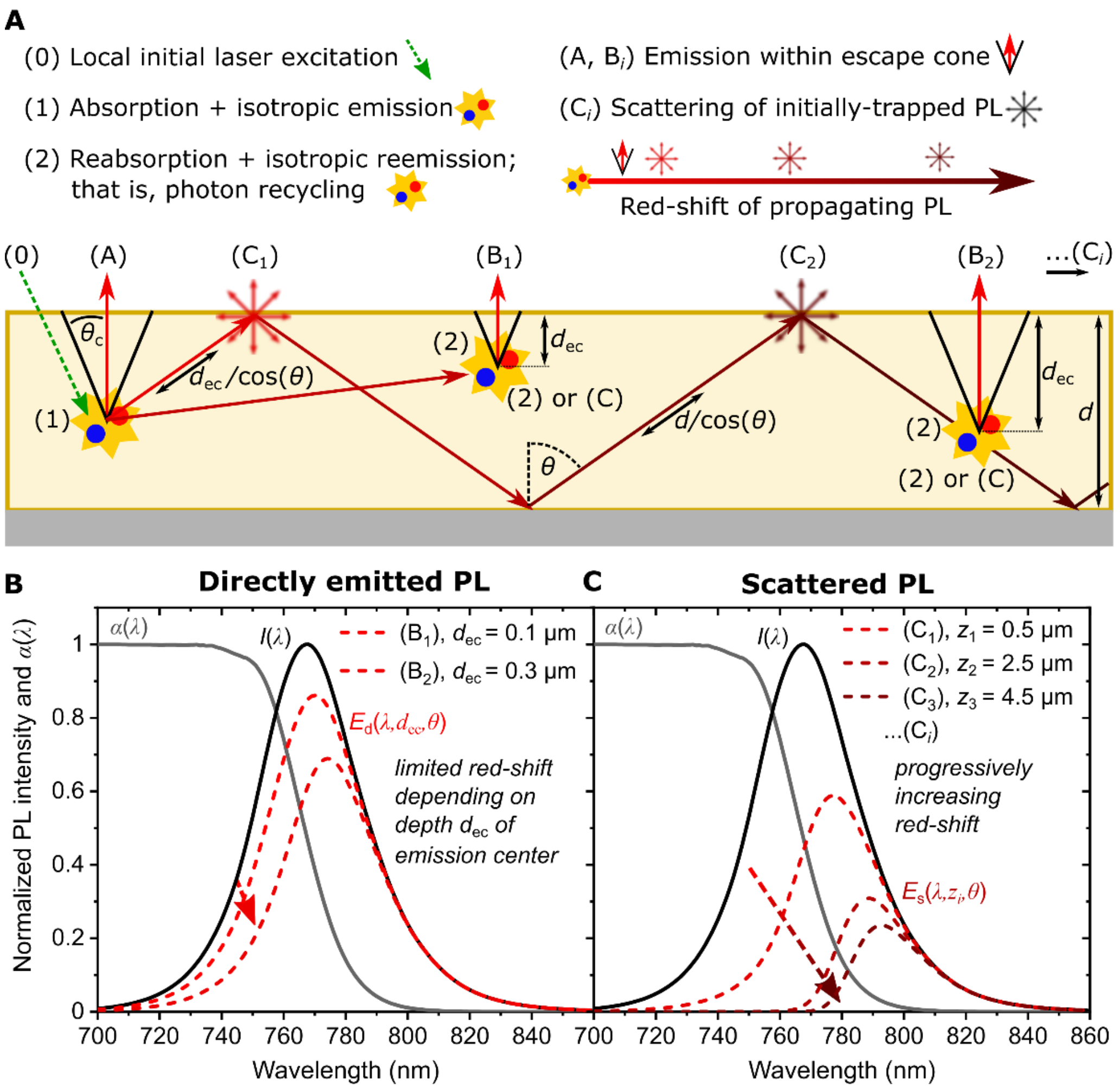

**Figure 2. Photoluminescence emission from perovskite films. (A)** Illustration of the various radiative processes (numbers) and PL emission paths (letters) in a perovskite film: (0) local initial laser excitation; (1) isotropic internal emission; (A) PL emitted within the escape cone ($\theta \lessapprox \theta_c$); (2) reabsorption and isotropic reemission, that is, photon recycling; (B$_i$) PL emitted within the escape cone following (2); and (C$_i$) PL emitted at angles $\theta \gtrapprox \theta_c$ (initially trapped in the film by total internal reflections) coupling out upon scattering (scattered PL). The color change from red to brown denotes the progressively increasing red-shift of PL propagating in the perovskite film considering the much stronger reabsorption of high-energy photons. Without loss of generality only front-face emission is visualized here by assuming a perfect back-reflector at the bottom (see more general case in Figure S16). **(B) and (C)** Internal PL spectrum $I(\lambda)$ (black) and absorption coefficient $\alpha(\lambda)$ (grey) of the MAPI films as determined in this work (Note S3; Figure S10). **(B)** Directly emitted PL ($E_d(\lambda, d_{ec}, \theta \lessapprox \theta_c)$; Equation (5)) exhibits a small red-shift depending on the finite depth $d_{ec}$ of the emission center within the film. **(C)** Scattered PL ($E_s(\lambda, z_i, \theta \lessapprox \theta_c < 90°)$; Equation (6)) progressively red-shifts with increasing number $i$ of total internal reflections prior to a scattering event with total propagation distance $z_i$.

**Curve fitting model to determine the effective escape probability**

The curve fitting model introduced in this work fits PL spectra measured with an IS setup ($E_{\text{tot}}^{\text{exp}}(\lambda)$) with $I(\lambda)$ and $\alpha(\lambda)$ determined from confocal PL measurements as input (see Note S3 and Figure S10). It allows to accurately determine $\bar{p}_{\text{e}}$ and thereby $Q_{\text{i}}^{\text{lum}}$ of perovskite films *via* Equation (3) using measured values of $Q_{\text{e}}^{\text{lum}}$ as input. We provide a user-friendly open-source Matlab app for performing the fitting (user interface shown in Figure S21). The model is visualized in Figure 4A and described in the following.

The effective escape probability is simply the ratio of total externally emitted to total internally emitted photons[13–15]

$$\bar{p}_{\text{e}} = \bar{p}_{\text{e-d}} + \bar{p}_{\text{e-s}} = \frac{\int E_{\text{tot}}^{\text{exp}}(\lambda)}{\int I_{\text{tot}}(\lambda)} \quad (7),$$

where $\bar{p}_{\text{e-d}} = \frac{\int E_{\text{d-tot}}(\lambda)}{\int I_{\text{tot}}(\lambda)}$ and $\bar{p}_{\text{e-s}} = \frac{\int E_{\text{s-tot}}(\lambda)}{\int I_{\text{tot}}(\lambda)}$ denote the probabilities for direct emission and scattering-induced outcoupling, respectively. With $E_{\text{tot}}^{\text{exp}}(\lambda)$ measured experimentally, we need to derive $I_{\text{tot}}(\lambda)$. In steady-state, $I_{\text{tot}}(\lambda)$ is composed of initially-emitted and reemitted PL ($I_0(\lambda)$ and $I_{\text{PR}}(\lambda)$; events (1) and (2) in Figure 2A) and relates to $I(\lambda)$ by introducing the scaling parameters $a_{\text{tot}}$, $a_0$ and $a_{\text{PR}}$:

$$I_{\text{tot}}(\lambda) = a_{\text{tot}} \cdot I(\lambda) = I_0(\lambda) + I_{\text{PR}}(\lambda) = (a_0 + a_{\text{PR}}) \cdot I(\lambda) \quad (8)$$

$E_{\text{tot}}^{\text{exp}}(\lambda)$ is a superposition of directly emitted PL ($E_{\text{d-tot}}(\lambda)$; events (A) and (B$_i$) in Figure 2A) and scattered PL ($E_{\text{s-tot}}(\lambda)$; events (C$_i$) in Figure 2A) and relates to $I_{\text{tot}}(\lambda)$ by considering wavelength-dependent probabilities for direct and scattering-induced escape ($p_{\text{e-d}}(\lambda)$ and $p_{\text{e-s}}(\lambda)$, respectively) as:

$$E_{\text{tot}}^{\text{exp}}(\lambda) = I_{\text{tot}}(\lambda) \cdot p_{\text{e}}(\lambda) = E_{\text{d-tot}}(\lambda) + E_{\text{s-tot}}(\lambda) = a_{\text{tot}} \cdot I(\lambda) \cdot \left(p_{\text{e-d}}(\lambda) + p_{\text{e-s}}(\lambda)\right) \quad (9)$$

$p_{\text{e-d}}(\lambda)$ accounts for the fact that photons that would otherwise directly escape the film (with probability $p_{\text{e-d}}$) are partially reabsorbed (compare Figure 2B and Equation (5)), which is modelled by introducing a Lambert-Beer term with effective film thickness $d_{\text{eff}}$ (see discussion in Note S6; fits to confocal and simulated PL spectra in Figure S22 and Table S2):

$$p_{\text{e-d}}(\lambda) = p_{\text{e-d}} \cdot e^{-\alpha(\lambda) d_{\text{eff}}} \quad (10)$$

All remaining PL is initially trapped with probability $(1 - p_{\text{e-d}})$ and we consider the following to model the effect of scattering: a small fraction $P_{\text{s}}$ of trapped PL couples out after an average propagation distance $z_{\text{av}}$ within the film as visualized in Figure 4A (compare Equation (6)). The same continues and the next fraction $P_{\text{s}}$ of the remaining PL couples out after another average distance $z_{\text{av}}$. This continuous sequence can be expressed with a geometrical series by introducing a Lambert-Beer term as

$$p_{\text{e-s}}(\lambda) = (1 - p_{\text{e-d}}) \cdot e^{-\alpha(\lambda) z_{\text{av}}} \cdot P_{\text{s}} \cdot \sum_{i=0}^{\infty} \left( (1 - P_{\text{s}}) \cdot e^{-\alpha(\lambda) z_{\text{av}}} \right)^{i} = \frac{(1-p_{\text{e-d}}) \cdot e^{-\alpha(\lambda) \cdot z_{\text{av}}} \cdot P_{\text{s}}}{1-(1-P_{\text{s}}) \cdot e^{-\alpha(\lambda) \cdot z_{\text{av}}}} \quad (11),$$

which leads to the following final expression

$$\mathrm{E}_{\text{tot}}^{\text{exp}}(\lambda) = a_{\text{tot}} \cdot I(\lambda) \cdot \left( p_{\text{e-d}} \cdot e^{-\alpha(\lambda) \cdot d_{\text{eff}}} + \frac{(1 - p_{\text{e-d}}) \cdot P_{\text{s}} \cdot e^{-\alpha(\lambda) \cdot z_{\text{av}}}}{1 - (1 - P_{\text{s}}) \cdot e^{-\alpha(\lambda) \cdot z_{\text{av}}}} \right) \quad (12)$$

with fitting parameters $a_{\text{tot}}$, $p_{\text{e-d}}$, $d_{\text{eff}}$, $P_{\text{s}}$ and $z_{\text{av}}$. We emphasize that Equation (11) describes an 'average scattering' of PL either trapped in the perovskite film or substrate (compare Figure S16), and, therefore, $P_{\text{s}}$ and $z_{\text{av}}$ have no direct physical meaning. We decided to fix $P_{\text{s}}$ for fitting given that it makes the fit more stable (see discussion in Note S6). However, as discussed in more detail later, the fitted value of $z_{\text{av}}$ can be related to the film thickness in order to compare the magnitude of scattering for different samples. Fitting with Equation (12) allows the determination of $\bar{p}_{\text{e}}$, $\bar{p}_{\text{e-d}}$ and $\bar{p}_{\text{e_s}}$ and thereby of $Q_{\text{i}}^{\text{lum}}$ *via* Equation (3) using the experimentally measured $Q_{\text{e}}^{\text{lum}}$ as input. Furthermore, the relative contributions of initially generated PL ($I_0(\lambda)$) and recycled PL ($I_{\mathrm{PR}}(\lambda)$) can be derived upon calculating $a_0 = \frac{Q_{\mathrm{i}}^{\mathrm{lum}} \cdot \int E_{\mathrm{tot}}(\lambda)}{Q_{\mathrm{e}}^{\mathrm{lum}} \cdot \int I(\lambda)}$ and $a_{\mathrm{PR}} = a_{\mathrm{tot}} - a_0$ using the value determined for $Q_{\mathrm{i}}^{\mathrm{lum}}$ (compare Equation (8)).

To validate our curve fitting model, we employ a Monte Carlo approach to simulate 'externally-observed' PL spectra with all other parameters known. The Monte Carlo simulation (see Experimental Procedures and detailed description in Note S5) considers all radiative processes and subsequent event-chains illustrated in Figure 2A as well as non-radiative recombination processes that occur subsequent to an absorption event (see decision tree and output parameters in Figure S19). For a given $Q_{\mathrm{i}}^{\mathrm{lum}}$, values for $Q_{\mathrm{e}}^{\mathrm{lum}}$, $\bar{p}_{\mathrm{e}}$, $\bar{p}_{\mathrm{e_d}}$ and $\bar{p}_{\mathrm{e_}s}$ are determined in a simulation as well as the shape of the corresponding 'external' PL spectra ($E_{\mathrm{tot}}^{\mathrm{sim}}(\lambda)$, $E_{\mathrm{d\text{-}tot}}^{\mathrm{sim}}(\lambda)$ and $E_{\mathrm{s\text{-}tot}}^{\mathrm{sim}}(\lambda)$). This way, we can fit $E_{\mathrm{tot}}^{\mathrm{sim}}(\lambda)$ with Equation (12) and validate if the fit results agree with the known values. Figure 3B shows a fit to a representative $E_{\mathrm{tot}}^{\mathrm{sim}}(\lambda)$ with an asymmetric and broadened shape similar to that of the 260 nm-thick MAPI film in Figure 1A. The values of $Q_{\mathrm{i}}^{\mathrm{lum}}$ and $\bar{p}_{\mathrm{e}}$ are reproduced with close to perfect agreement for a large range of fixed $P_{\mathrm{s}}$, while $\bar{p}_{\mathrm{e_d}}$ and $\bar{p}_{\mathrm{e_s}}$ are sensitive to its choice and agree best for the smallest possible $P_{\mathrm{s}}$ (see related input parameters and fit results for various values of $P_{\mathrm{s}}$ in Table S3).

These results corroborate four important aspects: (i) the proposed conception of PL emission (Figure 2A) indeed results in a variety of different PL spectra (red-shifted/broadened/asymmetric) depending on the magnitude of scattering, which is set by the inverse scattering length ($l_{\mathrm{s}}^{-1}$) in the simulations (see further examples in Figure S20 and Table S4); (ii) in line with intensity dependent measurements of $\mathrm{E}_{\mathrm{tot}}^{\mathrm{exp}}(\lambda)$ (Figure S18), the spectral shape of $E_{\mathrm{tot}}^{\mathrm{sim}}(\lambda)$ is independent of $Q_{\mathrm{i}}^{\mathrm{lum}}$ and thus is the value of $\bar{p}_{\mathrm{e}}$ (see discussion in Note S4); (iii) $E_{\mathrm{tot}}^{\mathrm{exp}}(\lambda)$ can be fitted with Equation (12) to accurately determine $Q_{\mathrm{i}}^{\mathrm{lum}}$ and $\bar{p}_{\mathrm{e}}$ and to give a good approximation for $\bar{p}_{\mathrm{e_d}}$ and $\bar{p}_{\mathrm{e_s}}$; and (iv) the fitted value of $z_{\mathrm{av}}$ is inversely proportional to $l_s$ and thereby, considering scattering mainly takes place at interfaces, allows comparing the magnitude of scattering for different samples by relating it to the film thickness via $M_\mathrm{s} = \frac{d}{z_\mathrm{av}}$.

**Revealing the internal luminescence quantum efficiency**

We show the excellent agreement of fits with Equation (12) to $E_\mathrm{tot}^\mathrm{exp}(\lambda)$ of MAPI films with different thickness in Figure 3C, with $I_0^\mathrm{fit}(\lambda)$ and $I_\mathrm{PR}^\mathrm{fit}(\lambda)$ being highlighted in the upper panel and $E_\mathrm{tot}^\mathrm{fit}(\lambda)$, $E_\mathrm{d\text{-}tot}^\mathrm{fit}(\lambda)$ and $E_\mathrm{s\text{-}tot}^\mathrm{fit}(\lambda)$ in the lower panel. The corresponding film properties and fit results are summarized in Table 1. With increasing film thickness, we measure champion $Q_\mathrm{e\text{-}max}^\mathrm{lum}$ of 21.2%, 42% and 47.4% at an intensity close to an AM1.5G (1 Sun, 100 mW cm$^{-2}$) steady-state absorbed photon flux (see Experimental Procedures and Note S7 for details). To the best of our knowledge the latter value represents the highest reported $Q_\mathrm{e}^\mathrm{lum}$ for a MAPI film measured with an IS setup at ~1 Sun and room temperature to date. We attribute the slightly reduced $Q_\mathrm{e}^\mathrm{lum}$ for the thinner films to the higher surface to bulk ratio, being well-established that most defects are located at the perovskite film surface.[3,25,26,38,56] The strongly reduced grain size for the thinner films (Figure S4) might play an additional role in reducing $Q_\mathrm{e}^\mathrm{lum}$, however, we note that while there is no general consensus yet if grain boundaries are benign,[64] recent reports indicate that the impact of grain boundaries on the optoelectronic properties of high-quality perovskite films is negligible.[28,52,53,65,66] Most importantly, our model reveals that $Q_\mathrm{e}^\mathrm{lum}$ of 47.4% for the 260 nm-thick MAPI film correlates to $Q_\mathrm{i}^\mathrm{lum}$ of 78.0 $\pm$ 0.5%, setting a new real benchmark for $Q_\mathrm{i}^\mathrm{lum}$ of perovskite films that is more than 10% lower compared to previous reports for similar values of $Q_\mathrm{e}^\mathrm{lum}$. This is due to a high $\bar{p}_\mathrm{e}$ of 25.4 $\pm$ 0.7%, which is more than 10% higher in absolute terms compared to earlier assumptions for $\bar{p}_\mathrm{e}$ in the range of ~5-13% that were based on simple geometric optics or more rigorous derivations developed for perfectly flat films (see Note S3). In comparison, $Q_\mathrm{e}^\mathrm{lum}$ of 42% (21.2%) for the 160 (80) nm-thick MAPI film correlates to $Q_\mathrm{i}^\mathrm{lum}$ of 72.2 $\pm$ 0.6% (46.4 $\pm$ 0.7%) given the even higher $\bar{p}_\mathrm{e}$ of 28.0 $\pm$ 0.8% (30.9 $\pm$ 0.9%). As expected for perovskite films with a very low trap density, $Q_e^{lum}$ and the corresponding $Q_i^{lum}$ of the 260 (160) nm-thick MAPI film increase only slightly from a remarkable ~33.8% (~30.6%) and ~66.8% (61.1%) at ~0.1 Sun up to ~52.2% (~46.6%) and ~81.4% (~75.6%) at ~3 Sun (Figure S23), respectively.[8,9,27,67] Despite the very high $\bar{p}_e$, our analysis reveals that PR still accounts for an impressive $\frac{a_{PR}}{a_{tot}}$ ~58% of the total emission for the 260 nm-thick MAPI film (see upper panels in Figure 3C; ~52% and ~32% for the 160 nm and 80 nm- thick films, respectively) which is comparable to values recently simulated for the best state-of-the-art PeLEDs.[30] In addition, our curve fitting model also for the first time allows to estimate values for $\bar{p}_{e_d}$ of perovskite films, and, interestingly, it strongly increases from 6.0 $\pm$ 0.5% to 17.4 $\pm$ 0.7% with decreasing film thickness, which we attribute to the suppression of waveguide modes resulting in a larger fraction of direct emission.[30,48,63] Since at the same time the magnitude of scattering ($M_s$) exhibits decreasing values of 7.3 $\pm$ 0.4, 3.3 $\pm$ 0.2 and 1.4 $\pm$ 0.1, which is in line with the strong reduction in surface roughness with decreasing film thickness,[19,20] $\bar{p}_{e_s}$ decreases from a very high 19.4 $\pm$ 0.5% to 13.5 $\pm$ 0.3%.

Comparing $Q_i^{lum}$ determined in our work with $Q_i^{lum}$ ~90% reported by Braly *et al.* and Brenes *et al.* for MAPI films with comparable $Q_e^{lum}$ and thickness, it becomes apparent how crucial the accurate determination of $\bar{p}_e$ is (their results and assumptions are included in Table 1).[25,27] Given that the MAPI films in Brenes *et al.* work (reported $Q_i^{lum}$~89%) exhibit a similar surface roughness as our films, we estimate $\bar{p}_e \approx 26\%$ for their slightly thinner ~250 nm-thick films which would correlate to $Q_i^{lum}$ ~79.4% (measured at ~2.2 Sun). This ~10% absolute difference in the derived $Q_i^{lum}$ accounts for a significant underestimation of non-radiative recombination by a factor of about two. In fact, to really reach a value of $Q_i^{lum}$ ~90%, the 260 nm-thick MAPI film with $\bar{p}_e$ of 25.4% would need to exhibit an extraordinary high $Q_e^{lum}$ of ~70% at 1 Sun (compare Figure S1), revealing there is much more scope for future material optimization and surface passivation of perovskite films.

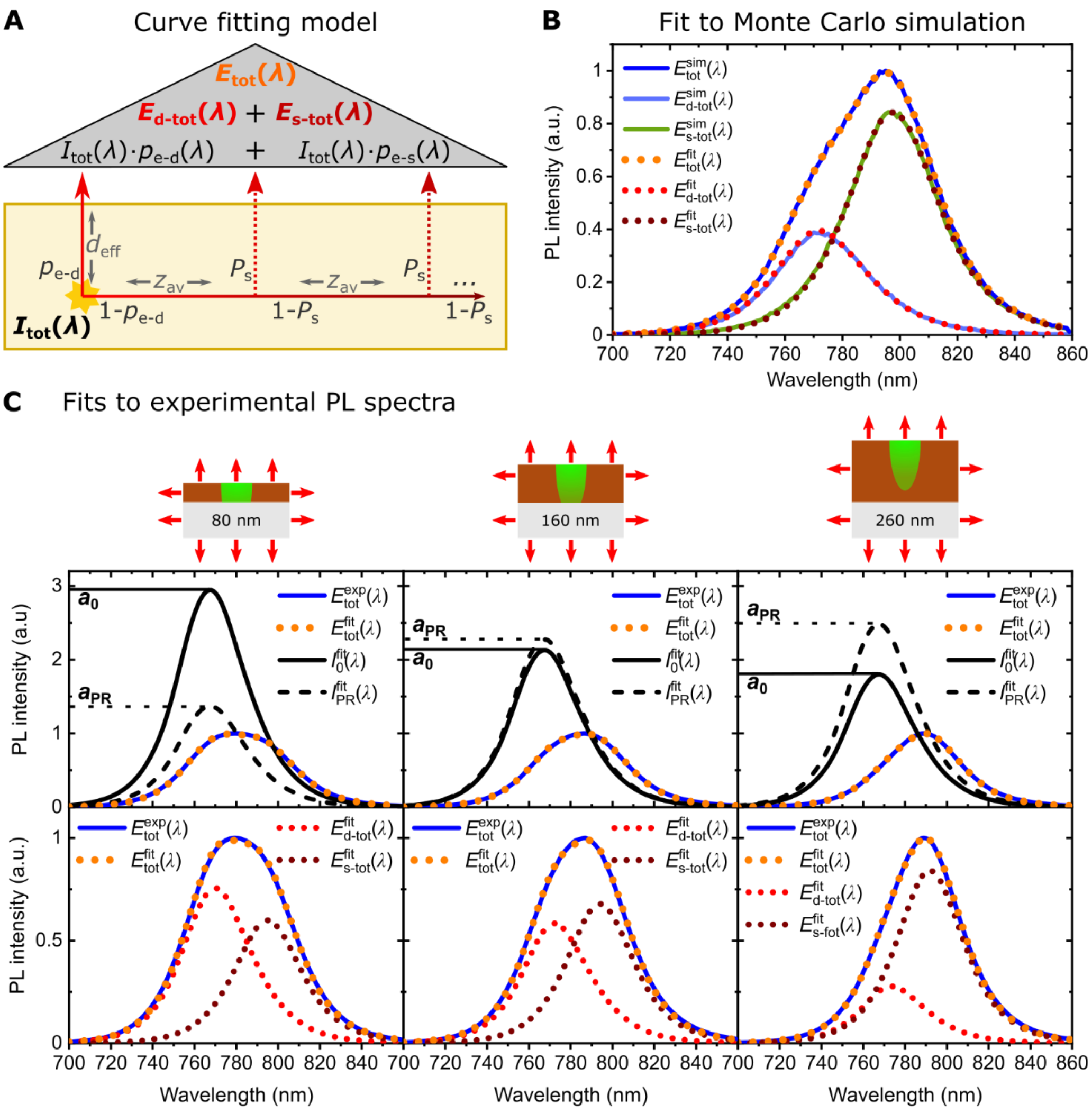

**Figure 3. Curve fitting model to determine the effective photon escape probability and internal luminescence quantum efficiency. (A)** Visualization of the curve fitting model: in steady state, the total internal PL $I_{\text{tot}}(\lambda)$ (composed of initially emitted and reemitted PL; Equation (8)) is either (i) directly emitted ($E_{\text{d-tot}}(\lambda) = I_{\text{tot}}(\lambda) \cdot p_{\text{e-d}}(\lambda)$; Equations (9) and (10)), whereby partial reabsorption is described by an effective film thickness $d_{\text{eff}}$ following the Lambert-Beer law or (ii) initially trapped with probability $1 - p_{\text{e-d}}$. Small fractions $P_{\text{s}}$ of initially-trapped PL couple out after every average propagation distance $z_{\text{av}}$ within the film, until all PL is either coupled out or reabsorbed considering the Lambert-Beer law ($E_{\text{s-tot}}(\lambda) = I_{\text{tot}}(\lambda) \cdot p_{\text{e-s}}(\lambda)$; Equation (11)). The expression for the total externally emitted PL $E_{\text{tot}}(\lambda)$ is given by Equation (12). **(B)** Fit with Equation (12) to a representative simulated 'external PL spectrum' ($E_{\text{tot}}^{\text{sim}}(\lambda)$) using a Monte Carlo approach (the corresponding parameters and fit results are summarized in Table S4). **(C)** Fits with Equation (12) to experimental PL spectra of MAPI films with different thicknesses with $I_0^{\text{fit}}(\lambda)$, $I_{\text{PR}}^{\text{fit}}(\lambda)$ being highlighted in the upper panel and $E_{\text{tot}}^{\text{fit}}(\lambda)$, $E_{\text{d-tot}}^{\text{fit}}(\lambda)$ and $E_{\text{s-tot}}^{\text{fit}}(\lambda)$ in the lower panel. The corresponding input parameters and fit results are summarized in Table 1.

Finally, we also apply our curve fitting model to two other commonly employed perovskite compositions, namely a triple-cation ($Cs_{0.05}(MA_{0.17}FA_{0.83})_{0.95}Pb(I_{0.83}Br_{0.17})_3$) and a double-cation ($FA_{1-x}MA_xPb(I_{0.97}Br_{0.03})_3$) perovskite film, both of which exhibit a much larger film thickness compared to the MAPI films (see Experimental Procedures). The corresponding confocal and IS PL spectra are discussed in detail in Figures S5 and S24. We measure $Q_\mathrm{e}^\mathrm{lum}$ up to 5% for TC and 19.6% for DC films, respectively. Importantly, despite the much larger film thicknesses that might indicate a higher reabsorption probability, we find very high $\bar{p}_\mathrm{e}$ of ~27.1% and ~20.1% with corresponding $Q_\mathrm{i}^\mathrm{lum}$ of ~18.5% and ~54.1%, respectively (see fit results in Figure S25 and Table S5). This confirms that the results discussed in detail for the archetypical MAPI films are exemplary for spin-coated perovskite films.

**The effects of photon recycling and parasitic absorption**

We proceed in analyzing the effects of photon recycling by plotting the wavelength-dependence of the escape probabilities ($p_\mathrm{e}(\lambda)$, $p_\mathrm{e\text{-}d}(\lambda)$ and $p_\mathrm{e\text{-}s}(\lambda)$), the reabsorption probability $p_\mathrm{r}(\lambda) = 1 - p_\mathrm{e}(\lambda)$ and $Q_\mathrm{e}^\mathrm{lum}(\lambda) = \frac{E_\mathrm{tot}^\mathrm{exp}(\lambda)}{I_0^\mathrm{fit}(\lambda)}$ for the 260 nm-thick MAPI film in Figure 4A (see results for other thicknesses in Figure S26). Above the bandgap ($\lambda \ll 767$ nm), reabsorption is dominant resulting in the values of $p_\mathrm{e}(\lambda)$ and $Q_\mathrm{e}^\mathrm{lum}(\lambda)$ being below 5% and 20%, respectively. Impressively, $Q_\mathrm{e}^\mathrm{lum}(\lambda)$ already exceeds a value of 100% (250%) for $\lambda \gtrsim 786$ nm (830 nm) together with $p_\mathrm{r}(\lambda)$ and $p_\mathrm{e}(\lambda)$ approaching 0% and 100% in that spectral range, respectively. The underlying reasons are the exponential drop in absorption (compare Figures S10 and S17) combined with the herein discussed scattering properties of perovskite films. We stress that even the rather low surface roughness of the 80 nm-thick MAPI film (~10 nm; Figure S4) is sufficient to couple out the low-energy photons before reabsorption can take place (Figure S26b).

To discuss the implications of our results on PSCs, we calculate the potential $\Delta V_{\mathrm{OC}}^{\mathrm{PR}}$ *via* Equation (4) using the values that we have determined for $Q_{\mathrm{i}}^{\mathrm{lum}}$ and $\bar{p}_{\mathrm{e}}$ and making the optimistic assumption of negligible parasitic absorption ($\bar{p}_{\mathrm{a}} = 0$).[14] We find $\Delta V_{\mathrm{OC}}^{\mathrm{PR}} = 22.5$ mV for the 260 nm-thick MAPI film, much smaller than $\Delta V_{\mathrm{OC}}^{\mathrm{PR}} = 49.1$ mV (39.8 mV) deduced using an underestimated $\bar{p}_{\mathrm{e}}$ by Braly *et al.* (Brenes *et al.*) (see Table 1 and compare Figure S2). More critically, $Q_{\mathrm{e}}^{\mathrm{lum}}$ of state-of-the-art PSCs is still severely limited by interfacial recombination[3,56,68] and parasitic absorption losses.[2,14,30,68] Even considering the highest reported $Q_{\mathrm{e}}^{\mathrm{lum}}$ of 5% for a complete PSC to date[69] and neglecting parasitic absorption, $\bar{p}_{\mathrm{e}} =$ 25.4% relates to $Q_{\mathrm{i}}^{\mathrm{lum}} = 18.9\%$ and $\Delta V_{\mathrm{OC}}^{\mathrm{PR}} = 4.2$ mV, revealing that the effect of PR on enhancing the PCE of state-of-the-art PSCs is still small. We analyze the dependence of $\Delta V_{\mathrm{OC}}^{\mathrm{PR}}$ on $Q_{\mathrm{e}}^{\mathrm{lum}}$ as well as of the reduction in $V_{\mathrm{OC}}$ due to non-radiative recombination ($\Delta V_{\mathrm{OC}}^{\mathrm{nr}} = -\frac{kT}{q} \ln Q_{\mathrm{e}}^{\mathrm{lum}}$; compare Equation (1)) on $Q_{\mathrm{i}}^{\mathrm{lum}}$ for various realistic values of $\bar{p}_{\mathrm{e}}$ and $\bar{p}_{\mathrm{a}}$ in Figures S27a and S27b, respectively. It shows that, considering the range of escape probabilities found in this study ($\bar{p}_{\mathrm{e}}$ ~20-30%), $\Delta V_{\mathrm{OC}}^{\mathrm{PR}}$ is limited to values around ~30-40 mV even in case $Q_{\mathrm{i}}^{\mathrm{lum}}$ approaches the radiative limit and $\bar{p}_{\mathrm{a}} = 0$ (see also Figure S2). We note while $\bar{p}_{\mathrm{r}}$ and thus $\Delta V_{\mathrm{OC}}^{\mathrm{PR}}$ might in principle be increased by reducing the magnitude of scattering, for example by imprinting[57,58,60] or fabricating perovskite films with a very low surface roughness,[20,30] this strategy is not beneficial for the overall PCE of PSCs, because a reduction of $\bar{p}_{\mathrm{e}}$ always results in a reduced $Q_{\mathrm{e}}^{\mathrm{lum}}$ and thus lower total $V_{\mathrm{OC}}$ (Equation (1); Figures S1 abd S27b).[2,5,8,13,16,70,71] Therefore, $\bar{p}_{\mathrm{e}}$ should always be maximized without comprising absorption and the optoelectronic properties of a PSC, for example by employing nanostructures,[59,63,70,72] even if this is at the expense of PR as recently discussed by Bowman *et al.*[70,71] In fact, our study reveals that scattering-induced outcoupling of initially-trapped PL is an inherent property of state-of-the-art perovskite films that strongly increases $\bar{p}_{\mathrm{e}}$ and therefore $Q_{\mathrm{e}}^{\mathrm{lum}}$ and $V_{\mathrm{OC}}$ (Figures S1 and S27b). At the same time, it limits the effect that PR has on enhancing the PCE of PSCs (Figure

S27a). Therefore, the first step to further enhance the PCE of PSCs is to increase $Q_{\mathrm{i}}^{\mathrm{lum}}$ by minimizing non-radiative recombination,[2,3,16,56,68,73] which is regularly done in recent literature by chemical passivation and interface engineering approaches as well as by employing optimized charge transport layers.[56,74–79] At a stage for which $Q_{\mathrm{i}}^{\mathrm{lum}}$ of PSCs exceeds ~40%, minimizing $\bar{p}_{\mathrm{a}}$ becomes crucial in order to maximize $Q_{\mathrm{e}}^{\mathrm{lum}}$ and $V_{\mathrm{OC}}$ (Figures S27a and S27b).[2,7,13,14,30,68]

In the following, we make use of the scattering properties of perovskite films and propose a new procedure to analyze parasitic absorption in a layer stack and thereby reveal the importance of minimizing any kind of parasitic absorption in a PSC even in case $Q_{\mathrm{i}}^{\mathrm{lum}} \ll 40\%$. To show the concept, we employ a ~130 nm-thick commercial indium-tin-oxide (ITO) layer as used in state-of-the-art PSCs,[79,80] which exhibits low (yet non-negligible) absorption in the spectral range of the MAPI PL emission (Figure S28a). In Figure 4B, we compare the PL spectra (measured with an IS setup) of a 160 nm-thick MAPI film spin-coated either on a glass substrate, on the backside of an ITO-coated glass substrate or directly on top of ITO. We do not expect a strong effect of parasitic absorption on directly emitted PL (that is, $\bar{p}_{\mathrm{e\text{-}d}}$) considering PL emitted within the bottom escape cone only propagates through the ITO layer once, which is expected to result in a relative absorption of only ~3.2% and no peak shift considering the Lambert-Beer law (Figure S28b). However, the presence of ITO in the layer stack in fact strongly blue-shifts the observed PL peak, with the relative shift being much larger in case MAPI is directly spin-coated on top of ITO. This effect is attributed to efficient parasitic absorption of initially-trapped PL by ITO before it can couple out upon scattering. Thus, this parasitic absorption significantly reduces the emitted fraction of scattered PL (that is, $\bar{p}_{\mathrm{e\text{-}s}}$) and thereby the relative PL red-shift as compared to confocal PL measurements (compare Figure 1C). In the case that ITO is at the backside of the substrate, only PL propagating in the substrate mode can be parasitically absorbed, while in the case that MAPI is directly on top of ITO both the waveguide

**Table 1. Results of fits with Equation (12) to experimental PL spectra of MAPI films (see Figure 3) and comparison to literature.**

| | This work | | | Braly *et al.*[27] | Brenes *et al.*[25] |
|---|---|---|---|---|---|
| Setup | IS | IS | IS | IS | IS + fiber-coupled |
| Intensity[a] [Sun] | ~0.84 | ~1.00 | ~0.95 | ~0.90 | ~2.24 |
| $d$ [nm] | 80 | 160 | 260 | 250 | 250 |
| $R_q$ [nm] | $10.7 \pm 0.7$ | $19.0 \pm 0.8$ | $44.0 \pm 1.0$ | ~16 | ~44.5 |
| $Q_{e\text{-max}}^{lum}$ [%] | 21.1% | 41.9% | 47.4% | 36.7%[b]/45.6%[c] | ~50%[d] |
| $P_s$ | 0.005 | 0.005 | 0.005 | - | - |
| $d_{eff}$ [nm] | $88.1 \pm 14.9$ | $185.5 \pm 14.6$ | $241.7 \pm 9.7$ | - | - |
| $z_{av}$ [nm] | $56.3 \pm 2.7$ | $48.7 \pm 2.4$ | $35.8 \pm 2.0$ | - | - |
| $\bar{p}_{e\text{-d}}$ [%] | $17.5 \pm 0.7$ | $13.0 \pm 0.5$ | $6.0 \pm 0.5$ | - | - |
| $\bar{p}_{e\text{-s}}$ [%] | $13.5 \pm 0.3$ | $15.0 \pm 0.5$ | $19.4 \pm 0.8$ | - | - |
| $\bar{p}_e$ [%] | $30.9 \pm 0.9$ | $28.0 \pm 0.8$ | $25.4 \pm 0.9$ | 7.4% ($\frac{1}{2n^2}$)[e] | 12.7%[e] |
| $a_{PR}/a_{tot}$ [%] | 32 | 52 | 58 | - | - |
| $M_s$ | $1.4 \pm 0.1$ | $3.3 \pm 0.2$ | $7.3 \pm 0.4$ | - | - |
| $Q_i^{lum}$ [%] | $46.4 \pm 0.7$ | $72.0 \pm 0.6$ | $78.0 \pm 0.5$ | $91.9 \pm 2.7\%$ | ~89% |
| $\Delta V_{OC}^{PR}$ [mV] | 10.0 | 19.0 | 22.5 | 49.1 | 38.7 |

$d$, film thickness; $R_q$, rms roughness (Figure S4); $Q_{e\text{-max}}^{lum}$, champion value of $Q_e^{lum}$
$Q_i^{lum}$, calculated via Equation (3) with $\bar{p}_e$ and $Q_{e\text{-max}}^{lum}$ as input; $\Delta V_{OC}^{PR}$, calculated via Equation (4) assuming $\bar{p}_r = 1 - \bar{p}_e$; $M_s = \frac{d}{z_{av}}$, magnitude of scattering; $a_{PR}/a_{tot}$, fraction of recycled PL to the external emission; All other parameters, see Equations (7) and (12).

[a]Intensity calculated as described in Note S7
[b]Experimental value measured in IS
[c]Indirectly determined by fitting
[d]PL measured by a front collection setup, $Q_{e\text{-max}}^{lum}$ extrapolated from IS measurement
[e]Assumption for the value of $\bar{p}_e$ used to determine $Q_i^{lum}$ (see Note S3)

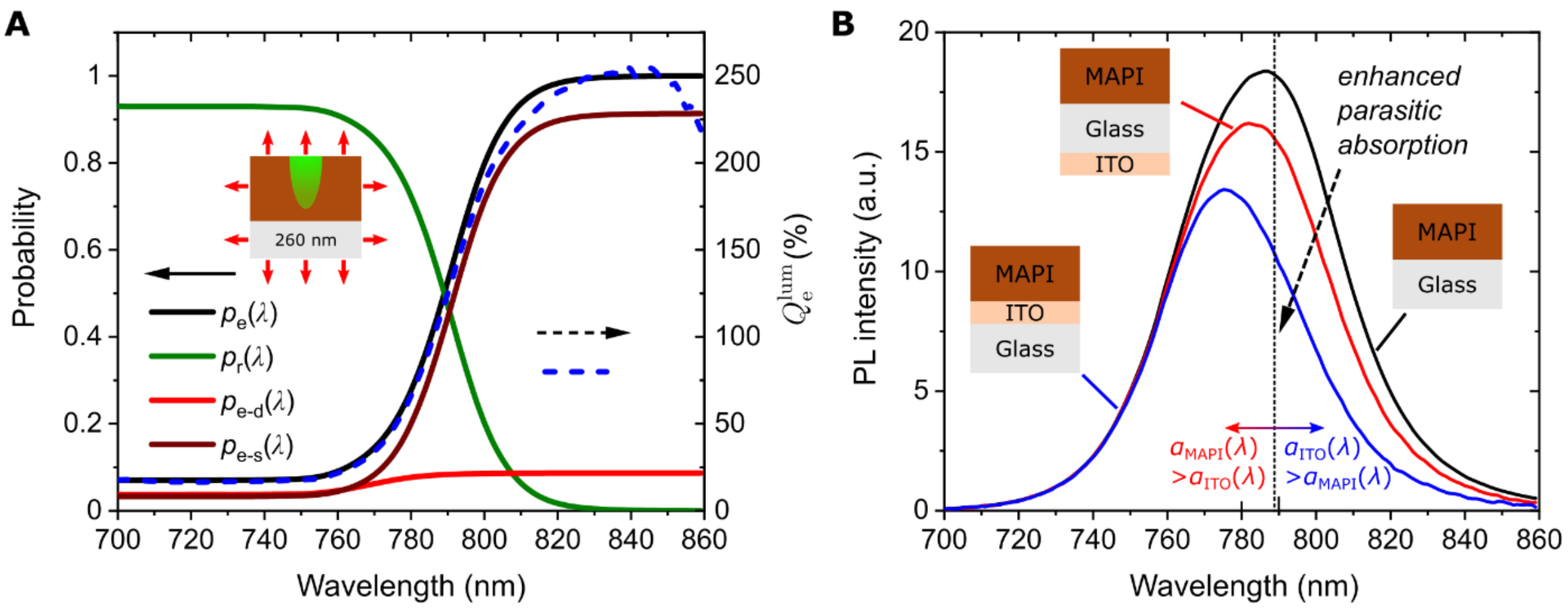

**Figure 4. The effects of photon recycling and parasitic absorption. (A)** Wavelength-dependent escape and reabsorption probabilities (left y-axis; see Equations (10) and (11), $p_e(\lambda) = p_{e\text{-}d}(\lambda) + p_{e\text{-}s}(\lambda)$, $p_r(\lambda) = 1 - p_e(\lambda)$) and external luminescence quantum efficiency (right y-axis; $Q_e^{lum}(\lambda) = \frac{E_{tot}^{exp}(\lambda)}{I_0^{fit}(\lambda)}$) for the 260 nm-thick $CH_3NH_3PbI_3$ (MAPI) film with the champion $Q_e^{lum}$ of 47.4% (see results for other thicknesses in Figure S20). **(B)** PL spectra measured with an IS setup of a 160-nm thick MAPI film spin-coated either on glass (black), on the backside of ITO-coated glass (Lumtec, 15 Ω/sq) or directly on top of ITO (blue) normalized to the values at 730 nm. The reduction in the red-shift compared to the confocal PL spectra (see Figure 1C) is attributed to parasitic absorption by ITO given that an increasing fraction of initially-trapped PL is parasitically absorbed before coupling out upon scattering (from black to blue; see discussion in main text). The dotted line denotes the wavelength at which the absorption coefficient of ITO ($\alpha_{ITO}(\lambda)$) dominates over that of MAPI ($\alpha_{MAPI}(\lambda)$) (see Figure S28a).

mode (via evanescent coupling) and substrate mode (via Lambert-Beer like absorption; Figure S28b) are affected (compare Figure S16); this explains the difference between the PL shifts for these two cases. This effect (which has been observed, but was not discussed in previous literature; examples in Figure S29) primarily increases $\bar{p}_a$ by reducing $\bar{p}_e$ and thereby lowers $Q_e^{lum}$; critically, this has a severe effect on $V_{OC}$ even in case of low $Q_i^{lum}$ and thus reduces the PCE state-of-the-art PSCs (Figures S1 and S28b).[2,5,13,14,68] In addition, $\bar{p}_r = 1 - \bar{p}_e - \bar{p}_a$ can be expected to be slightly reduced given that potentially less initially-trapped photons have a chance to be reabsorbed by the perovskite; this secondary effect, however, could only have a noticeable effect on $\Delta V_{\mathrm{OC}}^{\mathrm{PR}}$ in case of very high $Q_{\mathrm{i}}^{\mathrm{lum}}$ (Figures S2 and S28a) and thus can be neglected as potential cause for a reduced PCE of state-of-the-art PSCs. Such parasitic absorption can for example be reduced by employing alternative transparent electrodes with reduced near-infrared absorption such as indium-zinc-oxide or hydrogen-doped indium-oxide,[81] as well as by using either non-absorbing or very thin charge transport layers such as self-assembled monolayers.[79]

While it is beyond the scope of the present work, accurately quantifying the values of $\bar{p}_{\mathrm{e}}$, $\bar{p}_{\mathrm{a}}$, $Q_{\mathrm{i}}^{\mathrm{lum}}$ and $\Delta V_{\mathrm{OC}}^{\mathrm{PR}}$ for complete PSCs with various perovskite compositions, scattering properties and transport/electrode layers - by implementing the above described effects of parasitic absorption in our curve fitting model - is goal of ongoing work. Such detailed understanding of these 'hidden' internal parameters will be a crucial step on the way of PSCs towards the radiative efficiency limit.

## Conclusion

In this work, we establish a curve fitting model that allows accurate determination of the effective photon escape probability ($\bar{p}_{\mathrm{e}}$) and thereby of the internal luminescence quantum efficiency ($Q_{\mathrm{i}}^{\mathrm{lum}}$) of perovskite films. We reveal that it is of paramount importance to account for initially-trapped PL that scatters out of the films before being reabsorbed, confirming that the large variety of PL spectral shapes reported in the literature for perovskite films can be explained by simply considering light propagation and scattering. To date, not accounting for scattering has led to a significant underestimation of $\bar{p}_{\mathrm{e}}$ by more than 10% in absolute terms and thus to an overestimation of $Q_{\mathrm{i}}^{\mathrm{lum}}$ and the effect that PR has on enhancing the PCE of PSCs. Applying our model to $CH_3NH_3PbI_3$ films with exceptionally high external luminescence quantum efficiencies up to 47.4% sets a real benchmark for $Q_{\mathrm{i}}^{\mathrm{lum}}$ at $78.0 \pm 0.5\%$, revealing there is beyond a factor of two more scope for minimizing non-radiative recombination in perovskite films than previously thought. Finally, we propose a new procedure to analyze parasitic absorption in a layer stack and highlight the importance of further enhancing $Q_\mathrm{i}^\mathrm{lum}$ and minimizing parasitic absorption in complete PSCs. This work provides important new insights into the photophysics of perovskite films and establishes a novel method to accurately determine $Q_\mathrm{i}^\mathrm{lum}$ that will help develop PSCs towards the radiative efficiency limit.

## Experimental Procedures

### Fabrication of perovskite films

$CH_3NH_3PbI_3$ (MAPI) perovskite layers were prepared in dry air (<1% relative humidity) using a lead acetate trihydrate ($Pb(OAc)_2$, Sigma Aldrich, 99.999%) precursor with the addition of hypophosphorous acid (HPA, Sigma Aldrich, 50wt% in $H_2O$) [40]. In short, methylammonium iodide (MAI, GreatCell Solar) and $Pb(OAc)_2$ in the intended molar ratio and concentration were dissolved in anhydrous N,N-dimethylformamide (DMF, Sigma Aldrich, anhydrous, 99.8%) after which HPA (1.7 µl/100 mg MAI) was added shortly before using the precursor solution. Based upon previous work, an optimized stoichiometry $y$ = MAI : $Pb(OAc)_2$ = 2.99 was employed which yields stable and pin-hole free films with high $Q_\mathrm{e}^\mathrm{lum}$ [38,39]. To fabricate MAPI films with thicknesses of 80/125 nm and 160/260 nm, precursor solutions with concentrations of 25 wt% and 40 wt% were spin-coated on cleaned (ultrasonic bath in acetone and isopropanol for 10 min) and oxygen plasma treated microscope slides (16x16 $mm^2$) at 2000/6000 rpm for 60 s, respectively. Directly after spin-coating, a flow of dry air was directed towards the samples for ~30 s before annealing them at 100 °C for 5 min.

Triple-cation perovskite films ($Cs_{0.05}(FA_{0.83}MA_{0.17})_{0.95}Pb(I_{0.83}Br_{0.17})_3$) were prepared by dissolving 1 M formamidinium iodide (FAI, GreatCell Solar), 0.2 M methylammonium bromide (MABr, GreatCell Solar), 0.2 M lead bromide ($PbBr_2$, TCI), and 1.1 M lead iodide ($PbI_2$, TCI) in a 4:1 (v/v) mixture of DMF and dimethyl sulfoxide (DMSO, Sigma Aldrich, anhydrous, ≥99.9%). Afterwards, 42.1 µl of a 1.5 M cesium iodide (CsI, Alpha Aesar) stock solution in DMSO was added in the precursor solution in a 5:95 (v/v) ratio. The perovskite films were spin-coated by a two-step process in a $N_2$ filled glovebox: (1) 1000 rpm for 10 s and (2) 6000 rpm for 20 s. 100 µl CB was poured on the spinning substrate 10 s prior to the end of the second step. Afterwards, the samples were annealed at 100 °C for 60 min in an inert atmosphere.

Double-cation perovskite films ($FA_{1-x}MA_xPb(I_{0.97}Br_{0.03})_3$) were prepared *via* a two-step spin-coating process. A 1.3M $PbI_2$ (Sigma-Aldrich, 99.999%) precursor solution was prepared by dissolving $PbI_2$ in a 9.5:0.5 (v/v) solvent mixture of DMF and DMSO. First, the $PbI_2$ solution was spin-coated at 1500 rpm for 30 s in a $N_2$ filled glovebox followed by annealing at 70 °C for 1 min. In the second step, a solution of 60 mg FAI, 6 mg MABr and 6 mg methylammonium chloride (MACl, Lumtec) in 1 ml IPA was loaded onto the $PbI_2$ thin films and spin-coated at 1300 rpm for 30 s followed by a thermal annealing step at 150 °C for 15 min in ambient conditions (~30–40% humidity).

**Confocal PL measurements**

The samples were excited by the spectrally filtered output of a pulsed supercontinuum laser source (Fianium WhiteLase SC400, 20 MHz repetition rate, ~10 ps pulse width). The excitation wavelength was 530 nm and measurements were performed at various intensities between ~0.4 $W/cm^2$ and ~$4\cdot10^4$ $W/cm^2$ to study the effect of intensity (and thus film temperature) on the PL spectral shape and Urbach Energy. The confocal spectra used for determining $\alpha(\lambda)$ and $I(\lambda)$ (Figure S10) - which are used for Equation (12) - were measured at ~10-50 $W/cm^2$ (~0.5-2.5 $\mu J/cm^2$/pulse). These intensities resulted in a good signal-to-noise ratio and film temperatures around ~305-315 K (compare Figures S7 and S11), which are similar to the temperatures determined from the integrating sphere PL spectra. Considering the much higher irradiation intensities as compared to the integrating sphere measurements, a ~100 nm PMMA layer was spin-coated on top of most MAPI films (40 mg/ml at 4000 rpm without annealing step) to protect the films from quick light and moisture-induced degradation. We verified that the additional PMMA layer has a negligible impact on the measured PL spectral shape (Figure S7). An infinity-corrected 50× objective (Olympus, N.A. 0.7, focal length ~3.0 mm) was used to focus the excitation laser onto the sample with a spot size of ~2 μm. The PL was collected through the same objective and focused onto the entrance slit of an Acton SpectraPro SP2358 grating spectrograph (150 lines mm-1, blaze 1200 nm) by a tube lens with 150 mm focal length (see setup schematic in Figure S6). The following geometric considerations define the field-of-view (FOV) for photon collection from the sample: The detection in the direction parallel to the photodiode array is limited by the entrance slit width of 300 μm, whereas the detection in the direction perpendicular to the photodiode array is limited by the sensor pixel height of 500 μm. Hence, we approximate the detected area in the focus of the tube lens as circular with diameter ~300 μm. Due to the 50× magnification of the microscope ($\frac{f_{\text{tube lens}}}{f_{\text{objective}}} = \frac{150\text{ mm}}{3\text{ mm}} = 50$), this image translates to a FOV of ~6 μm diameter from which photons are collected from the sample. The emission was detected by a liquid nitrogen cooled InGaAs line camera (Princeton Instruments OMA V). Scattered laser light was blocked by a longpass filter with 700 nm cut-on wavelength. A broadband light source (Thorlabs SLS201/M, 300-2600 nm) with known spectral power distribution was used to correct the emission spectra for the wavelength-dependent detection efficiency and absorption due to optical components.

For deriving confocal PL spectra as reported in this work, measurements were performed on ~200 spots on each sample for the respective laser intensity using a XYZ Nano-LP200 piezo-stage and subsequently the data was averaged, smoothed, fitted and extrapolated (see detailed procedure in Figures S7 and S11).

**Integrating sphere measurements**

Integrating sphere measurements were carried out inside an integrating sphere (LabSphere, 15 cm diameter) in ambient air (relative humidity < 40%). For cross-check, measurements were performed at two different setups at Heidelberg University and KIT. A green laser (OBIS 532 nm CW LS 50 mW from Coherent or LD-515-10MG from Roithner Lasertechnik, respectively) was directed into the sphere through a small entrance port (4 mm diameter). An optical fiber was used to collect the emission from the exit port of the integrating sphere and transfer it to the spectrometer (QE65 Pro from Ocean Optics or AvaSpec-ULS2048x64TEC from Avantes, respectively). The spectral response of both setups was calibrated using a calibration lamp (HL-3plus-INT-Cal from Ocean Optics). All raw measured spectra were recalculated to give power spectrums using the integration time. The external luminescence quantum efficiency ($Q_{\mathrm{e}}^{\mathrm{lum}}$) was determined using the method described by de Mello *et al.*[82] The samples were placed at an angle of ~15° with respect to the laser beam to avoid specular reflectance towards the entrance port. The power density absorbed by the sample was calculated based on the laser intensity shone into the sphere (continuously measured by a power meter calibrated towards the power shone into the sphere using a 90/10 beam splitter), the laser spot size (3 $\mathrm{mm}^2$ or 4.5 $\mathrm{mm}^2$, respectively) and the absorptance $A$ of the samples which is determined during the analysis[82] (see Note S7 for more details). Both setups yielded very similar spectral and absolute results. As previously reported by us and others[25,38,83], oxygen (or atmosphere with low levels of humidity) and light can have a positive effect on the value of $Q_{\mathrm{e}}^{\mathrm{lum}}$ for MAPI films. After such a 'treatment', $Q_{\mathrm{e}}^{\mathrm{lum}}$ typically remains at a high value for months even when the samples are subsequently stored in $N_2$. Specifically, our best samples did not exhibit any severe light-soaking effects and showed very high $Q_{\mathrm{e}}^{\mathrm{lum}}$ directly after turning on the laser. Therefore, the champion values of $Q_{\mathrm{e}}^{\mathrm{lum}}$ reported in Table 1 have been measured after a maximum light-soaking time of ~5 min. In general, the measured PL spectral shapes did not exhibit any changes with time or intensity (see Figure S18).

**Monte Carlo method**

The Monte Carlo code for simulating 'externally-observed' PL spectra has been deposited and is freely available at https://github.com/pfassl/FaMi-model. The code in the repository is associated with a GNU General Public License (GPL 3.0). The Monte Carlo approach is discussed in detail in Note S5 and visualized in Figure S19.

**Curve fitting model**

The curve fitting procedure was performed using the curve fitting toolbox in Matlab using the least squares method (lsqcurvefit). A user-friendly Matlab app for performing the fitting (based upon Equation (12)) will be freely available at https://github.com/pfassl/FaMi-model soon. The app in the repository is associated with a GNU General Public License (GPL 3.0).

## Acknowledgements

The authors thank Dmitry Busko for help and fruitful discussions regarding the integrating sphere setup and Mahdi Malekshahi Byranvand for help with fabrication of the two-step double-cation perovskite films. The authors acknowledge the financial supports by the Initiating and Networking funding of the Helmholtz Association (HYIG of Dr. U.W. Paetzold [VH-NG-1148]; Recruitment Initiative of Prof. B.S. Richards; the Helmholtz Energy Materials Foundry (HEMF); PEROSEED [ZT-0024]; Innovationpool); the Helmholtz Association – through the program "Science and Technology of Nanosystems (STN)"; the KIT Young Investigator Network; and the German Federal Ministry of Education and Research (grants: PRINTPERO [03SF0557A]). F.J.B. and J.Z. acknowledge funding from the European Research Council (ERC) under the European Union's Horizon 2020 research and innovation programme (Grant agreement No. 817494 "TRIFECTs"). Y. V. acknowledges funding from the European Research Council under the European Union's Horizon 2020 research and innovation programme (ERC Grant Agreement n° 714067, ENERGYMAPS) as well as the Deutsche Forschungsgemeinschaft (DFG) in the framework of the Special Priority Program (SPP 2196) project PERFECT PVs (#424216076). The authors gratefully acknowledge the help and support of the Karlsruhe School of Optics & Photonics (KSOP) and Max Planck School of Photonics (MPSP), respectively.

## Author contributions

P.F. conceived the initial idea for this study and developed it further with support of U.W.P. and I.A.H. P.F. designed the experiments, fabricated the perovskite films, performed the integrating sphere measurements and analyzed all data. P.F. and V.L. developed the curve fitting model with helpful advice from L.F. and U.W.P. V.L. and P.F. wrote the code for the provided open-source Matlab app. I.A.H. initiated the idea for conducting Monte Carlo simulation and wrote the Matlab code. F.B. and P.F. performed and J.Z. supervised the work on the confocal PL microscopy measurements. I.A.H., Y.V., B.S.R and U.W.P. supervised the whole project. P.F. and U.W.P drafted the manuscript with support of I.A.H and all authors reviewed and commented on the paper.

## Declaration of Interests

The authors declare no competing interests.

# Supplemental Information

## Revealing the internal luminescence quantum efficiency of perovskite films via accurate quantification of photon recycling

Paul Fassl[1,2,5*], Vincent Lami[5], Felix J. Berger[4], Lukas M. Falk, Bryce S. Richard[1,2], Jana Zaumseil[4], Ian A. Howard[1,2], Yana Vaynzof[3,5], Ulrich W. Paetzold[1,2*]

[1]Institute of Microstructure Technology (IMT), Karlsruhe Institute of Technology, Eggenstein-Leopoldshafen, Germany

[2]Light Technology Institute (LTI), Karlsruhe Institute of Technology, Karlsruhe, Germany

[3]Integrated Center for Applied Physics and Photonic Materials (IAPP) and Centre for Advancing Electronics Dresden (cfaed), Dresden University of Technology, Dresden, Germany

[4]Institute for Physical Chemistry (PCI) and Centre for Advanced Materials (CAM), Heidelberg University, Heidelberg, Germany

[5]Kirchhoff-Institute for Physics (KIP) and Centre for Advanced Materials (CAM), Heidelberg University, Heidelberg, Germany

[*]e-mail: paul.fassl@kit.edu; ulrich.paetzold@kit.edu

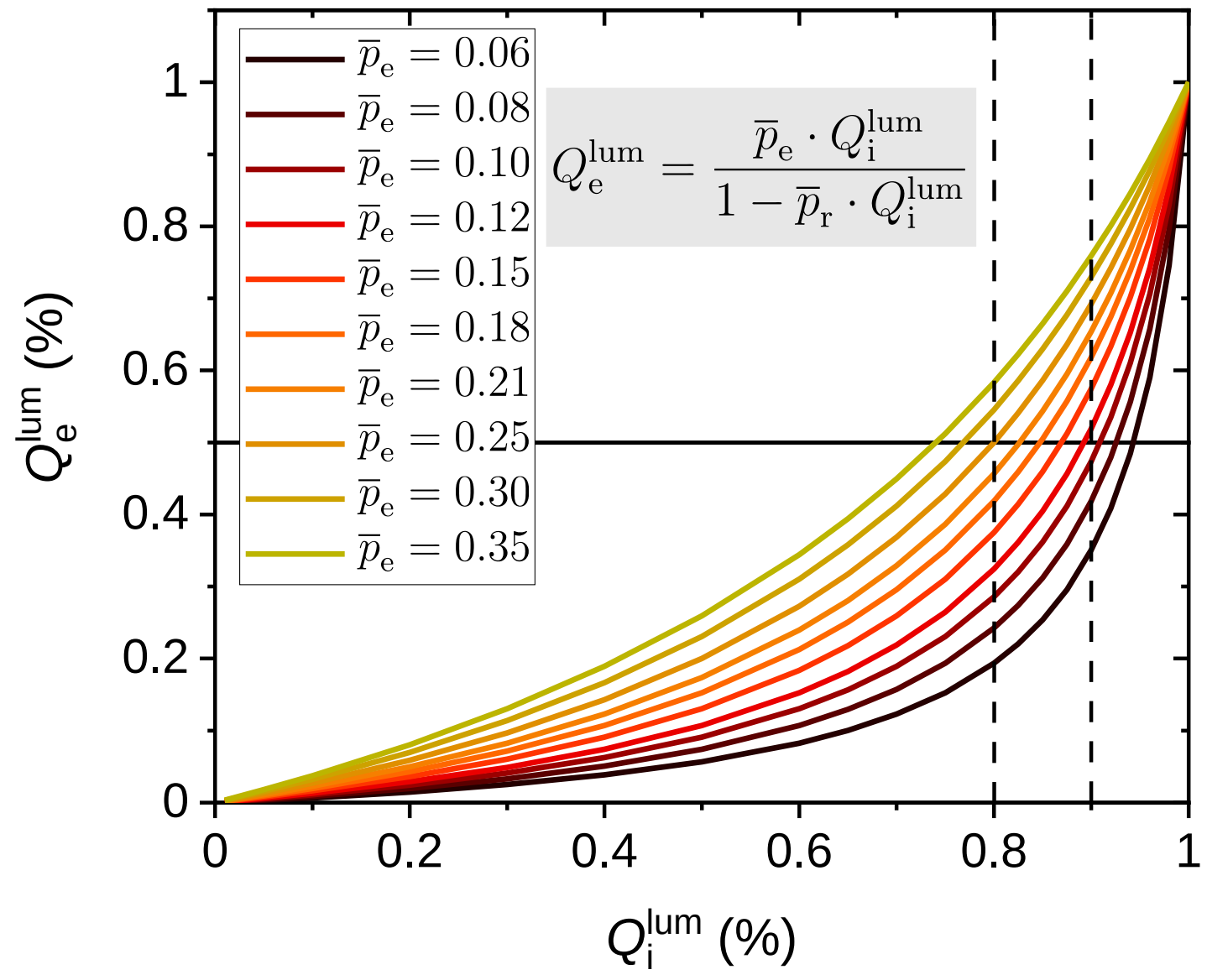

**Figure S1.** Relation of external ($Q_e^{lum}$) to internal ($Q_i^{lum}$) luminescence quantum efficiency considering photon recycling for various photon escape probabilities ($\bar{p}_e$) and neglecting parasitic absorption (reabsorption probability $\bar{p}_r = 1 - \bar{p}_e - \bar{p}_a \xrightarrow{\bar{p}_a = 0} \bar{p}_r = 1 - \bar{p}_e$). The solid and dotted black lines visualize values of $Q_e^{lum} = 50\%$ (around the highest reported value for perovskite films[1–3]) and $Q_i^{lum} = 80\%$ (90%), respectively, and overlap at $\bar{p}_e \approx 25\%$ (11%). This shows that the value of $Q_i^{lum}$ derived from experimentally measured values of $Q_e^{lum}$ (Equation (3) in the main text) sensitively depends on $\bar{p}_e$, emphasizing the paramount importance of its accurate determination. For example, to reach $Q_i^{lum} =$ 90% (50%) in case $\bar{p}_e = 25\%$, a high $Q_e^{lum} = 69.2\%$ (20%) is necessary, while for $\bar{p}_e = 10\%$ it is $Q_e^{lum} =$ 47.4% (9.1%).

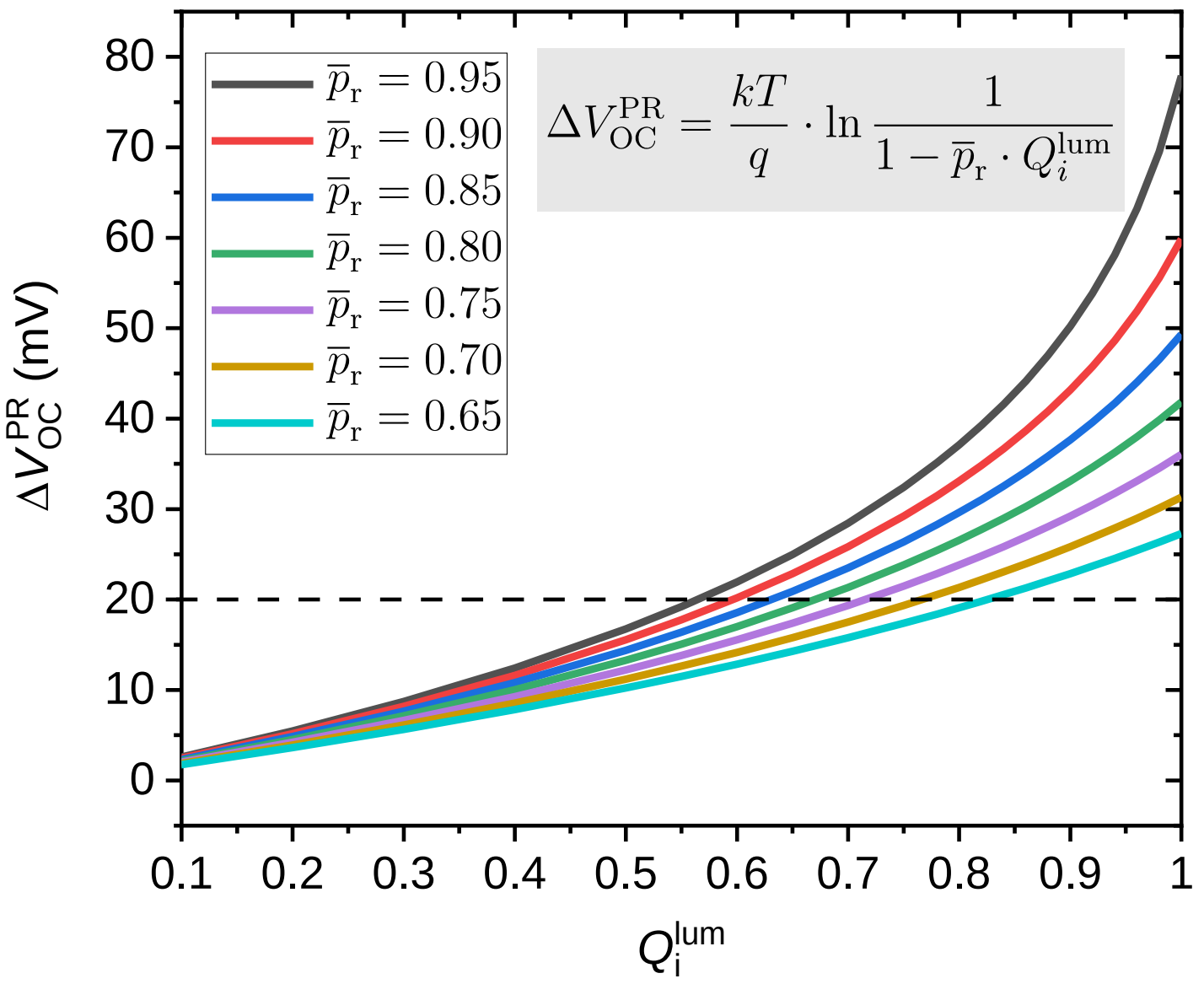

**Figure S2.** Increase of the open-circuit voltage due to photon recycling ($\Delta V_{OC}^{PR}$) over internal luminescence quantum efficiency ($Q_i^{lum}$) for various photon reabsorption probabilities ($\bar{p}_r = 1 - \bar{p}_e - \bar{p}_a$) as derived by Abebe *et al.*[4] For low $Q_i^{lum} \lesssim 40\%$, $\Delta V_{OC}^{PR}$ is low and similar for all $\bar{p}_r$, while for higher $Q_i^{lum}$

the value of $\Delta V_{OC}^{PR}$ more sensitively depends on $\bar{p}_r$. The dotted line visualizes a 'noticeable' value of $\Delta V_{OC}^{PR} = 20$ mV.

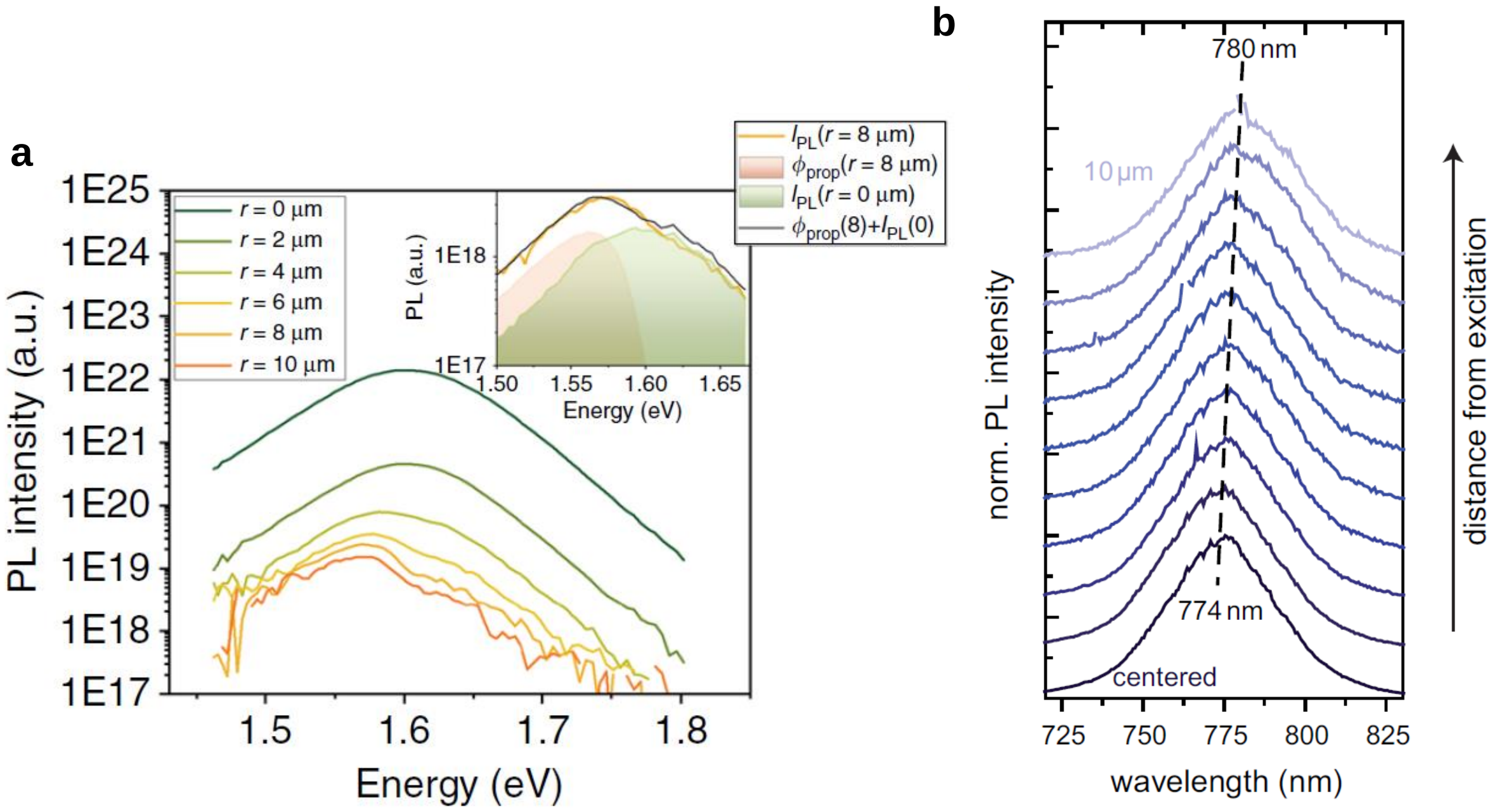

**Figure S3.** Confocal PL measurements reported by **a,** Bercegol *et al*. (Reproduced with permission from [5]. Copyright 2019, Nature Publishing Group) and **b,** Ciesielski *et al*. (Reprinted with permission from [6]. Copyright 2018, American Chemical Society). In both cases, the emission was collected at various distances (0-10 µm) close to the small (~< 1 µm) illumination spot. With increasing distance, the detected PL spectral shape exhibits a slight overall red-shift and/or broadening with the appearance of a 'shoulder' at the low-energy side. Both studies attributed this observation to initially-trapped PL that scatters out of the films towards the front detector. Bercegol *et al*. modelled the observed spectral shapes by a linear combination of reabsorbed/reemitted PL (that is, photon recycling, with the spectrum being equal to that of the direct emission at $r = 0$ µm due to phonon-assisted upconversion[7]) and initially-trapped PL that scatters out of the film (see discussion in the main text).

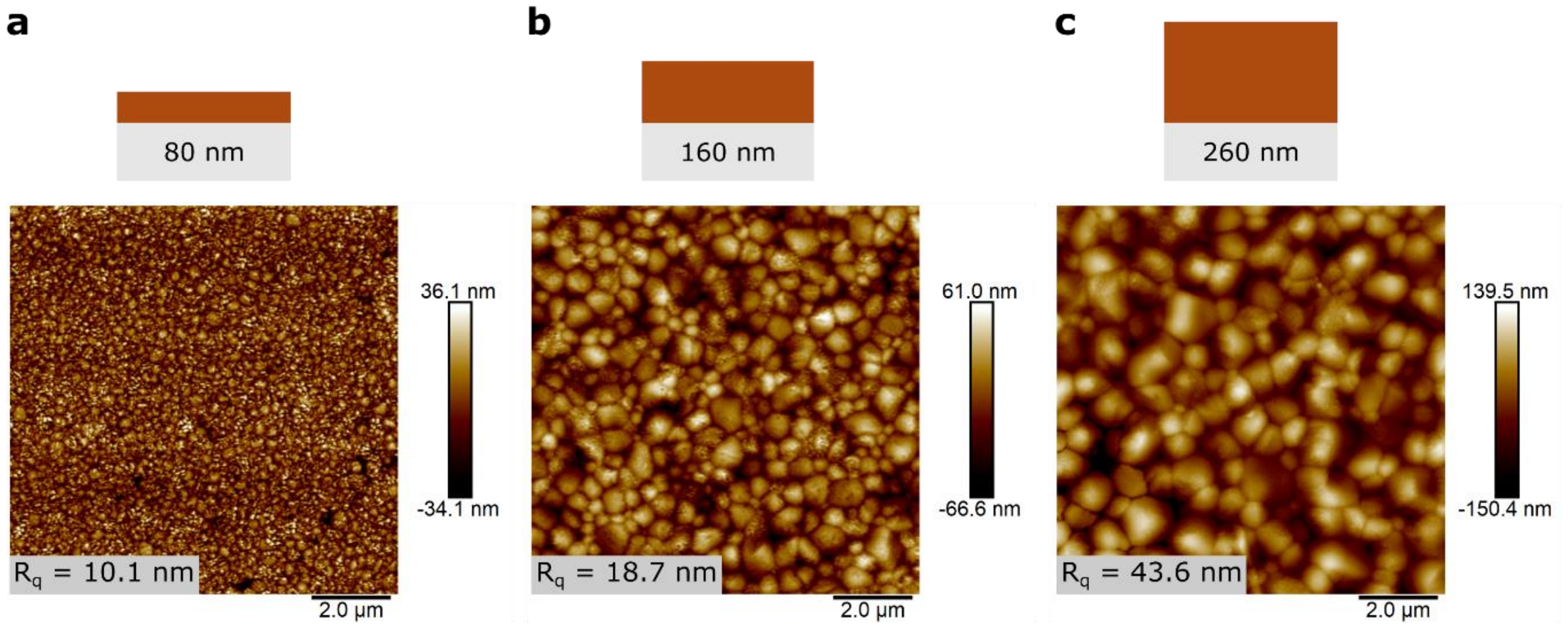

**Figure S4. Representative atomic force microscopy (AFM) images of bare MAPI films with different thicknesses on glass. a**, 80 nm-thick film with a root-mean-square roughness ($R_q$) of 10.1 nm

(average (10.7 ± 0.7) nm). **b**, 160 nm-thick film with $R_q$ = 18.7 nm (average (19 ± 0.8) nm). **c**, 260 nm-thick film with $R_q$ = 43.6 nm (average (44 ± 1.1) nm).

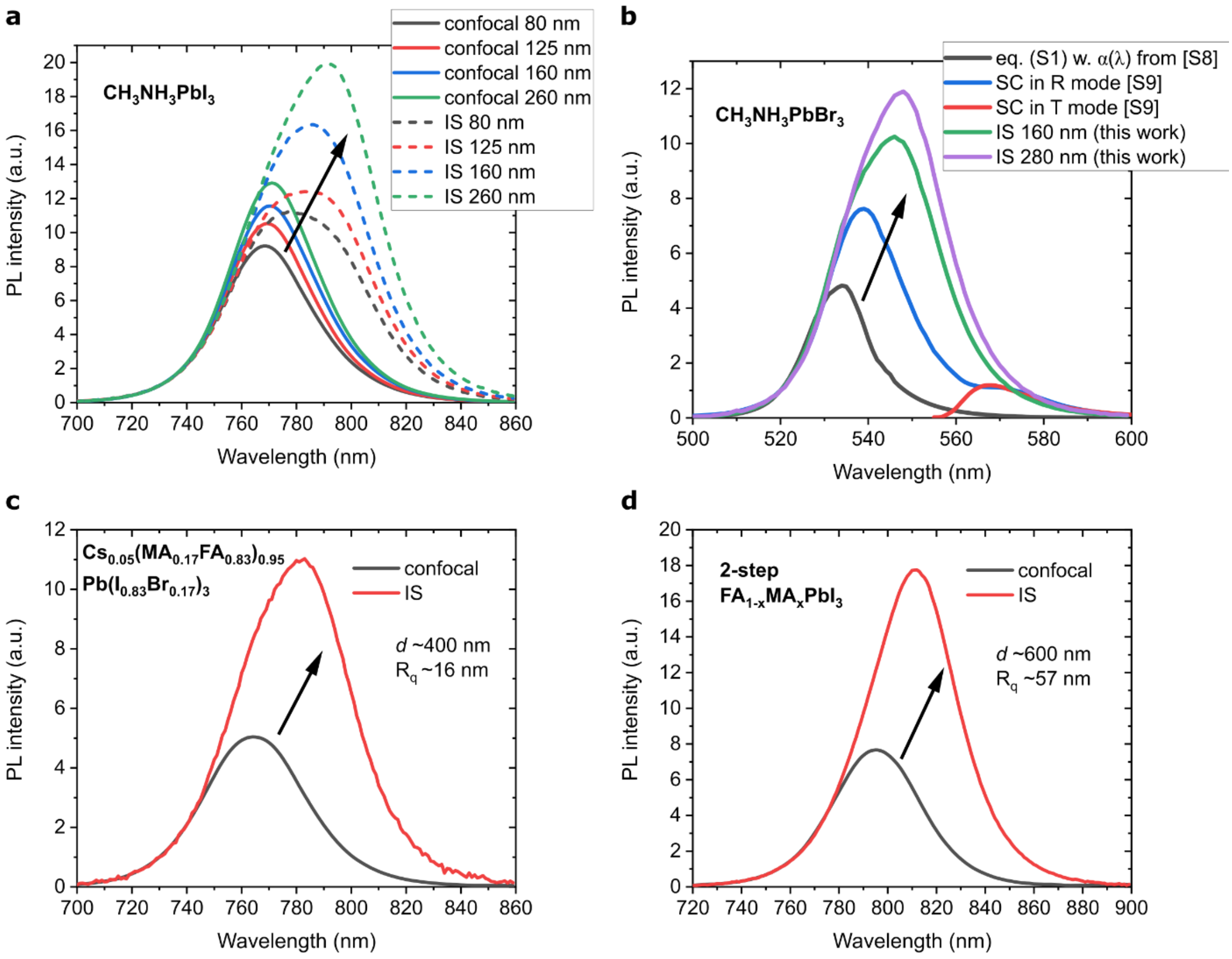

**Figure S5. Generality of the results in Figure 1 for various perovskite compositions. a**, PL spectra of $CH_3NH_3PbI_3$ films with four different thicknesses normalized to the value at 730 nm measured with a confocal microscope (solid lines) and with an integrating sphere (IS) setup (dashed lines) in this work. **b**, Internal PL spectra calculated with Equation (S2) using absorption efficient spectra ($\alpha(E)$) of $CH_3NH_3PbBr_3$ perovskite reported by Wenger *at al*. (black)[8]; PL spectra of $CH_3NH_3PbBr_3$ single crystals (SC) measured in reflection (R) mode (blue) and transmission (T) mode (red) by Fang *et al*.[9]; IS PL spectra of $CH_3NH_3PbBr_3$ films with two different thicknesses measured in this work. All spectra (apart of the T-mode spectra) are normalized to the value at 520 nm. **c**, Confocal and IS PL spectra of a ~400 nm-thick triple-cation ($Cs_{0.05}(MA_{0.17}FA_{0.83})_{0.95}Pb(I_{0.83}Br_{0.17})_3$) thin film with a surface roughness of $R_q$ ~16 nm measured in this work and normalized to the value at 730 nm. **d**, Confocal and IS PL spectra for a ~600 nm-thick double cation ($FA_{1-x}MA_xPb(I_{0.97}Br_{0.03})_3$) thin film with a surface roughness of $R_q$ ~57 nm prepared *via* the two-step method measured in this work and normalized to the value at 750 nm[10]. The arrows denote the increasing fraction of initially-trapped scattered PL to the measured signal.

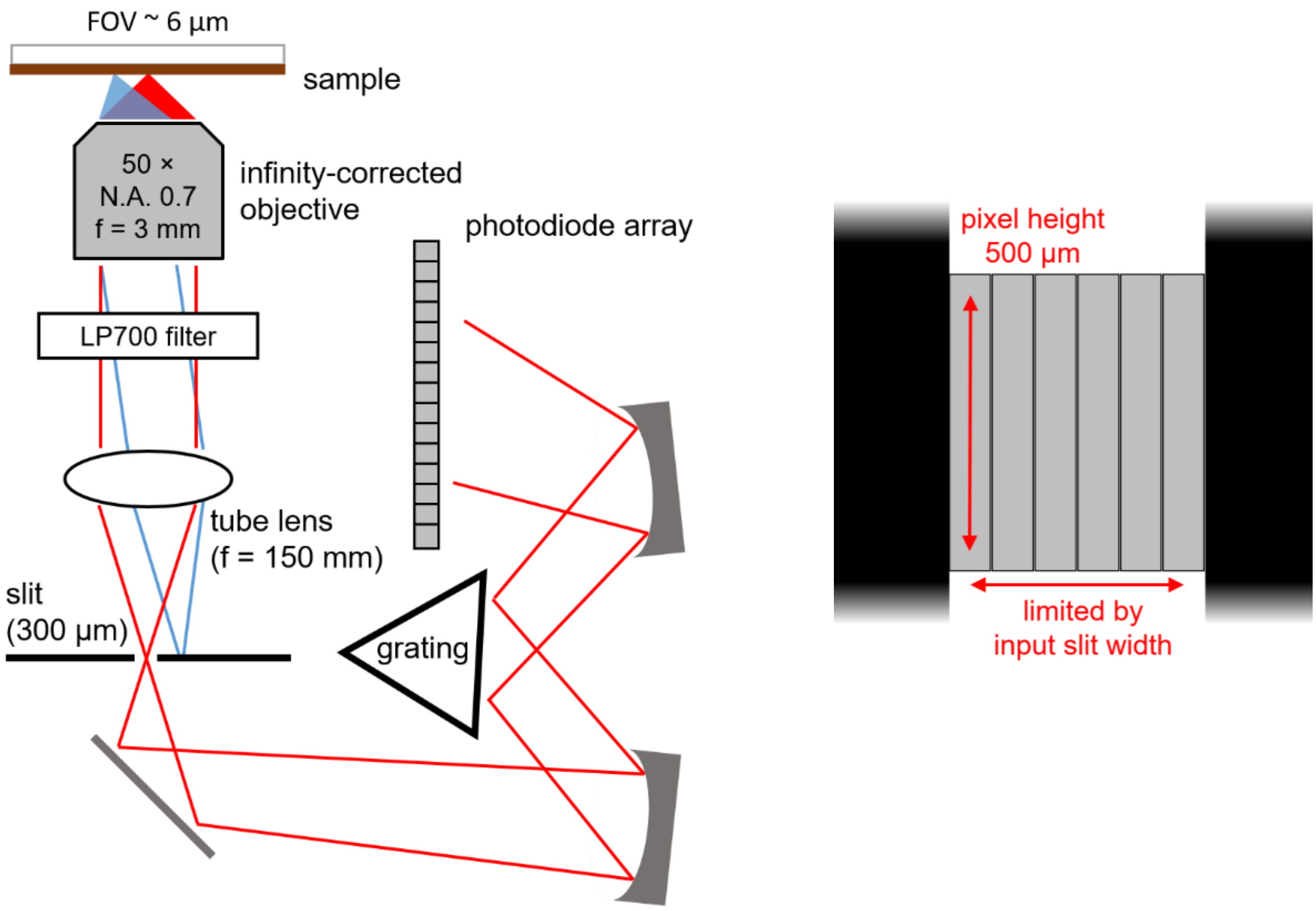

**Figure S6. Schematic of the confocal PL setup employed in this work.** The field-of-view (FOV) from which photons emitted from the sample are detected is limited by the input slit width of the detector in horizontal direction and the pixel height of the photodiode array in vertical direction. We estimate the FOV to be ~6 µm in diameter. More details can be found in the Experimental Procedures.

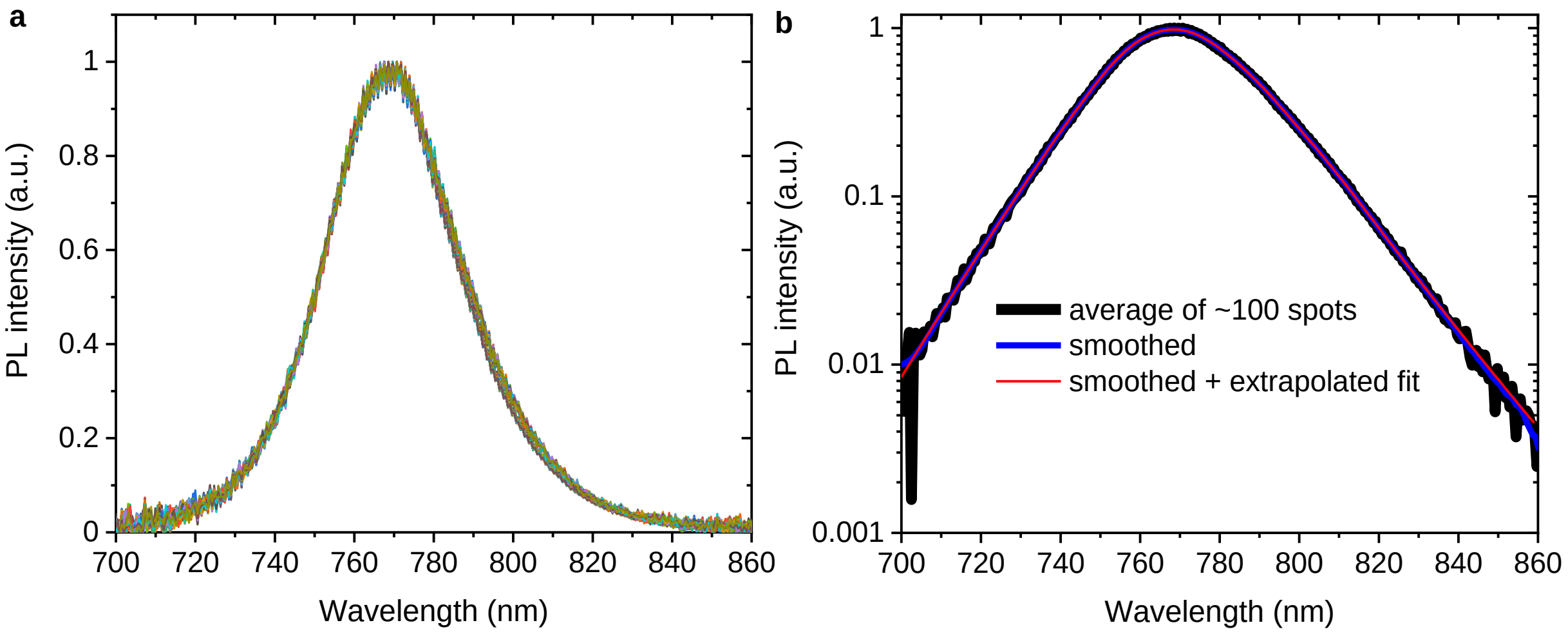

**Figure S7. a,** Representative confocal PL spectra of two 80 nm-thick MAPI films on glass prepared in the same batch and measured at 200 different spots per sample at ~20 W/cm$^2$ intensity (see Experimental Procedures for more details). One of the samples is covered with a 100 nm-thick PMMA film to suppress light- and moisture-induced degradation, revealing that PMMA has a negligible effect on the measured PL spectral shape. **b**, Average of ~200 confocal PL spectra on a 80 nm-thick MAPI film (black) and the corresponding smoothed data (blue). The red curve is a combination of the smoothed data and corresponding fits to the data in the spectral regions around ~730 nm and ~820 nm (see Figure S11 for details regarding the fitting procedure) that are extrapolated to 700 nm and 860 nm, respectively. The smoothed data and extrapolated fits only slightly deviate at very low PL intensities. Considering the smoothed data points around 700 nm and 860 nm are slightly affected by a limited signal-to-noise ratio, the smoothed + extrapolated fit dataset (red curve) is employed in the following to extract the absorption coefficient of MAPI films with different thickness with Equation (S9) or (S11) (see Figures S8 and S10).

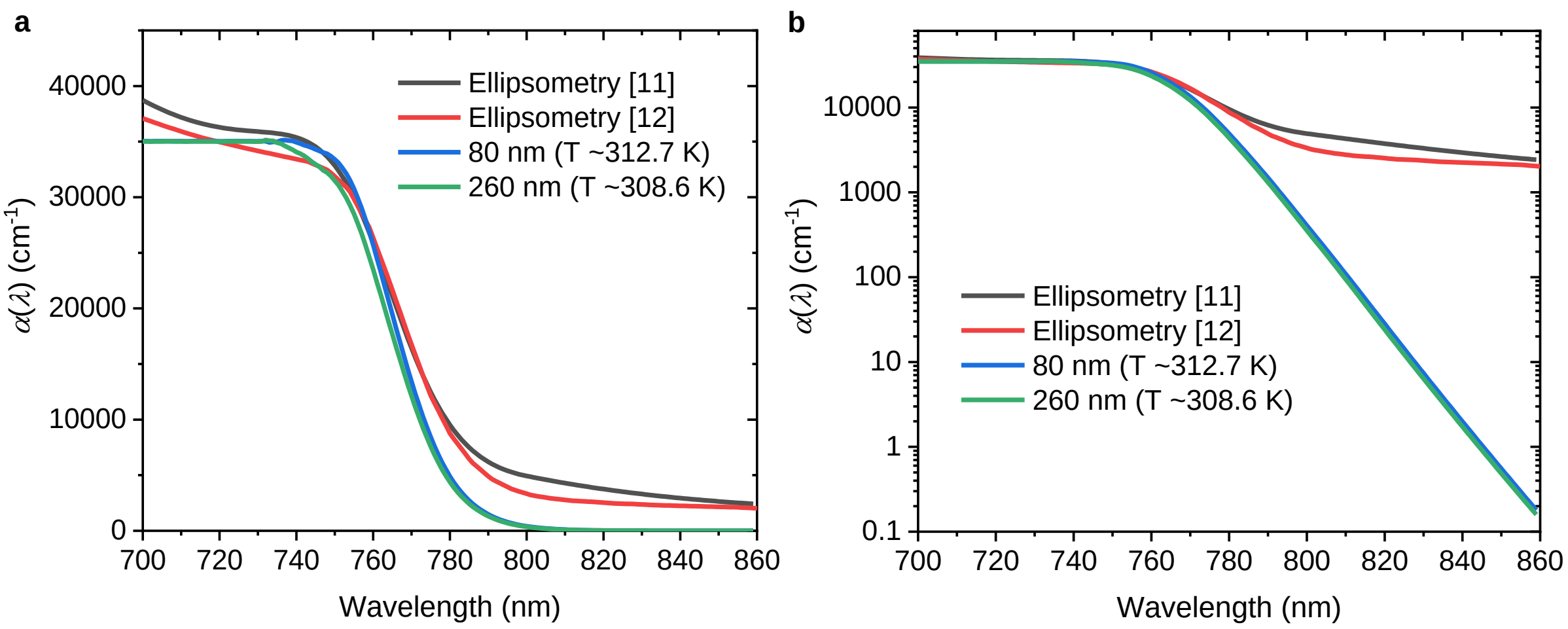

**Figure S8.** Absorption coefficient ($\alpha(\lambda, T)$) in **a,** linear and **b**, logarithmic scale determined from two representative confocal PL spectra of a 80 nm- and 260 nm-thick $CH_3NH_3PbI_3$ film *via* Equation (S11) which are scaled at $\lambda$ = 730 nm to the average of the two $\alpha(\lambda)$-values of high-resolution ellipsometry data by Phillips *et al*.[11] and Soufiani *et al*.[12] (see Note S3 for more details regarding the procedure). Note that no Jacobian transformation must be performed when converting $\alpha(E, T)$ (see Equation (S11)) to $\alpha(\lambda, T)$, that is, only the x-axis should be rescaled.[13]

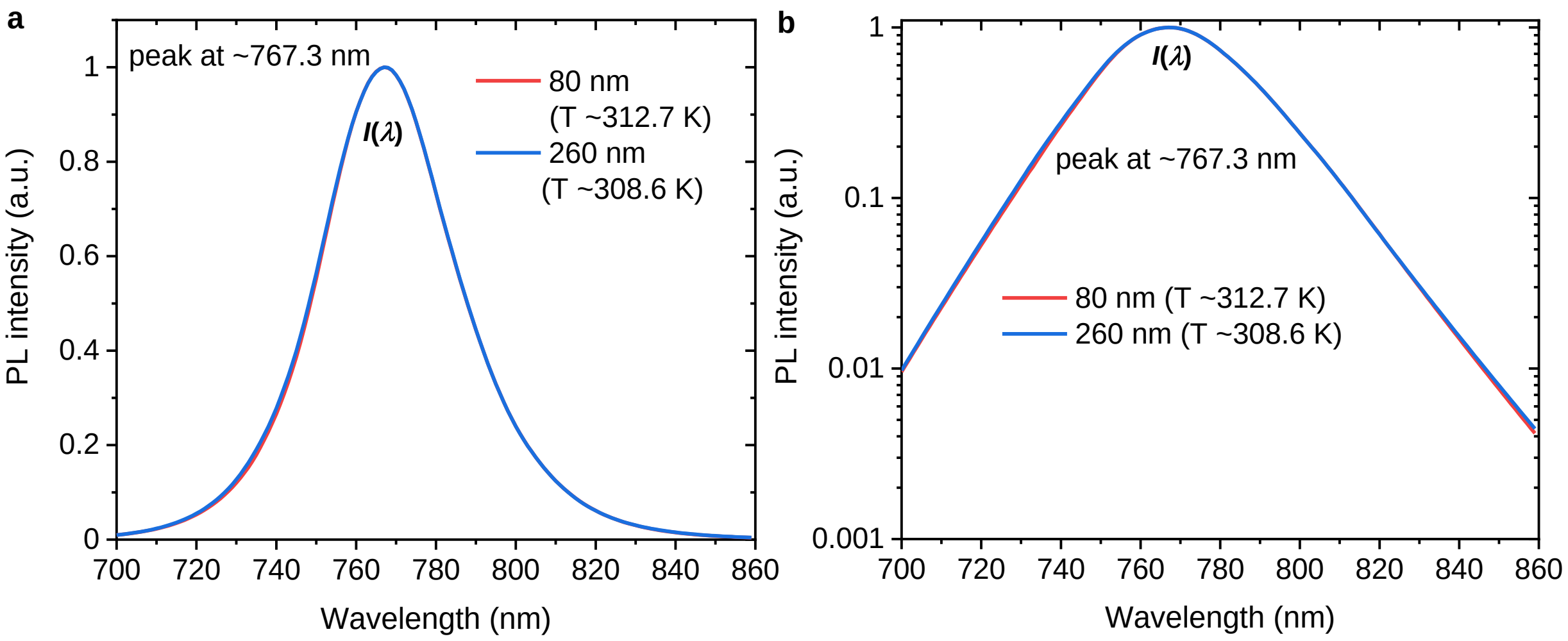

**Figure S9.** Internal emission spectrum ($I(\lambda)$) in **a**, linear and **b**, logarithmic scale calculated *via* Equation (S2) from the exemplary absorption coefficient spectra of $CH_3NH_3PbI_3$ films with two different thicknesses shown in Figure S8 (see Note S3

for more details). We emphasize that for fitting the integrating sphere PL spectra with Equation (12) in the main text, it is important to extract $\alpha(\lambda, T)$ (and subsequently determine $I(\lambda)$) from confocal PL spectra which have a similar slope at the high-energy side of the PL peak (that is, a similar temperature). Considering that the temperatures extracted from the integrating sphere measurements was found to be in the range ~305-315 K (note that in the studied intensity range of ~0.1–3 Sun the measured spectral shape is unchanged, see Figure S18) as well as to provide an error estimation, $I(\lambda)$ was determined from eight confocal spectra of different samples in total, all with extracted temperatures around ~305-315 K (see Figures S10 and S11).

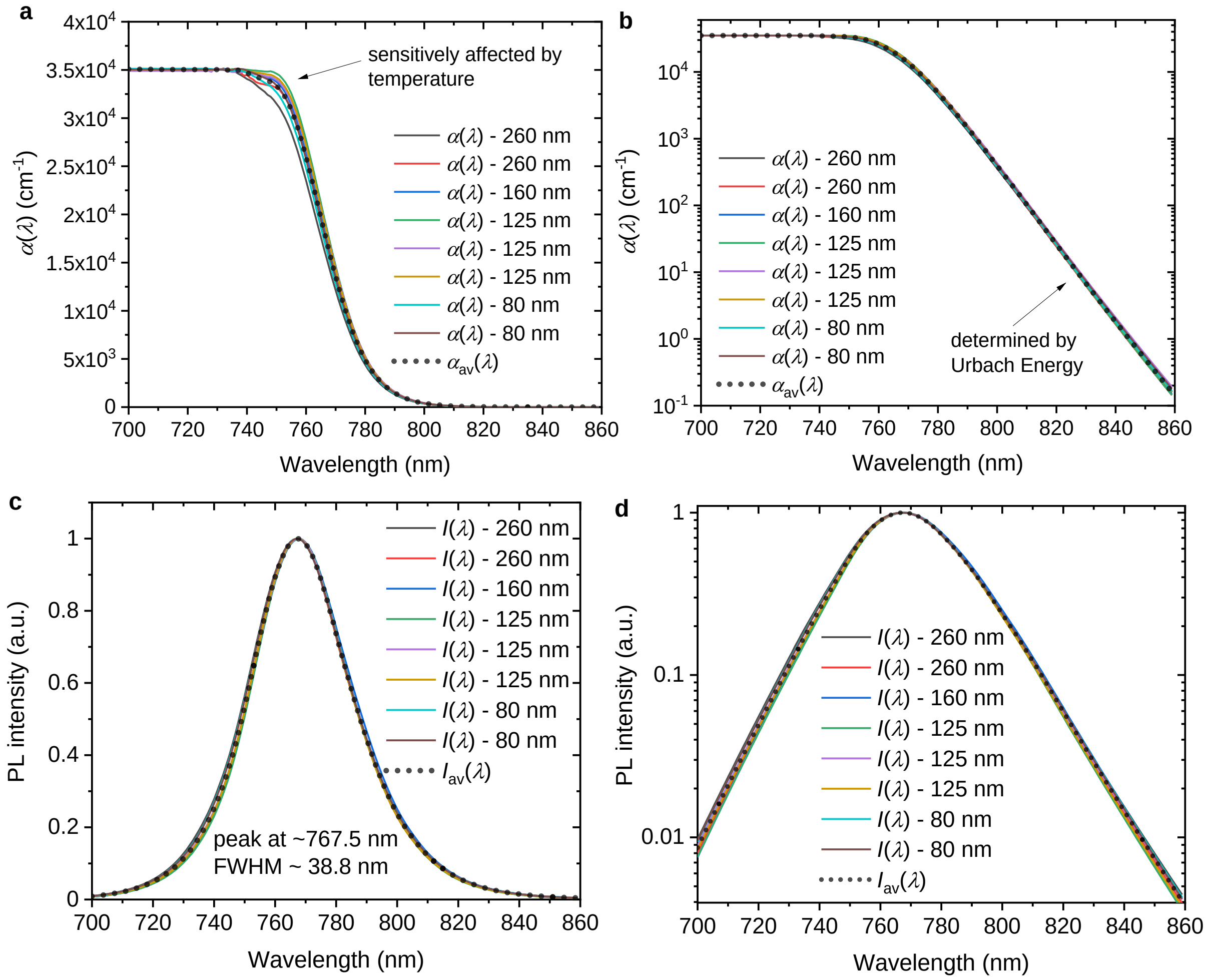

**Figure S10. a** and **b,** Absorption coefficient spectra ($\alpha(\lambda)$) determined from eight different confocal PL (all with extracted sample temperatures ~305-315 K, compare Figures S9 and S11) spectra of MAPI films with various thicknesses *via* Equation (S9) or (S11) (as described in Note S3) together with their average ($\alpha_{\mathrm{av}}(\lambda)$) in **a**, linear and **b**, logarithmic scale. **c** and **d**, The corresponding internal emission spectra ($I(\lambda)$) calculated from the absorption coefficient spectra in **a** and **b** *via* Equation (S2) together with their average ($I_{\mathrm{av}}(\lambda)$) in **c**, linear and **d**, logarithmic scale. The experimental integrating sphere PL spectra (See Figure 3 in main text) are fitted *via* Equation (12) with all eight $\alpha(\lambda)$ and $I(\lambda)$ spectra as input in order to provide an error estimation of the derived parameters (see Table 1 in main text). We compare $\alpha_{\mathrm{av}}(\lambda)$ determined in this work to literature data in Note S1 (Figure S13).

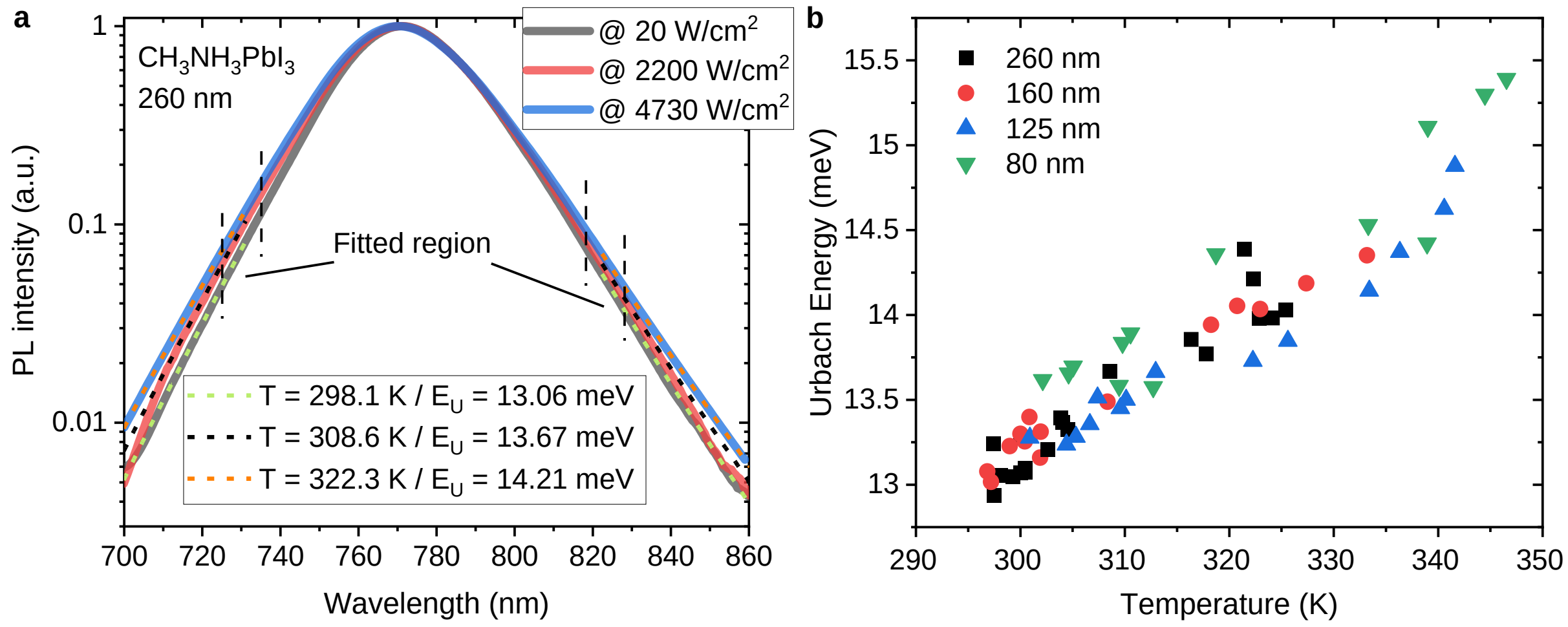

**Figure S11. Fitting procedure to determine the temperature and Urbach Energy and to extrapolate the data. a,** Exemplary fits to confocal PL spectra of a 260 nm-thick $CH_3NH_3PbI_3$ film (average of 200 spots on one sample which has been smoothed; see Figure S7) measured at three different irradiation intensities. First, the high-energy part (where the absorptance $a(E,T)$ is in good approximation constant[14,15]) is fitted by $E_{PL}^{n}(\lambda) \propto \frac{c}{\lambda^4} \cdot (e^{\left(-\frac{1239.85}{\lambda \cdot kT}\right)} - 1)^{-1}$ to extract the temperature (with fitting parameters $c$ and $T$, and Boltzmann constant $k$ and wavelength $\lambda$ (in nm); compare Equation (S7) for which a Jacobian transformation has been performed[16]). Second, to determine the Urbach Energy ($E_U$), the extracted temperature is used as input to fit the low-energy part (where $a(E,T)$ is proportional to the absorption coefficient $\alpha(E,T)$[14,15] which exponentially depends on $E_U$[17]) by $E_{PL}^{n}(\lambda) \propto \frac{c}{\lambda^4} \cdot e^{\left(-\frac{1239.85}{\lambda \cdot E_U}\right)} \cdot (e^{\left(-\frac{1239.85}{\lambda \cdot kT}\right)} - 1)^{-1}$ (with fitting parameters $c$ and $E_U$). The fits are extrapolated to 700 nm and 860 nm, respectively, in order to represent the data in that spectral range (see dotted lines) (compare Figure S7). **b,** Dependence of the Urbach energy of $CH_3NH_3PbI_3$ films on the extracted film temperatures. Measurements were performed for films with various thicknesses, at multiple spots and at various irradiation intensities. The results show that $E_U$ linearly scales with temperature, in line with recent results by Ledinsky *et al.*[18] At room temperature (300 K), we find $E_U = (13.2 \pm 0.1)$ meV.

## Note S1 - Literature analysis

*Variety of reported PL spectral shapes:*

In Figure S12 we show exemplary PL spectra of MAPI films reported in literature for which fabrication is based on a lead acetate precursor[2,6,27–33,19–26]. For simplicity, other fabrication routes are not included to avoid influences of the recipe on the exact bandgap (i.e. PL peak position) and/or film properties, but we stress that the discussed varieties in the reported PL spectral shape are general for perovskite thin films, independent of recipe and composition. The apparent peak position and full-width-half-maximum (FWHM) (sorted by measurement setup) extracted from the spectra in Figure S12 are summarized in Table S1. The peak position ranges from as low as 767.3 nm up to 792.9 nm, and the FWHM from as low as 38 nm up to 59.3 nm. We note that for both, fiber-coupled and Fluorometer setups, we do not exactly know the experimental details, e.g. the probed field-of-view of the sample or if the employed Fluorometer has an integrated integrating sphere for spectral measurements. This analysis, however, clearly show the large variability of the PL peak position, shape and FWHM for very similar, or even exactly the same, MAPI fabrication routes. As discussed in this work, these differences can be explained by simply considering varying film properties (e.g. film thickness, roughness, pin-hole density etc.) and differences in the measurements setups (e.g. probed field-of-view on the sample, front collection setup, integrating sphere etc.), which leads to varying fractions of initially-trapped scattered PL being measured (see Figures 1 and 2 in main text).

Finally, we want to note that differences in (or even non-existing) spectral calibration of PL setups also can result in variations of reported spectral shapes. However, considering that commercial Fluorometers and home-build confocal and integrating sphere setups are typically spectrally calibrated (such information is often not provided), we do not believe that it has a severe effect on the comparison presented in Figure S12.

*Comparison of absorption coefficient spectra reported in the literature with that determined in this work:*

A comparison of reported literature spectra for $\alpha(\lambda)$ with that determined in this work (see $\alpha_{\mathrm{av}}(\lambda)$ in Figure S10) is shown in Figure S13[2,11,40–46,12,17,34–39]. A large variety of spectra is observed, which to a large extent can be attributed to the limited resolution of the employed setups in different spectral regions. For example, ellipsometry has a low resolution for low absorption values, while diffuse reflectance spectroscopy is not suited well for large absorption values. The data extracted in this work is scaled to high-resolution ellipsometry data of MAPI films by Phillips *et al.*[11] and Soufiani *et al.*[12] at 730 nm (see details in Note S3 and Figure S8) and lies 'in the middle' of all other literature data at the high-energy side, while showing one of the steepest slopes at the low-energy side given the very low Urbach Energy of our samples (see Figure S11).

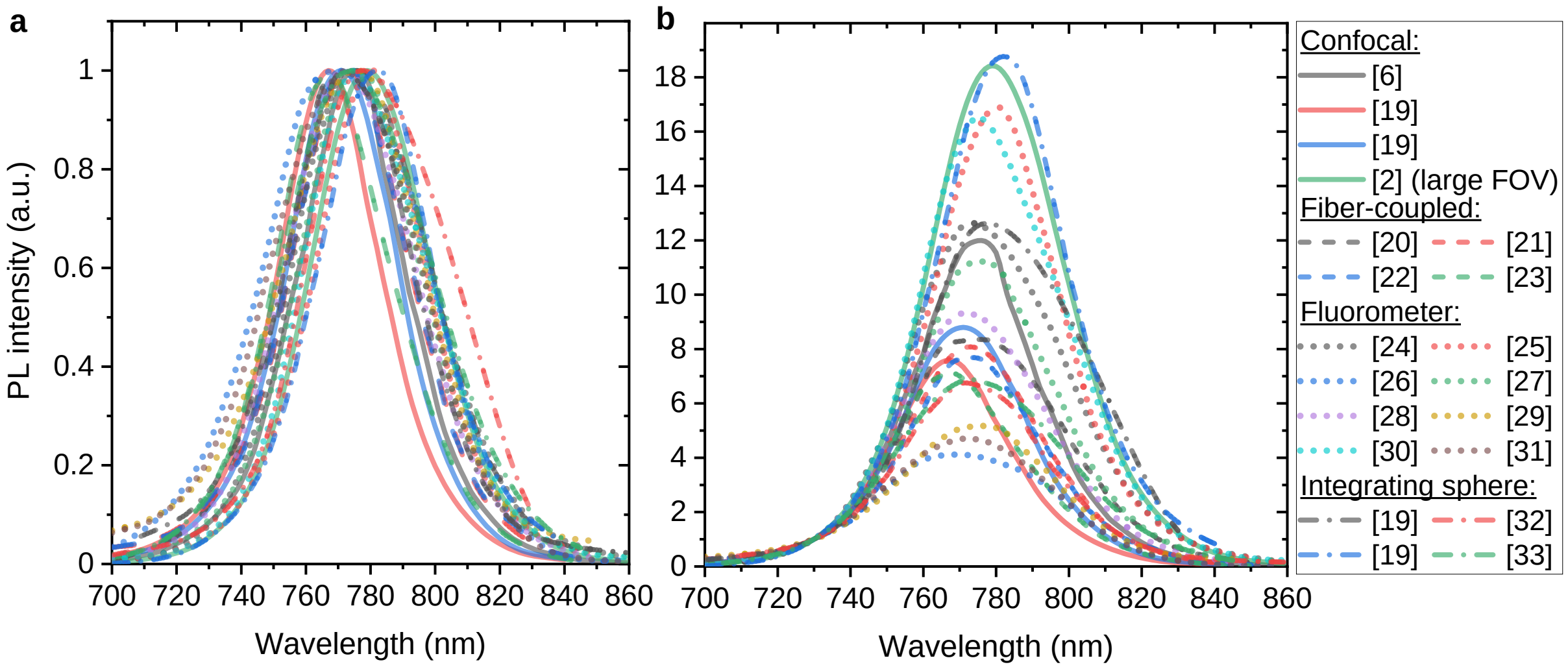

**Figure S12. PL spectra of MAPI reported in literature. a**, Normalized PL spectra of MAPI films (fabrication routes all based on lead acetate as precursor) measured with various PL setups (stated in the legend). The data was extracted using WebPlotDigitizer. **b**, The same peaks as in **a**, but normalized to their value at 730 nm to better highlight the dissimilarities in PL peak shape (compare Figure S5). The apparent peak positions and FWHM of all peaks are summarized in Table S1. Note that most of the peaks cannot be fitted satisfactory with a Gaussian or a Voigt profile (as nearly all peaks are a superposition of directly emitted and scattered PL), so this data was extracted by hand.

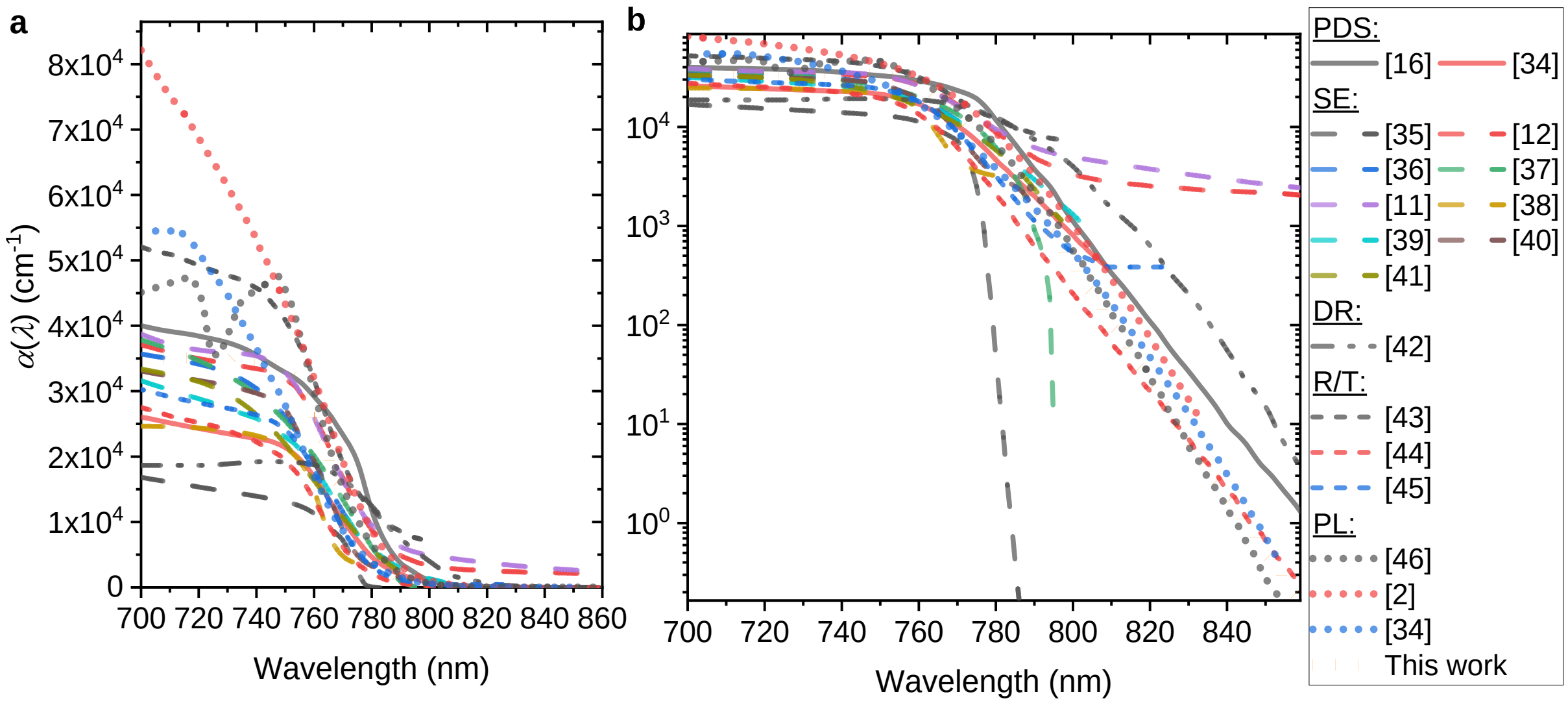

**Figure S13.** Absorption coefficients of MAPI extracted from literature data (using WebPlotDigitizer or original data provided in the manuscripts) in linear (left) and logarithmic (right) scale. The measurement setups used to determine $\alpha(\lambda)$ are stated in the legend: photothermal-deflection spectroscopy (PDS); spectroscopic ellipsometry (SE); diffuse reflectance spectroscopy (DR); reflectance/transmittance measurements (R/T); photoluminescence spectroscopy (PL).

**Table S1.** Measurement setups and data extracted from the PL spectra of MAPI films reported in the literature (all fabricated with lead acetate as precursor) as shown in Figure S12. Any employed passivation/additive is stated.

| Measurement setup | Peak position [nm] | FWHM [nm] | Reference | Passivation/Additive |
|---|---|---|---|---|
| Confocal | 775.4 | 39.8 | [6] | None |
| Confocal | 767.3 | 38 | [19] | None |
| Confocal | 770.9 | 40.4 | [19] | TOPO/None |
| Confocal (large FOV) | 779.2 | 44.1 | [2] | TOPO/None |
| Fiber-coupled | 792.9 | 51.2 | [20] | None |
| Fiber-coupled | 772.9 | 43.8 | [21] | None/HPA |
| Fiber-coupled | 772.9 | 44.2 | [22] | None/HPA |
| Fiber-coupled | 767.2 | 42.9 | [23] | TOPO/None |
| Fluorometer | 773.3 | 49.5 | [24] | None/HPA |
| Fluorometer | 780.4 | 40.7 | [25] | None |
| Fluorometer | 769.0 | 59.3 | [26] | None |
| Fluorometer | 775.5 | 44.6 | [27] | None/MABr |
| Fluorometer | 771.6 | 47.7 | [28] | None/DMSO |
| Fluorometer | 775.8 | 51.4 | [29] | None/PbCl2 |
| Fluorometer | 775.0 | 45.7 | [30] | None/HPA |
| Fluorometer | 772.8 | 54.7 | [31] | None |
| Integrating sphere | 772.5 | 47.0 | [19] | None |
| Integrating sphere | 774.2 | 55.1 | [32] | None/PbCl2 |
| Integrating sphere | 777.5 | 53.5 | [19] | TOPO/None |
| Integrating sphere | 782 | 42.7 | [33] | HPA |

## Note S2 - Comparison of PL spectra measured with different setups

We analyze the absolute photon flux of the 260 nm-thick MAPI film with the highest $Q_e^{lum}$ in this work in Figure S14a and estimate the contribution of edge emission to the total signal by subtracting the measurement with the edges being covered from the measurement of a bare film and estimate that ~50% of the total PL is emitted from the edges (see orange dots). In addition, this spectrum exhibits a strongly asymmetric and red-shifted shape, which entails that edge emission, as expected, is mainly composed of initially-trapped PL and not directly emitted PL (see Figure 2 and Figure S16). In Figure S14b, we compare the absolute photon flux to that of a ~250 nm-thick MAPI film (similar fabrication route) reported by Braly *et al*.[2] measured with an absolute intensity confocal microscope setup.

In Figure S15, we further in detail compare the PL spectral shapes of the 260 nm-thick MAPI film measured with 4 different PL setups (as shown in the insets) with the spectra reported by Braly *et al*.[2] There are two notable differences in the experimental setup in the work by Braly *et al.* which possibly modifies the external spectrum $\boldsymbol{E}(\boldsymbol{\lambda})$ compared to the internal spectrum $\boldsymbol{I}(\boldsymbol{\lambda})$ (compare Figures 1B and 2 in main text and Note S3): (i) the gold back reflector increases the effective path length of photons emitted within the bottom escape cone as they propagate through the perovskite film once more before reaching the detector. This is expected to result in a slight additional red-shift as compared to our confocal setup (setup (1) in Figure S15); (ii) the much larger FOV (~2.2 mm[47]) compared to this study (~6 µm) strongly increases the fraction of scattered PL measured by the detector. The larger fraction of scattered PL is expected to result in a further red-shift as well as broadening of the PL spectra (see Figure 1 in main text).

Indeed, the spectrum of our MAPI film with the edges being covered (setup (3) in Figure S15) is very comparable to their reported data. This analysis clearly indicates that a considerable amount of initially-trapped, scattered PL has been detected in the work by Braly *et al*. which is very comparable to that for our 260 nm-thick MAPI film measured with setups (3) and (4).

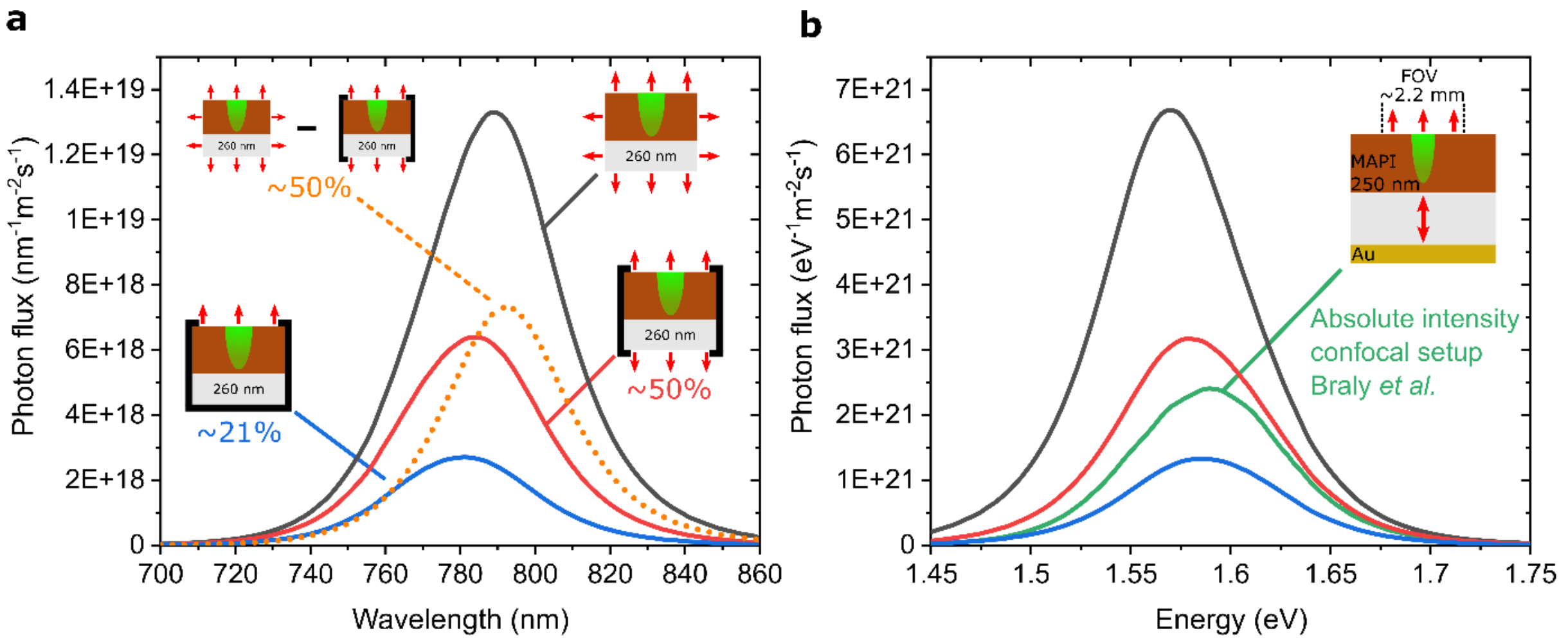

**Figure S14. a**, Absolute photon flux of the 260 nm-thick MAPI film with the highest $Q_e^{lum}$ measured in this work at an intensity close to an AM1.5G equivalent absorbed photon flux (~1 Sun; see Note S7 for details) measured with an IS setup. Data is shown for a bare film (black) and the same sample with the edges (red) or edges and backside (blue) being covered (see insets). Covering the edges reduces the total photon flux by ~50% (~79% for edges and backside), revealing the significant contribution (~50%) of red-shifted PL edge emission (orange dots; estimated by subtracting the red from the black data) for perovskite films on glass. Note that edge emission possibly is slightly overestimated given that the black tape within the integrating sphere will absorb a small fraction of emitted PL. **b**, The same data as in **a** but converted to energy scale by a Jacobian transformation and compared to literature results by Braly *et al.* (green) (data extracted using WebPlotDigitizer)[2]: in that work, absolute intensity PL spectra of MAPI films (also at ~1 Sun) with a similar thickness (~250 nm) were measured with a confocal setup with a large field-of-view (FOV) of ~2.2 mm in order to collect all photons emitted from the front[47]. Both the absolute photon flux ($Q_e^{lum}$ was ~20% in both cases) and the PL spectral shape are very similar. In Figure S15 we provide a comparison of the same four peaks to the confocal spectra measured in this work in wavelength-scale.

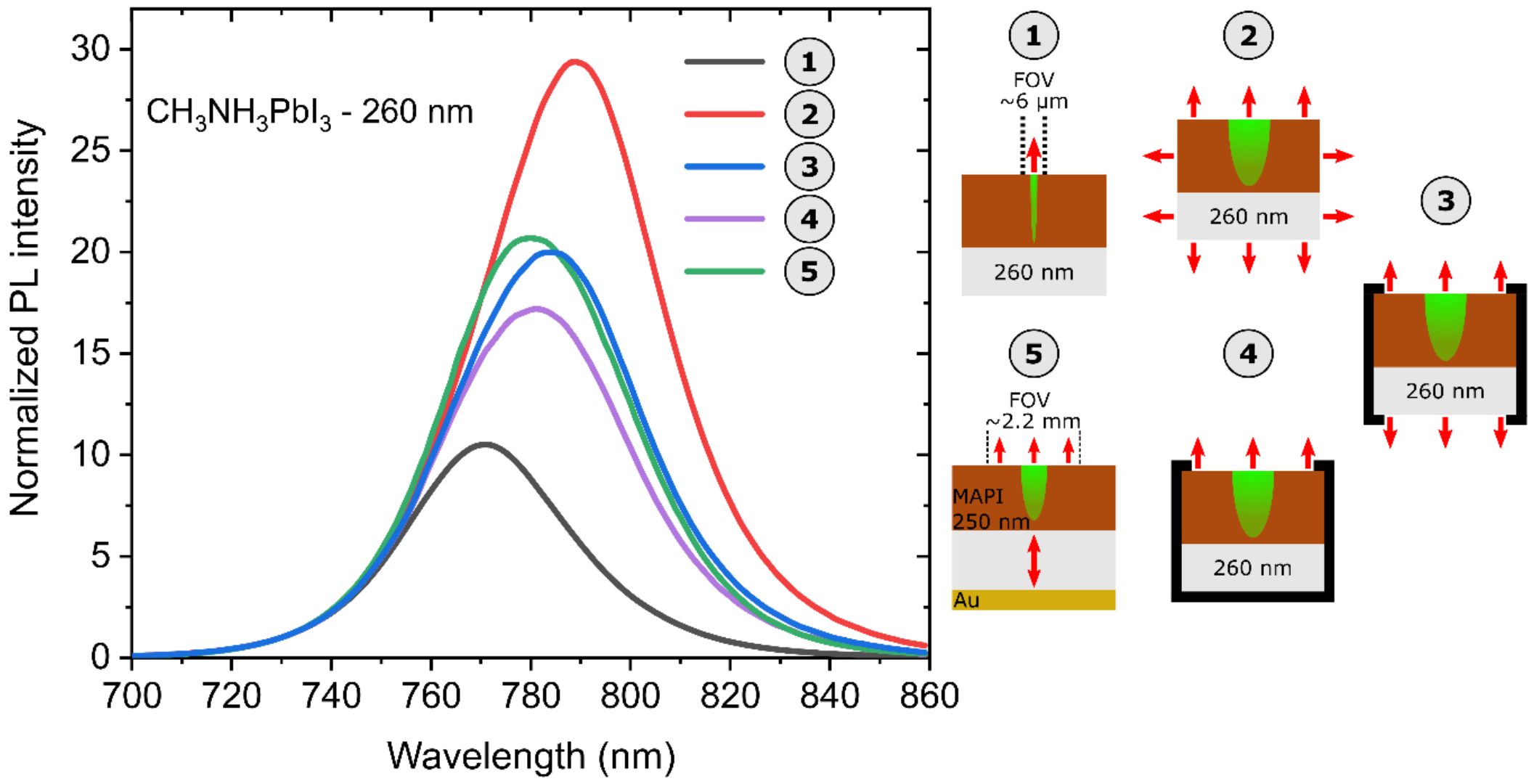

**Figure S15.** Comparison of PL spectra of a 260 nm-thick MAPI film measured in this work with different setups (1 - 4) (see insets on the right) with that of a ~250 nm-thick MAPI film reported by Braly *et al.*[2] (5) normalized to their value at 730 nm: (1) confocal setup with a small field-of-view (FOV) of ~6 µm; (2) standard integrating sphere setup; (3) modified IS setup to observe total front- and bottom emission; (4) modified IS setup to observe total front emission; (5) absolute intensity confocal setup with a large FOV of ~2.2 mm in order to collect all photons emitted from the front[2,47]. There is no large difference between setups (3) and (4), revealing that the reduced red-shift compared to (2) is mainly due to low-energy photons emitted from the edges of the substrate (see Figure S14a). The spectral shape reported by Braly *et al.* is 'in between' that measured with setups (3) and (4), indicating towards a strong and comparable contribution of red-shifted scattered PL in to the measured signal.

## Note S3 - Determination of $\alpha(\lambda)$ and $I(\lambda)$

*PL emission in semiconductor films:*

The following considerations are based on the detailed balance between absorption and emission[48] that holds for non-thermal luminescence in semiconductors as shown by Würfel[49]. The relevant theories are nicely summarized in recent reports by Rau and Kirchartz (refs. [34,50,51]**)** and these citations are not further stated in the following. In general, we assume full radiative recombination and isotropic internal emission of PL within the perovskite films[5,45,52]. Note, while for the sake of clarity we plot PL spectra in wavelength scale, the physically more meaningful energy scale is used for the relevant physical Equations. Note that when converting measured PL spectra from wavelength to energy scale, it is crucial to perform a Jacobian transformation[16], such that $E_{\mathrm{PL}}(E) = E_{\mathrm{PL}}(\lambda) \cdot \frac{\lambda^2}{hc}$.

The internal emission rate of photons within a semiconductor film (per unit volume, per energy interval d$E$) at energy $E$ and temperature $T$ is given by[49,53]:

$$r_{\mathrm{sp}}(E,T) = \left(\frac{2\pi E^2}{h^3c^2}\right) \cdot \frac{n^2(E)\cdot\alpha(E,T)}{e^{\left(\frac{E-\Delta E_F}{kT}\right)}-1} \approx \phi_{\mathrm{bb}}(E,T)\cdot n^2(E)\cdot\alpha(E,T)e^{\left(\frac{\Delta E_F}{kT}\right)} \quad \text{(Equation S1)}$$

Here, $\phi_{\mathrm{bb}}(E,T) = \left(\frac{2\pi E^2}{h^3c^2}\right)\cdot\frac{1}{e^{\left(\frac{E}{kT}\right)}-1}$ is the black body spectrum, $n(E)$ is the refractive index, $\alpha(E,T)$ the absorption coefficient spectrum and $\Delta E_F$ the quasi fermi level splitting of electrons and holes ($V_{\mathrm{OC}} = q\cdot\Delta E_F$ in an ideal solar cell). The approximation made in Equation (S1) is only valid in the case that the energy-range at which PL is emitted is $E - \Delta E_F \geq 3kT$, which is the case for perovskite films.[54] For simplicity of comparison with the externally observed spectra, we define the internal PL spectrum $I_{\mathrm{PL}}(E,T)$ as the spectrum that would be observed in case $n(E) = 1$ (to avoid a large shift in wavelengths, due to refractive index) such that:

$$I_{\mathrm{PL}}(E,T) \stackrel{\text{def}}{=} \phi_{\mathrm{bb}}(E,T)\cdot\alpha(E,T)\cdot e^{\left(\frac{\Delta E_F}{kT}\right)} \quad \text{(Equation S2),}$$

For a homogeneous distribution of electrons and holes, which is a good approximation for perovskite thin films considering that the diffusion length typically extends the film thickness and charges are quickly redistributed by photon reabsoption[44,52], the external PL spectrum of any semiconductor closely follows the generalized Planck's law[15,55]:

$$E_{\mathrm{PL}}(E,T) = \left(\frac{2\pi E^2}{h^3c^2}\right)\cdot\frac{a(E,T)}{e^{\left(\frac{E-\Delta E_F}{kT}\right)}-1} \approx \phi_{\mathrm{bb}}(E)a(E,T)e^{\left(\frac{\Delta E_F}{kT}\right)} \quad \text{(Equation S3),}$$

where the only difference to Equation (S2) is that the absorption coefficient is replaced with the absorptance $a(E,T)$.

Note that we for simplicity, we omit the notation 'PL' in the rest of the manuscript such that $I(\lambda) \stackrel{\text{def}}{=} I_{\mathrm{PL}}(E,T)\cdot\frac{E^2}{hc}$ and $E(\lambda) \stackrel{\text{def}}{=} E_{\mathrm{PL}}(E,T)\cdot\frac{E^2}{hc}$.

*Estimation of the photon escape probability of flat perovskite films:*

**(1) Geometric optics:**

Richter *et al*. estimated $\bar{p}_e$ based on simple geometric optics as summarized in the following (partly quoted from their SI):[56]

The probability for a photon to be transmitted through an interface between two media with refractive indices $n_1$ and $n_2$ was estimated as:[57]

$$p_{\mathrm{trans}} = \frac{n_2^3}{n_1 \cdot (n_1 + n_2)^2}$$

By considering refractive indices of 1, 1.5 and 2.7 for air, glass and MAPI, respectively, the transmission probabilities were estimated to be ~7.0% for the perovskite/glass interface and ~2.7% for the perovskite/air interface. Here, it was for simplicity assumed that photons entering the glass will not get back into the perovskite film but have escaped, which is not really accurate since a certain fraction of these photons will re-enter the perovskite film away from the illumination spot; this results in an overestimation of escape. In addition, photons that are reflected at the perovskite/air interface (propagating once more through the perovskite film) and with an emission angle in between the critical angles of the perovskite/glass and perovskite/air interfaces, were estimated to couple out at the perovskite/glass interface with (7%-2.7%) probability.

The films in Richter *et al.*'s work had an optical density of ~0.05 at the PL emission wavelength. Before reaching an interface, photons thus will on average have travelled through an optical density of ~0.025. Therefore, the total escape probability for MAPI based on the considerations above was estimated as:

$$\bar{p}_e = 10^{-0.025} \cdot \left(7\% + 2.7\% + 10^{-0.05} \cdot (7\% - 2.7\%)\right) = 12.7\%$$

This rough estimation of $\bar{p}_e$, which actually is a strong overestimation as compared to more rigorous considerations for flat films (see below), was used in various follow-up publications[1,3,58,59], for example by Brenes *et al*. (compare Table 1)[1]. Considering a refractive index of 2.6 for MAPI (instead of 2.7), the above estimation yields $\bar{p}_e$ ~14%.

**(2) 'Escape cone probability'**

Yablonovitch *et al*. used a straight-forward approach to estimate $\bar{p}_e$ of optically planar films by considering the amount of photons falling within the top and bottom escape cone via[60,61]:

$$\bar{p}_e = 2 \cdot \frac{1}{4n^2} = \frac{1}{2n^2} = 7.4\%$$

This estimation for $\bar{p}_e$ (assuming 2.6 as refractive index for MAPI) was for example used by Braly *et al*. (compare Table 1)[2].

**(3) Rigorous considerations**

A more rigorous calculation of $\bar{p}_{\mathrm{e}}$ was performed by Rau *et al.*[50]: isotropically emitted internal photons are either externally emitted (with escape probability $\bar{p}_{\mathrm{e}}$), internally reabsorbed (with probability $\bar{p}_{\mathrm{r}}$) or parasitically absorbed (with probability $\bar{p}_{\mathrm{a}}$) such that $1 = \bar{p}_{\mathrm{r}} + \bar{p}_{\mathrm{e}} + \bar{p}_{\mathrm{a}}$ must hold. The escape probability $\bar{p}_{\mathrm{e}}$ of isotropically emitted photons in a semiconductor film (into air) is the ratio of the externally emitted photon flux over the total internally emitted photon flux and can be written considering the absorptance $a(E,T)$ and differences in refractive index via[34,50]

$$\bar{p}_{\mathrm{e}} = \frac{c\int_0^\infty a(E,T)\phi_{\mathrm{bb}}(E,T)\,dE}{4\pi d\int_0^\infty \alpha(E,T)n^2(E)\phi_{\mathrm{bb}}(E,T)\,dE} \quad \text{(S4)},$$

where $d$ is the film thickness and $n^2(E)$ the refractive index of the absorber layer. The factor $c$ is correlated to the étendue and equals $2\pi$ if emission is possible in all directions (for example, a film on a glass substrate) and $\pi$ when emission is limited to the top hemisphere (for example, a solar cell with a perfect back reflector).

Considering $n(E)$ derived by Loeper *et al.*[36] together with $\alpha(E)$ and $a(E,T)$ determined in this work (see details in next sections; Equations (S8) and (S11); Figure (S10)), Equation (S4) yields values for $\bar{p}_{\mathrm{e}}$ of ~6.3%, ~5.9% and ~5.4% for our 80 nm-, 160 nm- and 260 nm-thick MAPI films, respectively.

The issue with the above estimations for $\bar{p}_{\mathrm{e}}$ is that all of them are valid only in case all PL not falling within the escape cone remains trapped in the films with no chance for subsequent escape, that is, it is considered completely reabsorbed. Therefore, while specifically Equation (S4) provides a mathematically correct description for near-atomically flat semiconductors such as GaAs, it cannot describe the effective escape probability $\bar{p}_{\mathrm{e}} = \bar{p}_{\mathrm{e\text{-}d}} + \bar{p}_{\mathrm{e\text{-}s}}$ of polycrystalline perovskite films with a non-negligible surface roughness, which is much higher compared to the above flat film estimations due to scattering of initially-trapped photons as discussed in detail in the main text.

*Absorptance in the active layer:*

The absorptance $a(E,T)$ of the active layer depends on the optical properties (that is, scattering, reflection etc.) and geometry of a film or multiple layer stack. For estimating the correlation between $a(E)$ and $\alpha(E)$ (for simplicity, we omit noting the temperature $T$ from here), typically two generic idealized scenarios are considered for solar cells:

(i) Lambert-Beer-case: At normal incidence, the absorptance of a flat solar cell with a perfect back-reflector (light propagates twice through the active layer) and negligible front reflectivity is expressed as

$$a(E) = 1 - e^{-2\alpha(E)d} \quad \text{(S5)}$$

considering the Lambert-Beer law with film thickness $d$.

(ii) Lambertian-Light-Trapping: The absorptance in the limiting case of Lambertian light trapping (for example, valid for textured Si solar cells) is expressed as

$$a(E) = \frac{4n(E)^2\alpha(E)d}{4n(E)^2\alpha(E)d + 1} \quad \text{(S6)}$$

where $n(E)$ is the refractive index of the film[51,61–63].

Given the rather smooth surface of typical spin-coated perovskite films (~10-40 nm Rq roughness), for simple analytical models the idealized Lambert-Beer-case is employed in many reports. However, especially for full solar cell stacks, interference effects must be considered in addition to accurately describe the absorptance over the whole energy range and for all incident angles, for example by employing the Transfer Matrix Method[4,64,65].

*Determination of absorptance and absorption coefficient:*

In the special case of normal incidence and detection which is limited to a very small field-of-view (FOV) (as approximately the case for the confocal microscope employed in this work), the absorptance can be expressed using the measured (normalized) external emission spectrum ($E_{\mathrm{PL}}^{\mathrm{n}}(E) = \frac{E_{\mathrm{PL}}(E)}{\mathrm{E}_{\mathrm{PL}}^{\mathrm{max}}}$; see Figure 1B in the main text) by rewriting Equation (S3) as

$$a(E,T) = \frac{E_{\mathrm{PL}}^{\mathrm{n}}(E)}{b \cdot \phi_{\mathrm{bb}}(E,T)} \quad \text{(S7)}$$

with dimensionless scaling parameter $b = \mathrm{E}_{\mathrm{PL}}^{\mathrm{max}} \cdot e^{\left(\frac{\Delta E_F}{kT}\right)}$.

In addition, the idealized Lambert-Beer-case (Equation (S5)) can be further 'improved' by considering the reflectivities at the front side ($R_{\mathrm{f}}$) and back side ($R_{\mathrm{b}}$) of the film[14,66,67]:

$$a(E) = (1 - R_{\mathrm{f}}) \cdot \frac{(1 - e^{-\alpha(E)d}) \cdot (1 + R_{\mathrm{b}} e^{-\alpha(E)d})}{(1 - R_{\mathrm{b}} R_{\mathrm{f}} e^{-2\alpha(E)d})} \quad \text{(S8),}$$

Equating Equations (S8) and (S7) and solving for $\alpha(E)$ yields the following analytical expression for $\alpha(E)$:

$$\alpha(E) = \mathrm{d}^{-1} \cdot \ln\left( \frac{-\sqrt{b^2 \cdot \phi_{\mathrm{bb}}(E,T)^2 \cdot (R_f \cdot R_b - R_f - R_b + 1)^2 + 4R_b \left(E_{\mathrm{PL}}^{\mathrm{n}}(E) + b \cdot \phi_{\mathrm{bb}}(E) \cdot (R_f - 1)\right) \cdot \left(E_{\mathrm{PL}}^{\mathrm{n}}(E) \cdot R_f + b \cdot \phi_{\mathrm{bb}}(E) \cdot (R_f - 1)\right)} - b \cdot \phi_{\mathrm{bb}}(E) \cdot (R_f \cdot R_b - R_f - R_b + 1)}{2(E_{\mathrm{PL}}^{\mathrm{n}}(E) + b \cdot \phi_{\mathrm{bb}}(E) \cdot (1 - R_f))} \right) \quad \text{(S9)}$$

A further simplification can be made in case $R_{\mathrm{f}} \approx R_{\mathrm{b}} = R$, which simplifies the above expressions to:

$$a(E) = (1 - \mathrm{R}) \cdot \frac{\left(1 - e^{-\alpha(E)d}\right)}{(1 - Re^{-\alpha(E)d})} \quad \text{(S10),}$$

and thus

$$\alpha(E) = \frac{\ln\left(\frac{E_{\mathrm{PL}}^{\mathrm{n}}(E) \cdot \mathrm{R} - (1 - R) \cdot b \cdot \phi_{\mathrm{bb}}(E)}{E_{\mathrm{PL}}^{\mathrm{n}}(E) - (1 - R) \cdot b \cdot \phi_{\mathrm{bb}}(E)}\right)}{d} \quad \text{(S11).}$$

For most confocal PL measurements, a ~100 nm thick PMMA layer was spin-coated on the perovskite films to protect the films from moisture- and light-induced degradation. In that case, with $n_{\mathrm{PMMA}} \approx n_{\mathrm{Glass}} \approx 1.5$ (at ~530 nm), we used Equation (S11) with $R = 0.1$ to determine $\alpha(E)$. In case there is no PMMA film on top, we used Equation (S9) with $R_{\mathrm{f}} = 0.2$ and $R_{\mathrm{b}} = 0.1$. To obtain absolute values of $\alpha(E)$, we scale the data to high-resolution ellipsometry data for $CH_3NH_3PbI_3$ films by Philips *et al.*[11] and Soufiani *et al.*[12] at the high-energy side of the spectrum (730 nm) *via* the scaling parameter $b$ and the results are shown in Figures S8 and S10[11,15].

In principle, the value of $b$, which is related to $\Delta E_F$ (compare Equation (S7)), can be determined experimentally by absolute photoluminescence intensity measurements, which for example has been employed in some recent reports[2,47,54,68]. However, there are pitfalls with this method when wanting to determine the shape of $a(E)$ and $\alpha(E)$: a typical absolute intensity PL setup has either a large field-of-view (as is the case for the study by Braly *et al.*[2]; Note S2; Figures S14a and S15) or is performed in an integrating sphere setup (as is the case for the study by Stolterfoht *et al.*[69]; Figure S28). Therefore, the measured PL spectra are red-shifted and broadened due to scattered PL as discussed in the main text. When fitting such experimentally measured absolute intensity PL spectra with Equation (S3) in order to obtain the shape of $a(E)$ (as for example done by Braly *et al.*[2,47]), this method results in erroneous extraction of the shape of $\alpha(E)$. Therefore, it is crucial to use the above described method only in case the measured PL spectral shape is not affected by initially-trapped scattered PL, which is possible with a confocal PL setup with a very small FOV as employed in this work.

## Note S4 – PL emission from perovskite films

The processes depicted in Figure 2 in the main text are described in more detail in the following and the more general case of a perovskite film on a glass substrate is visualized in Figure S16:

Initially absorbed and reabsorbed photons generate charge carriers which isotropically emit PL with internal spectrum $I(\lambda)$ (compare Equations (S1) and (S2)) from fully radiative band-to-band transitions with probability $Q_{\mathrm{i}}^{\mathrm{lum}}$ [45,70]. PL which is emitted within the narrow escape cone ($\theta_{\mathrm{c}} \lessapprox 22.6°$) would directly escape the film (events (A) and (B$_i$) in Figure 2A), however, a small fraction of high-energy photons will be internally reabsorbed depending on the finite depth $d_{\mathrm{ec}}$ the PL was generated within the film (compare Figure 1B and Equation (5))[44]. The rest of the PL is initially either trapped in the perovskite film (waveguide mode) or the glass substrate (substrate mode) (see Figure S16). Trapped PL will either be reabsorbed and isotropically be reemitted (exhibiting the original internal PL spectral shape $I(\lambda)$ due to phonon-assisted up-conversion[71,72]) with chance $Q_{\mathrm{i}}^{\mathrm{lum}}$ (events (2) in Figure 2), that is PR, be lost non-radiatively after reabsorption with probability $\left(1 - Q_{\mathrm{i}}^{\mathrm{lum}}\right)$ or be scattered with a small probability every time when impinging an interface. For simplicity, we assume that scattering is governed by the perovskite/air interface which exhibits the by far largest roughness. However, in principle scattering can also take place at the perovskite/glass or glass/air interface. Scattering can take place at an angle $\theta$ such that (see Figure S16): (i) PL couples out from the top surface (events ($C_1$) and ($C_2$) in Figure 2); (ii) PL escapes via the bottom escape cone; (iii) PL becomes trapped in the substrate mode; or (iv) PL remains trapped in the waveguide mode but under a different angle $\theta$ (not shown in Figure S16). These events take place until all emitted photons have been either reabsorbed, coupled out of the film upon scattering or have recombined non-radiatively.

It is important to note that the PL spectral shape measured in an integrating sphere ($E_{\mathrm{tot}}^{\mathrm{exp}}(\lambda)$) is independent of the value of $Q_{\mathrm{i}}^{\mathrm{lum}}$. This can be rationalized by analyzing the various contributions to the externally measured PL signal: any signals from directly escaping PL (events (A) and (B*i*) in Figure 2A; with total direct escape probability $\bar{p}_{\mathrm{e_d}}$) as well as scattered PL (events (C*i*) in Figure 2A; with total scattering probability $\bar{p}_{\mathrm{e_s}}$) must originate from band-to-band emission events (1) and (2). Hence, given the isotropic character of the PL emission, the number of every single event (A), (B) and (C) is directly proportional to the number of events (1) and (2). Therefore, the ratio of scattered PL to the total measured signal (that is, $\frac{(C)}{(A)+(B)+(C)}$) - which determines the red-shift of the total external PL spectrum - is always the same. This is in line with literature results[5,73], found experimentally in intensity dependent integrating sphere measurements (see Figure S18) and validated by the simulated 'externally-observed' PL spectra (Note S5). We note, however, that a higher $Q_{\mathrm{i}}^{\mathrm{lum}}$ contributes to more charge carriers being generated further away from the initial point of illumination, given the more efficient PR as an optical mechanism of redistributing charge carriers[5,74]. Thereby, a higher $Q_{\mathrm{i}}^{\mathrm{lum}}$ potentially impacts, amongst others, spatially resolved PL measurements and measurements of the external bimolecular recombination constant as discussed in several recent studies.[5,44,45,52,56,62,74,75]

Finally, we note that because for very high excitation fluences (>> 100 Sun, that is, charge carrier densities above $10^{17}$ $\mathrm{cm}^{-3}$) the perovskite film considerably heats up, $E(\lambda)$ exhibits (i) an increase in bandgap (that is, a blue-shift in the PL peak position) and (ii) an additional temperature induced broadening, given the changes in the high-energy slope of the PL spectrum and an increase of the Urbach Energy (compare Equation (S3) and Figure S11)[18,76,77]. In such a case, a modification of the PL spectral shape measured with other setups (e.g., in an integrating sphere) would also be expected. However, the integrating sphere measurements in this work were performed at rather low intensity CW illumination (intensities around ~0.01-3 Sun) and we observe an unmodified $E_{\mathrm{tot}}^{\mathrm{exp}}(\lambda)$ in intensity dependent measurements (Figure S18).

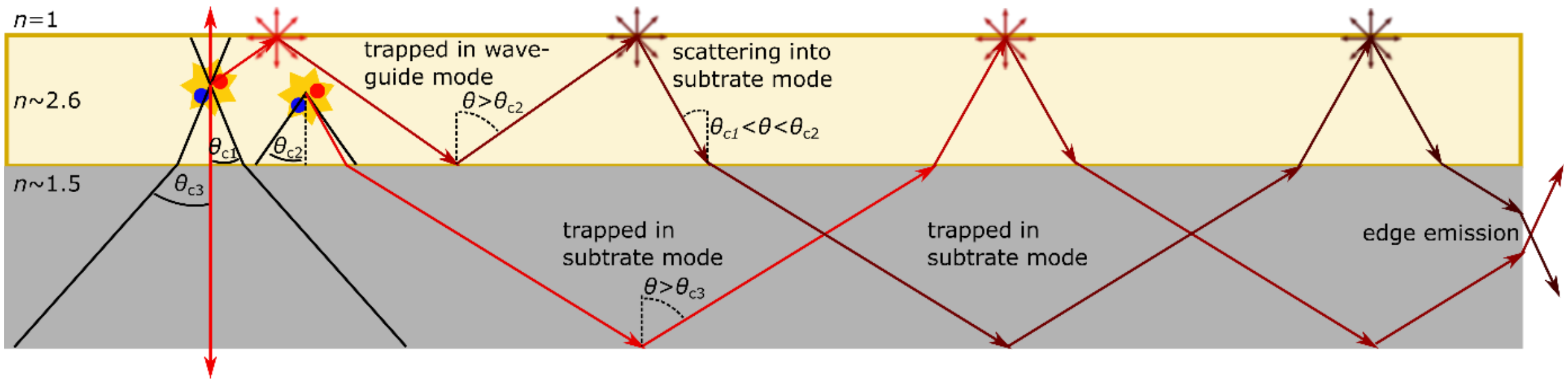

**Figure S16. Photoluminescence emission from perovskite thin films.** Visualization of the more general case of a MAPI film ($n_{\mathrm{MAPI}}$ ~2.6) on a glass substrate ($n_{\mathrm{glass}}$ ~1.5). PL emitted within the escape cone of the MAPI/air interface ($\theta < \theta_{\mathrm{c1}} = \sin^{-1}\left(\frac{1}{2.6}\right) \approx$ 22.6°) is directly emitted from the top or bottom side. In case $\theta_{\mathrm{c1}} < \theta < \theta_{\mathrm{c2}} = \sin^{-1}\left(\frac{1.5}{2.6}\right) \approx$ 35.2°, PL is initially trapped in the substrate mode and travels a long distance within the thick (~1.1 mm) glass until it reenters the perovskite film at location away from the initial illumination spot. It then is either (i), reabsorbed, (ii) scattered such that it escapes through the top or bottom, or (iii) remains trapped in the substrate mode until it is (potentially) emitted from the edge. Edge emission happens either in case the angle $\theta$ is within the escape cone of the glass/air interface or upon scattering at the (typically non-polished) edges of the glass substrate. In case $\theta > \theta_{c2}$, PL is initially trapped in the waveguide mode until it is reabsorbed, scattered such that it escapes from the top or bottom or couples into the substrate mode. Note that PL trapped in the waveguide mode can also be scattered such that it remains trapped in the waveguide mode under a different angle $\theta > \theta_{c2}$. The color change from red to brown denotes the progressively increasing red-shift of PL propagating in the film.

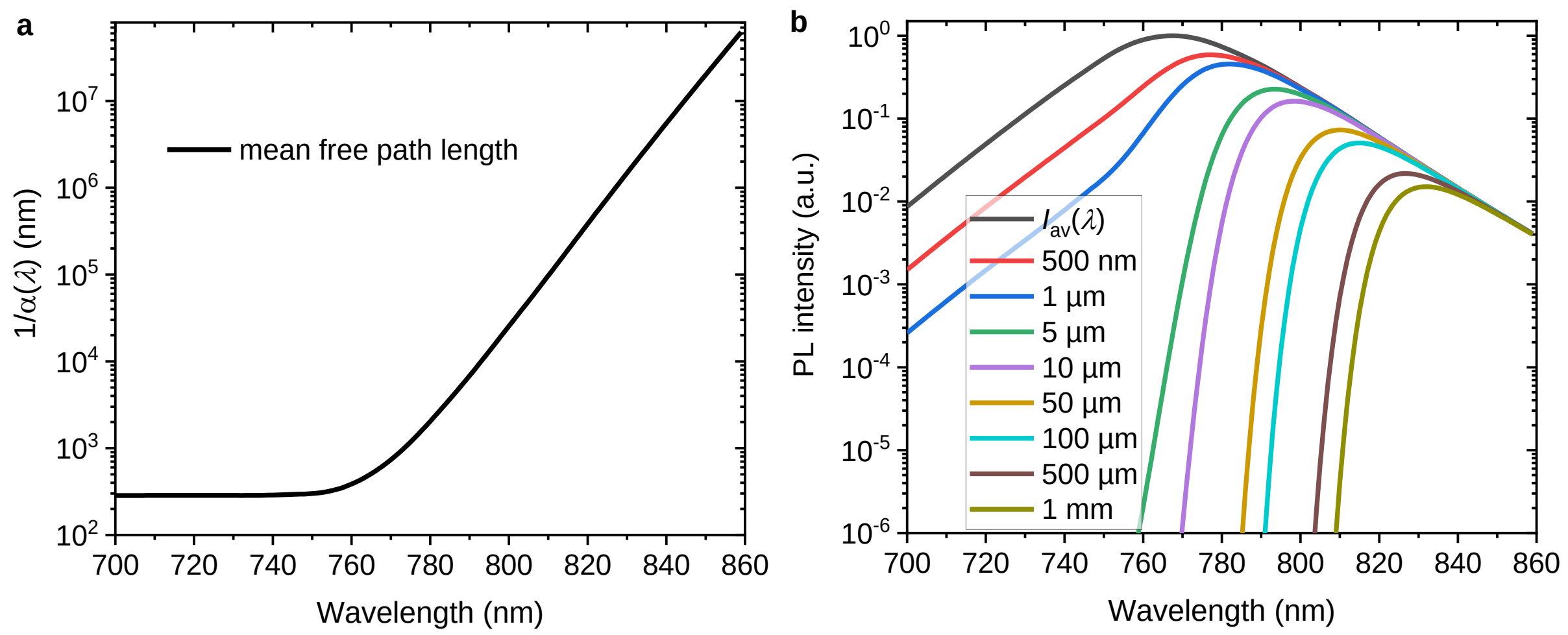

**Figure S17. a**, Mean free path for photons ($\frac{1}{\alpha_{av}(\lambda)}$) considering the Lambert-Beer law, showing that low-energy photons can propagate very long distances within the perovskite film before being reabsorbed given the exponential drop in absorption. **b**, Change of $I_{av}(\lambda)$ for various photon propagation distances considering the Lambert-Beer law. Even for a propagation distance of 1 mm, the spectrum at λ > 840 nm is barely affected by reabsorption.

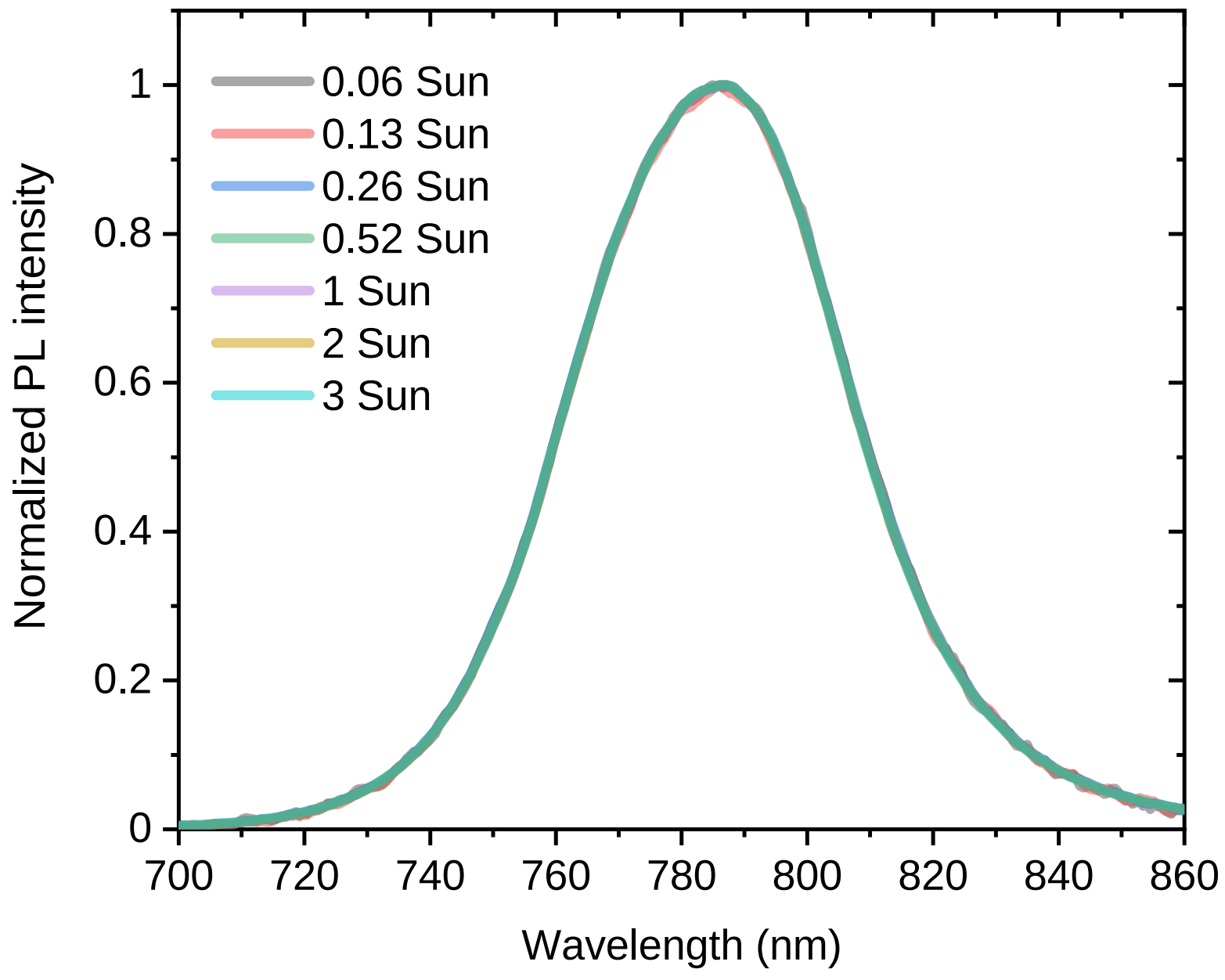

**Figure S18.** Intensity dependent measurements of the PL spectrum of a 160 nm-thick MAPI film measured within an integrating sphere, revealing that a change in within the studied intensity range of ~0.06-3 Sun - and thus, $Q_{\mathrm{i}}^{\mathrm{lum}}$ (compare Figure S23) - has no effect on the PL spectral shape and therefore the value of $\bar{p}_{\mathrm{e}}$ (see discussion in Note S4).

## Note S5 - Monte Carlo simulations

In order to validate the proposed conception of PL emission in a perovskite film as visualized in Figure 2A in the main text and in Figure S16, we developed a Monte Carlo code (available at https://github.com/pfassl/FaMi-model) based on a decision tree that is visualized in Figure S19. It offers a route to determine $Q_{\mathrm{e}}^{\mathrm{lum}}$ and the spectral shape of the external emission ($E_{\mathrm{tot}}^{\mathrm{sim}}(\lambda)$) assuming that the internal parameters are known. It considers all radiative recombination processes and subsequent event-chains illustrated in Figure 2A as well as non-radiative recombination processes that occur subsequent to an absorption event. The abortion criteria is defined when all photons have either recombined non-radiatively, directly escaped the film (with total probability $\bar{p}_{\mathrm{e_d}}$) or coupled out upon scattering (with total probability $\bar{p}_{\mathrm{e_s}}$). The magnitude of scattering is set by the inverse scattering length ($l_s^{-1}$), which assumes a fixed wavelength-independent scattering probability, such that the wavelength-dependent probability for a single reabsorption event is defined by $P_r(\lambda) = \frac{\propto(\lambda)}{\propto(\lambda)+l_s^{-1}}$ and the probability for a single scattering event is $P_s(\lambda) = 1 - P_r(\lambda) = \frac{l_s^{-1}}{\propto(\lambda)+l_s^{-1}}$. Applying the simulation - with $I_{\mathrm{av}}(\lambda)$ and $\propto_{\mathrm{av}}(\lambda)$ determined for our MAPI films as input (Note S3; Figure S10) - allows discriminating the initially absorbed number of photons ($\boldsymbol{n}_{\mathrm{abs}}$) into four contributions: reabsorbed ($\boldsymbol{n}_{\mathrm{abs\text{-}PR}}$), directly escaped ($\boldsymbol{n}_{\mathrm{esc\text{-}d}}$), scattered and escaped ($\boldsymbol{n}_{\mathrm{esc\text{-}s}}$) and lost due to non-radiative recombination. These quantities allow calculating the underlying $Q_{\mathrm{e}}^{\mathrm{lum}}$, $\bar{p}_{\mathrm{e}}$, $\bar{p}_{\mathrm{e_d}}$, $\bar{p}_{\mathrm{e_s}}$ and $E_{\mathrm{tot}}^{\mathrm{sim}}(\lambda)$ (see right panel in Figure S19). The simulation is summarized and linked to the various events shown in Figure 2A in the following:

- *(i)* $\mathrm{n}_{\mathrm{abs}}$ photons are absorbed following initial laser excitation (event (0) in Figure 2A)
- *(ii)* Absorbed photons are either emitted with probability $Q_{\mathrm{i}}^{\mathrm{lum}}$ and spectral shape of the internal PL spectrum $I_{\mathrm{av}}(\lambda)$ (event (1) in Figure 2A) or lost non-radiatively with probability $(1 - Q_{\mathrm{i}}^{\mathrm{lum}})$.
- *(iii)* Emitted photons either directly escape the film with probability $\boldsymbol{p}_{\mathrm{e\text{-}d}}(\lambda)$ ($\rightarrow \mathrm{n}_{\mathrm{e\text{-}d}}(\lambda) = \mathrm{n}_{\mathrm{e\text{-}d}}(\lambda) + 1$; events (A) and (B*i*) in Figure 2A), are reabsorbed 'during escape' with probability ($\boldsymbol{p}_{\mathrm{e\text{-}d}} - \boldsymbol{p}_{\mathrm{e\text{-}d}}(\lambda)$) or are initially trapped in the film with probability ($1 - \boldsymbol{p}_{\mathrm{e_d}}$). The wavelength-dependent probability for escape $\boldsymbol{p}_{\mathrm{e\text{-}d}}(\lambda)$ is simulated by assuming a random starting point for photons within the film (with thickness $d_{\mathrm{sim}}$ as an input parameter) and considering that 30% of the 'escaping' PL is reflected at interfaces (relating to ~20% at the perovskite/air interface and ~10% at the perovskite/glass/air interfaces) and thus travels through the perovskite film once more (secondary reflections are omitted). That is, 70% of the 'escaping' PL propagates a (randomly chosen) distance of $0 < d_{\mathrm{prop}} < d_{\mathrm{sim}}$ before escape and the other 30% propagates a distance of $d < d_{\text{prop}} < 2 \cdot d_{\text{sim}}$. Within this distance, reabsorption is governed by the Lambert-Beer law.

*(iv)* Initially-trapped photons are either reabsorbed with probability $\frac{\alpha(\lambda)}{\alpha(\lambda)+l_s^{-1}}$ ($\rightarrow \mathrm{n_{abs\text{-}PR}} = \mathrm{n_{abs\text{-}PR}} + 1$) or scattered such that they couple out from the film with probability $\frac{l_s^{-1}}{\alpha(\lambda)+l_s^{-1}}$ ($\rightarrow \mathrm{n_{e\text{-}s}}(\lambda) = \mathrm{n_{e\text{-}s}}(\lambda) + 1$; events ($C_i$) in Figure 2A).

*(v)* Reabsorbed photons are reemitted with probability $Q_i^{\text{lum}}$ (events (2) in Figure 2A) or lost non-radiatively with probability $(1 - Q_i^{\text{lum}})$.

*(vi)* Reemitted photons again either directly escape the film, are reabsorbed 'during escape' (see (iii)) or are initially trapped and subsequently scattered or reabsorbed (iv) and potentially reemitted again (v).

*(vii)* The simulation runs until all emitted photons have either escaped the film by direct emission (with total probability $\bar{p}_{\text{e_d}}$) or scattering (with total probability $\bar{p}_{\text{e_s}}$), or are lost non-radiatively.

Figure S20 shows illustrative examples of 'externally-observed' PL spectra ($E_{\text{tot}}^{\text{sim}}(\lambda) = E_{\text{d}}^{\text{sim}}(\lambda) + E_{\text{s}}^{\text{sim}}(\lambda) = \mathrm{n_{e\text{-}d}}(\lambda) + \mathrm{n_{e\text{-}s}}(\lambda)$) for two fixed values of $p_{\text{e-d}}$ (8% and 14%) and varying $l_s^{-1}$ (panel a) as well as two fixed values of $l_s^{-1}$ and varying $p_{\text{e-d}}$ (panel b), respectively. It can be clearly seen that varying these parameters results in a large variety of asymmetric spectral shapes (apparent double peak, red-shifted shoulder or broadened and red-shifted single peak) depending on the magnitude of scattering which is set by $l_s^{-1}$, in agreement with the experimental results (Figures 1 and 3 in the main text and Figure S5). Note that the shape of $E_{\text{tot}}^{\text{sim}}(\lambda)$ is also independent of the chosen input value of $Q_i^{\text{lum}}$, in line with the experimental results (Figure S18), since its shape is solely determined by the relative fraction of directly emitted to scattered PL (see discussion in Note S4).

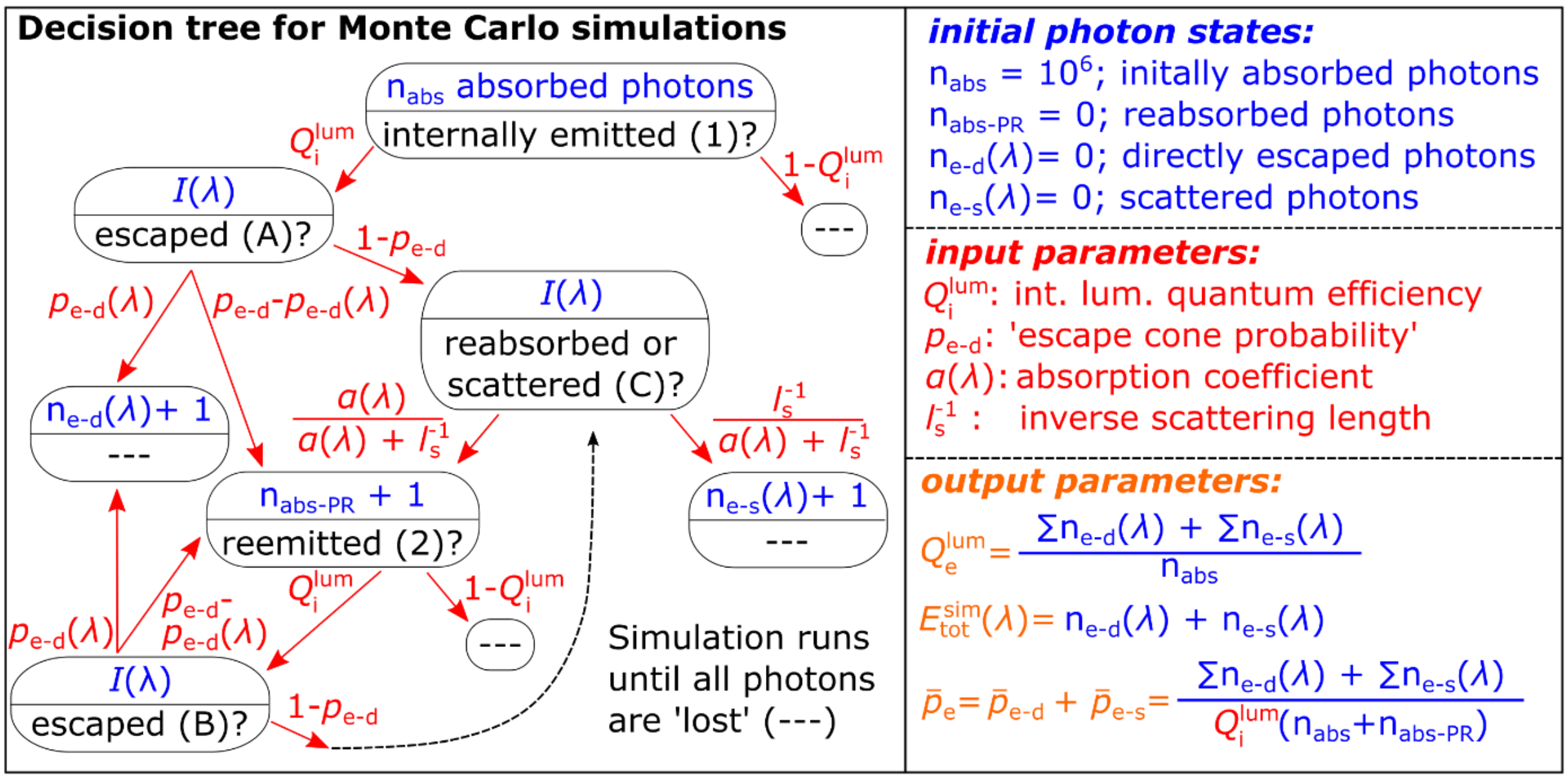

**Figure S19. Decision tree for Monte Carlo simulations describing the various radiative and non-radiative processes in a perovskite film (see details in Note S5).** The numbers and letters in brackets relate to the various events depicted in Figure 2A in the main text. The corresponding Matlab code is available at https://github.com/pfassl/FaMi-model.

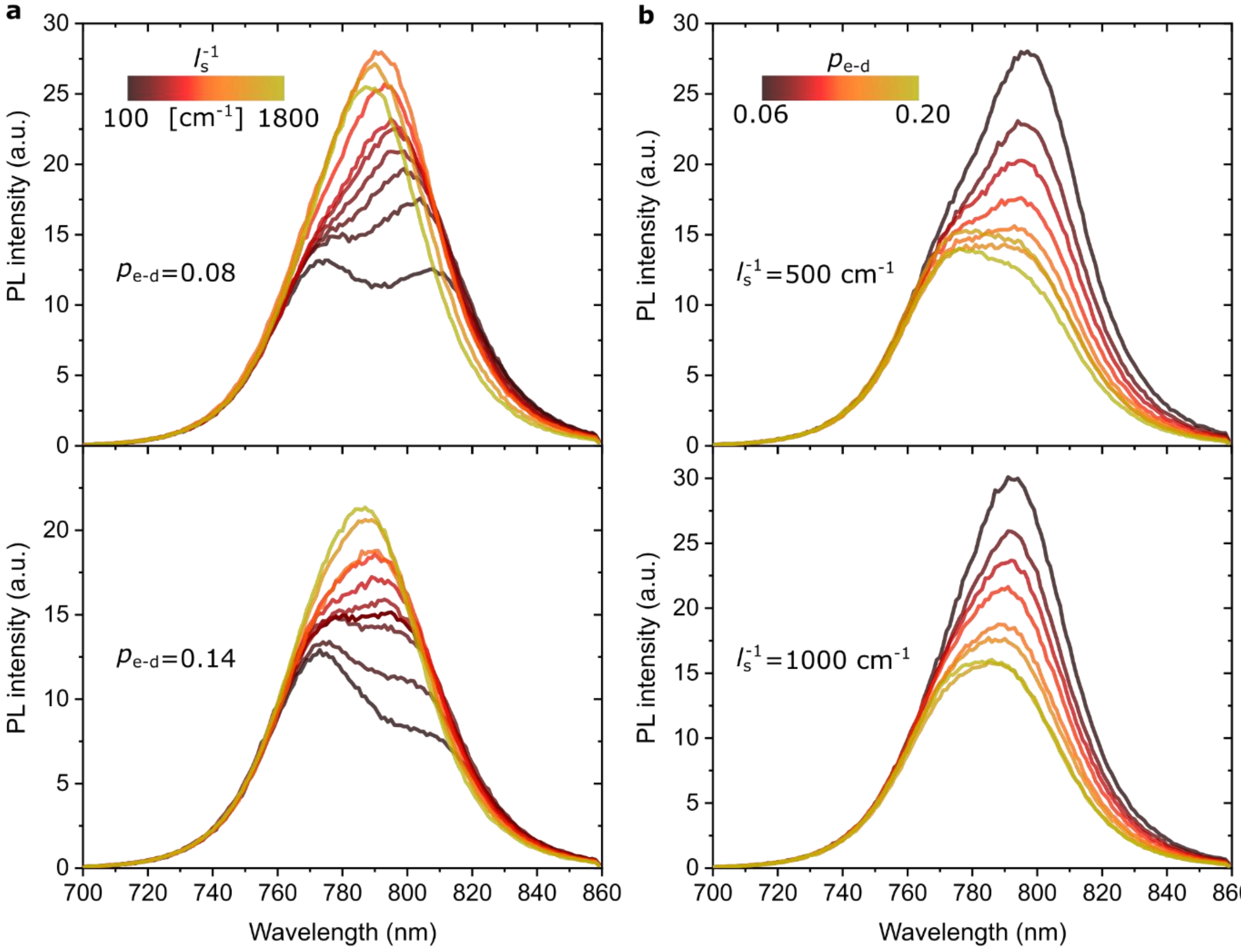

**Figure S20.** Further exemplary simulated 'externally observed' PL spectra for **a**, two fixed values of $p_{\text{e-d}}$ (8% and 14%) and varying magnitudes of scattering ($l_s^{-1}$) and **b**, two fixed values of $l_s^{-1}$ (500 cm$^{-1}$ and 1000 cm$^{-1}$) and varying $p_{\text{e-d}}$. A large variety of PL spectral shapes (asymmetric, broadened, red-shifted; apparent double-peak) can be observed, sensitively depending on the combination of both values.

## Note S6 – Details on the curve fitting model

While the use of an effective film thickness $d_{\text{eff}}$ to model the PL spectral shape of directly escaping PL (compare Equation (10)) is not an exact description of the problem, in Figure S22 and Table S2 we show that fitting simulated spectra as well as experimental confocal PL spectra with $E_{\text{d}}(\lambda) = b \cdot I(\lambda) \cdot e^{-\alpha(\lambda)d_{\text{eff}}}$ (where $b$ is a scaling parameter) describes the data very well (only for very thick simulated films the fits become a bit worse) and therefore, this approach is a valid simplification. The fitted value of $d_{\text{eff}}$ has no direct physical meaning as various mechanisms such as internal reflections at the interfaces, angle dependence and roughness can all affect the total direct escape probability and therefore $d_{\text{eff}}$.

Considering the strong cross-correlation between $P_{\text{s}}$ and $z_{\text{av}}$ in Equation (12) and that both values alone have no direct physical meaning (both are related to the magnitude of scattering of the films), we always fix $P_{\text{s}}$ for fitting, considering it makes the fit much more stable and faster. Note that a very good fit for $E_{\text{tot}}^{\text{sim}}(\lambda)$ (and therefore the extracted values of $Q_{\text{i}}^{\text{lum}}$ and $\bar{p}_{\text{e}}$) is possible for a large range of values for $P_{\text{s}}$, while the spectral shapes of $E_{\text{d-tot}}^{\text{sim}}(\lambda)$ and $E_{\text{s-tot}}^{\text{sim}}(\lambda)$ (and therefore the extracted values for $\bar{p}_{\text{e_d}}$ and $\bar{p}_{\text{e_s}}$) are reproduced best for small $P_{\text{s}}$ ~ 0.003. For even lower values of $P_{\text{s}}$ the fit becomes unstable (see Table S3). When fitting experimental data, we found that the fit sometimes became unstable for values of $P_{\text{s}}$ ~0.003-0.004. Therefore, we decided to fix $P_{\text{s}} = 0.005$ for all fits to experimental data shown in this work, which yielded only a minor overestimation of $\bar{p}_{\text{e_d}}$ when fitting simulated $E_{\text{tot}}^{\text{sim}}(\lambda)$ (see Figure 3B in the main text and Tables S3 and S4).

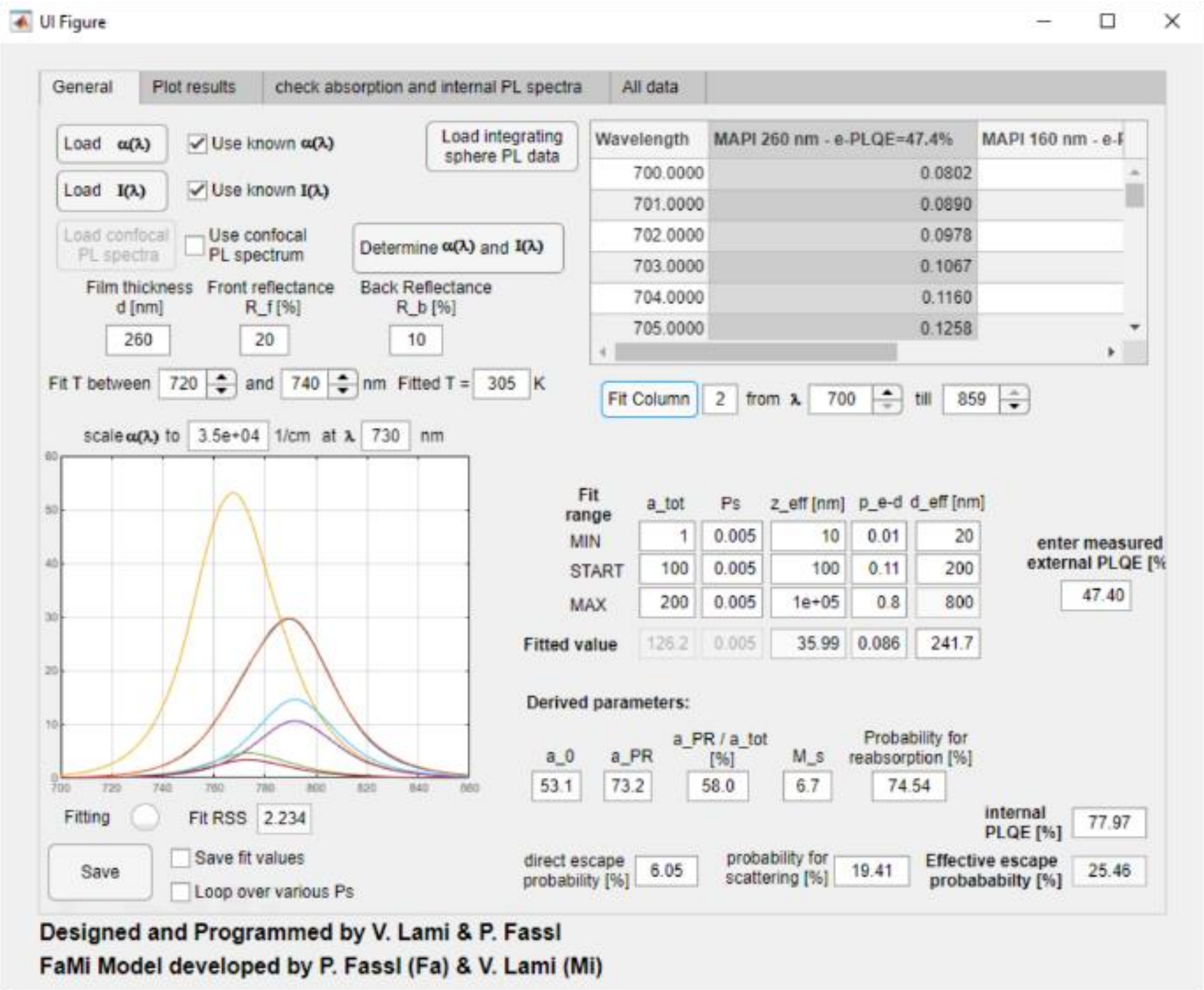

**Figure S21. User interface of the Matlab app for performing fitting with Equation (12)**. Example for the 260 nm-thick MAPI film with the results shown in Figure 3 and Table 1 in the main text. The app will be made available open-source on github (https://github.com/pfassl/FaMi-model) soon.

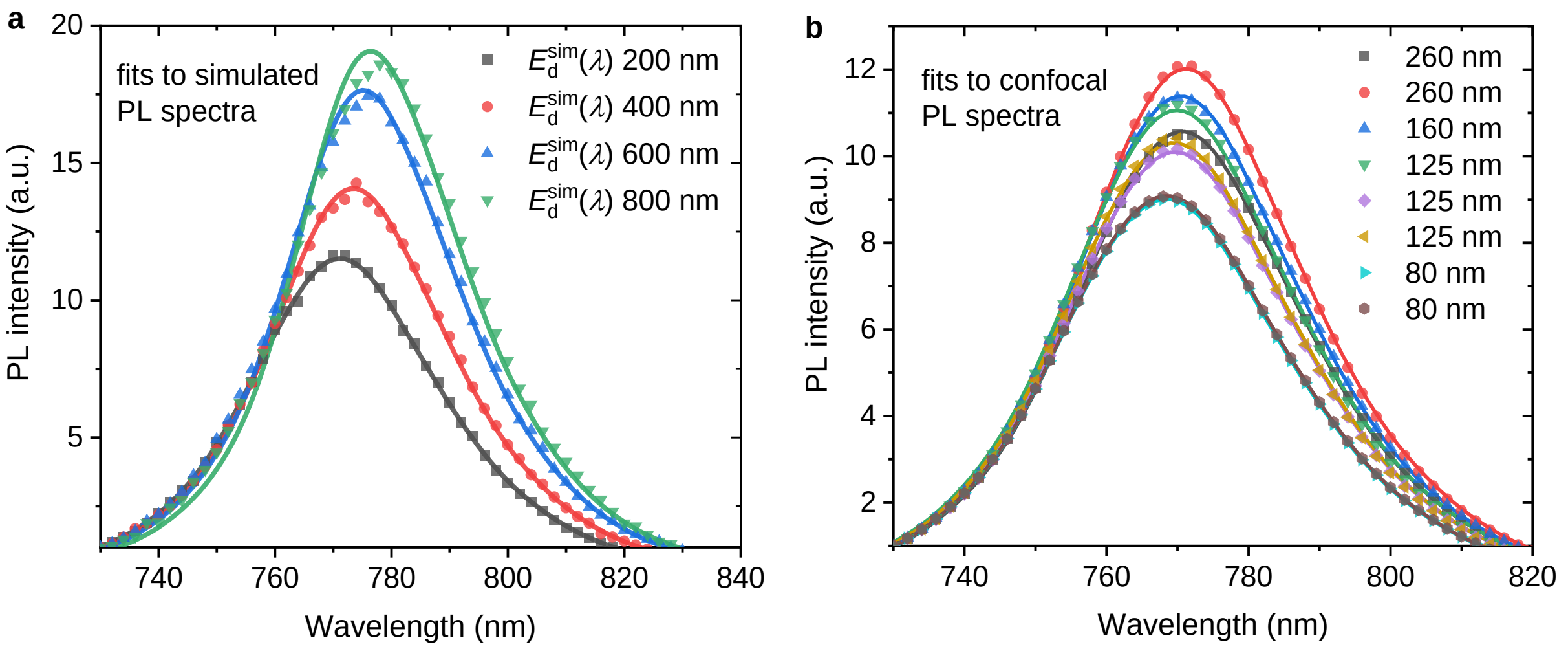

**Figure S22.** Fits with $E_\mathrm{d}(\lambda) = b \cdot I(\lambda) \cdot e^{-\alpha(\lambda) d_\mathrm{eff}}$ (solid lines) (compare Equation (10)) to; **a**, simulated PL spectra using a Monte Carlo approach for directly escaping PL with various $d_\mathrm{sim}$ as input. To make the simulations more relevant to a real film, it is assumed that 70% of PL directly escapes when reaching an interface, while 30% of PL are reflected once and travels through the film once more (see detailed description at (iii) in Note S5); and **b**, the 8 experimental confocal PL spectra used to determine $\alpha(\lambda)$ of our MAPI films (compare Figure S10). The fit results can be found in Table S2.

**Table S2. Validation for the use of an effective film thickness $d_{eff}$ in Equation (12) for modelling the spectral shape of directly escaping PL (compare Equation (5) and Figure 2).**

| **MAPI thickness** | $d_{eff}$ [nm] | Fit RSS |
|---|---|---|
| $E_d^{sim}(\lambda); d_{sim}$ = 200 nm | 152.6 | 0.55 |
| $E_d^{sim}(\lambda); d_{sim}$ = 400 nm | 263.8 | 1.36 |
| $E_d^{sim}(\lambda); d_{sim}$ = 600 nm | 366.3 | 8.25 |
| $E_d^{sim}(\lambda); d_{sim}$ = 800 nm | 450.6 | 20.0 |
| 260 nm confocal | 130.1 | 0.001 |
| 260 nm confocal | 152.9 | 0.001 |
| 160 nm confocal | 123.9 | 0.002 |
| 125 nm confocal | 100.4 | 0.011 |
| 125 nm confocal | 90 | 0.002 |
| 125 nm confocal | 83 | 0.006 |
| 80 nm confocal | 44.2 | <0.001 |
| 80 nm confocal | 46.7 | <0.001 |

Results of fits with $E_d(\lambda) = b \cdot I(\lambda) \cdot e^{-\alpha(\lambda) d_{eff}}$ (compare Equation (10)) to simulated spectra of directly escaping PL ($E_d^{sim}(\lambda)$; Note S5; Figure S19) for various simulated thicknesses ($d_{sim}$) as input as well as to the 8 experimental confocal PL spectra used to determine $\alpha(\lambda)$ and $I(\lambda)$ (Figure S10). The corresponding data and fits are shown in Figure S20. In case of $E_d^{sim}(\lambda)$, the fit describes the data well for small $d_{sim}$ and starts to slightly deviate for large $d_{sim}$, while in case of the experimental confocal PL spectra all fits describe the data well. Therefore, we conclude that the use of $d_{eff}$ in Equation (12) for the MAPI films studied in this work is a valid approach (see discussion in Note S6).

**Table S3. Results of fits with Equation (12) to the simulated PL spectrum shown in Figure 3B for various fixed values of $P_s$.**

| $P_s$ | $z_{av}$ [nm] | $d_{eff}$ [nm] | $\bar{p}_{e\text{-}d}$ [%] | $\bar{p}_{e_s}$ [%] | $\bar{p}_e$ [%] | $Q_i^{lum}$ [%] | Fit RSS |
|---|---|---|---|---|---|---|---|
| 5E-4 | 13.2 | 458 | 8.16 | 9.74 | 17.90 | 81.15 | 145.0 |
| 0.001 | 20.3 | 328.7 | 7.32 | 11.47 | 18.79 | 80.40 | 41.55 |
| 0.002 | 33 | 169.1 | 6.08 | 13.20 | 19.28 | 80.00 | 1.61 |
| 0.003 | 49.5 | 167.5 | 6.12 | 13.16 | 19.29 | 79.99 | 1.61 |
| 0.004 | 66 | 166.2 | 6.17 | 13.12 | 19.29 | 79.99 | 1.61 |
| 0.005 | 82.3 | 164 | 6.20 | 13.10 | 19.30 | 79.97 | 1.61 |
| 0.006 | 98.8 | 163.2 | 6.25 | 13.05 | 19.30 | 79.98 | 1.61 |
| 0.008 | 132 | 161.6 | 6.33 | 12.96 | 19.30 | 79.98 | 1.61 |
| 0.01 | 165.2 | 160.2 | 6.42 | 12.88 | 19.30 | 79.98 | 1.61 |
| 0.02 | 332.1 | 156.4 | 6.83 | 12.47 | 19.30 | 79.98 | 1.61 |
| 0.04 | 672.4 | 162.6 | 7.54 | 11.76 | 19.29 | 79.98 | 1.59 |
| 0.06 | 1020.2 | 179.3 | 8.12 | 11.17 | 19.29 | 79.99 | 1.59 |
| 0.08 | 1371.1 | 199.7 | 8.59 | 10.70 | 19.29 | 79.98 | 1.64 |
| 0.1 | 1718.5 | 219.7 | 8.97 | 10.34 | 19.31 | 79.97 | 1.79 |
| 0.12 | 2056.3 | 237.7 | 9.28 | 10.06 | 19.34 | 79.94 | 2.16 |
| 0.14 | 2380.9 | 253.2 | 9.54 | 9.85 | 19.39 | 79.90 | 2.81 |
| 0.16 | 2690.9 | 266.5 | 9.76 | 9.68 | 19.44 | 79.86 | 3.77 |
| 0.18 | 2986.3 | 277.8 | 9.95 | 9.55 | 19.50 | 79.81 | 5.05 |
| 0.2 | 3268 | 287.6 | 10.12 | 9.45 | 19.57 | 79.75 | 6.61 |

The input parameters for the Monte Carlo simulation (see Figure S19) are: $Q_i^{lum}$ = 80%; $p_{e\text{-}d}$ = 8%; $d_{sim}$ = 260 nm; and $l_s^{-1}$ = 600 cm$^{-1}$. The simulation results are $\bar{p}_{e\text{-}d}$ = 6.08%, $\bar{p}_{e\text{-}s}$ = 13.20%, $\bar{p}_e$ = 19.28% and $Q_e^{lum}$ = 43.53% and these the latter two values are used to derive $Q_i^{lum}$ *via* Equation (3). The goodness of the fits is compared via the residual sum of squares (RSS), which should be as small as possible. Note that the absolute value of RSS depends on the initial scaling of the input data (here the simulated spectra were normalized to 730 nm). The RSS for the fits to the simulated and experimental spectra is comparable (see example fit in Figure S21). The rows marked in red show that RSS strongly increases for very low $P_s \lesssim 0.002$ as well as for rather large $P_s \gtrsim 0.1$ and the simulated values for $\bar{p}_{e\text{-}d}$ are reproduced best for the smallest possible $P_s$. Given that fits to some experimental and simulated PL spectra became unstable for $P_s \lesssim$ 0.003-0.004, we decided to fix $P_s$ = 0.005 (grey background) for all further fits in this work.

**Table S4. Results of fits with Equation (12) to various simulated PL spectra shown in Figure S20.**

| Input \ Results | $Q_e^{lum}$ [%] | $z_{av}$ [nm] | $d_{eff}$ [nm] | $\bar{p}_{e\text{-}d}$ [%] | $\bar{p}_{e\text{-}s}$ [%] | $\bar{p}_e$ [%] | $Q_i^{lum}$ [%] | Fit RSS |
|---|---|---|---|---|---|---|---|---|
| $p_{e\text{-}d}$ = 8%; $l_s^{-1}$ = 200 cm$^{-1}$ | 35.41 | 247.8 | 170.6 | 6.05<br>6.20 | 7.65<br>7.49 | 13.71<br>13.70 | **80.00** | 1.56 |
| $p_{e\text{-}d}$ = 8%; $l_s^{-1}$ = 400 cm$^{-1}$ | 40.26 | 123.7 | 165.8 | 6.05<br>6.21 | 10.79<br>10.69 | 16.85<br>16.89 | **79.96** | 1.17 |
| $p_{e\text{-}d}$ = 8%; $l_s^{-1}$ = 600 cm$^{-1}$ | 43.53 | 82 | 163.2 | 6.05<br>6.10 | 13.22<br>13.14 | 19.27<br>19.24 | **80.03** | 1.68 |
| $p_{e\text{-}d}$ = 8%; $l_s^{-1}$ = 1000 cm$^{-1}$ | 47.81 | 49.8 | 164.3 | 6.04<br>6.12 | 16.85<br>16.77 | 22.89<br>22.88 | **80.01** | 1.36 |
| $p_{e\text{-}d}$ = 8%; $l_s^{-1}$ = 1400 cm$^{-1}$ | 50.8 | 35.4 | 159.3 | 6.00<br>6.13 | 19.77<br>19.72 | 25.83<br>25.85 | **79.98** | 1.23 |
| $p_{e\text{-}d}$ = 14%; $l_s^{-1}$ = 200 cm$^{-1}$ | 41.5 | 243.3 | 170.6 | 10.58<br>10.71 | 7.15<br>7.09 | 17.74<br>17.79 | **79.95** | 0.62 |
| $p_{e\text{-}d}$ = 14%; $l_s^{-1}$ = 400 cm$^{-1}$ | 45.34 | 119.9 | 167.0 | 10.60<br>10.63 | 10.13<br>10.18 | 20.73<br>20.81 | **79.94** | 0.88 |
| $p_{e\text{-}d}$ = 14%; $l_s^{-1}$ = 600 cm$^{-1}$ | 47.86 | 81.4 | 165.5 | 10.61<br>10.64 | 12.32<br>12.355 | 22.93<br>22.99 | **79.97** | 0.95 |
| $p_{e\text{-}d}$ = 14%; $l_s^{-1}$ = 1000 cm$^{-1}$ | 51.33 | 48.9 | 167.0 | 10.61<br>10.555 | 15.77<br>15.86 | 26.38<br>26.41 | **79.97** | 0.73 |
| $p_{e\text{-}d}$ = 14%; $l_s^{-1}$ = 1400 cm$^{-1}$ | 53.75 | 35.5 | 164.6 | 10.59<br>10.58 | 18.46<br>18.47 | 29.05<br>29.05 | **80.00** | 0.60 |

The fixed input parameters for all simulations are $Q_i^{lum}$ = 80% and $d$ = 260 nm, while the varying values for $p_{e\text{-}d}$ and $l_s^{-1}$ are shown in the first column (compare decision tree in Figure S19). The simulation results for $Q_e^{lum}$, $\bar{p}_{e\text{-}d}$, $\bar{p}_{e\text{-}s}$ and $\bar{p}_e$ and the fit results are shown in red and black, respectively, where $Q_i^{lum}$ is determined *via* Equation (3) with the simulated $Q_e^{lum}$ and fitted $\bar{p}_e$ as input. The goodness of the fits is compared via the residual sum of squares (RSS), which is low and comparable for all fits. All fits reproduce the simulated parameters with a very high accuracy.

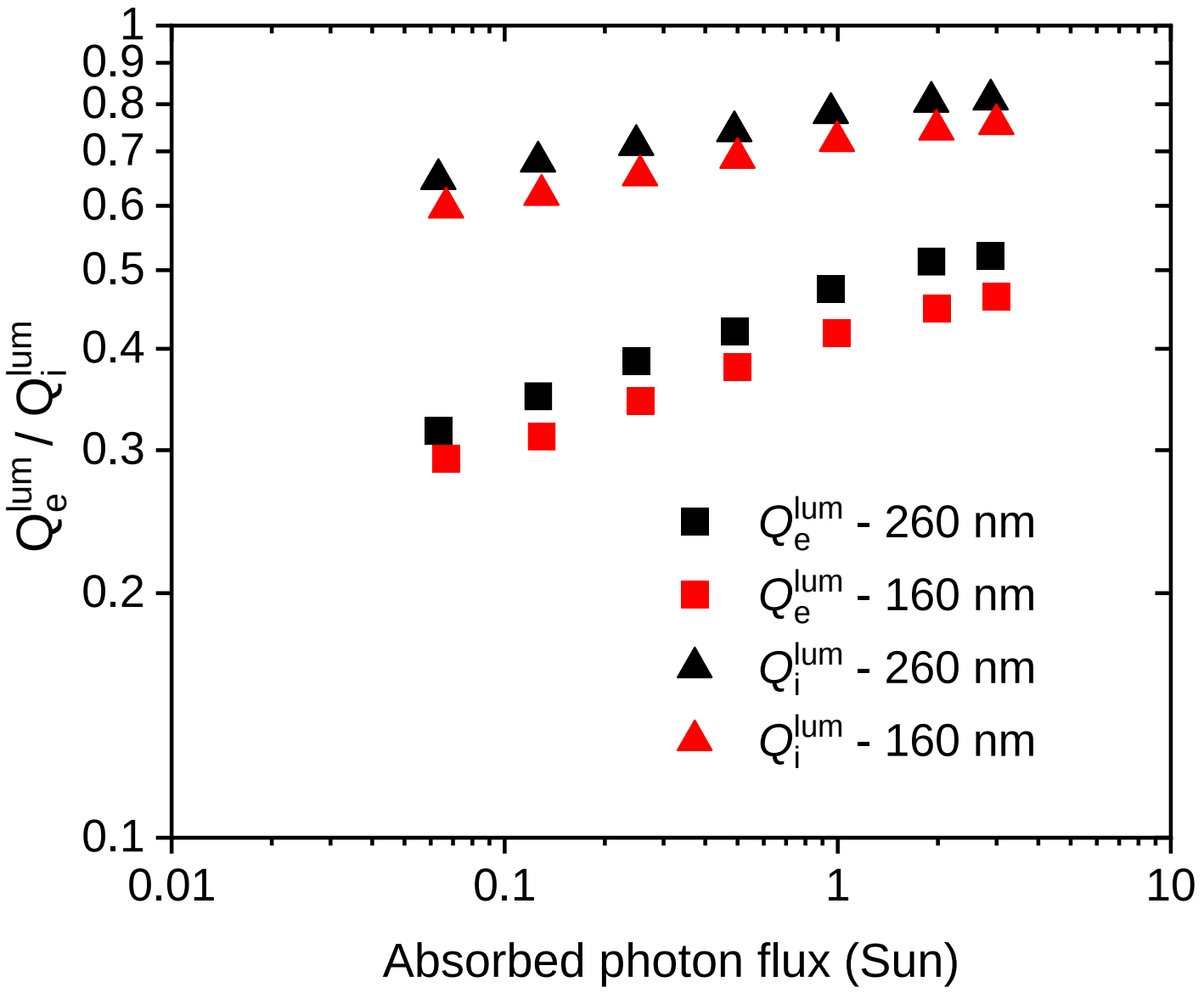

**Figure S23.** Intensity dependence of $Q_e^{lum}$ and $Q_i^{lum}$ for the 260 nm- and 160 nm-thick MAPI films in log-log scale. $Q_e^{lum}$ only slightly reduces to ~32.1% and ~29.1% for 260 nm and 160 nm at low intensities of ~0.07 Sun, respectively, emphasizing the high quality of our samples. At higher intensities, $Q_e^{lum}$ starts to plateau and reaches maximum values of 52% and 46.4% at ~3 Sun, correlating to $Q_i^{lum} = 81.1 \pm 0.5\%$ and $Q_i^{lum} = 75.6 \pm 0.5\%$, for the 260 nm- and 160 nm-thick film, respectively. At higher intensities, Auger recombination will begin to dominate and result in a reduction of $Q_e^{lum}$ as reported previously[2,56].

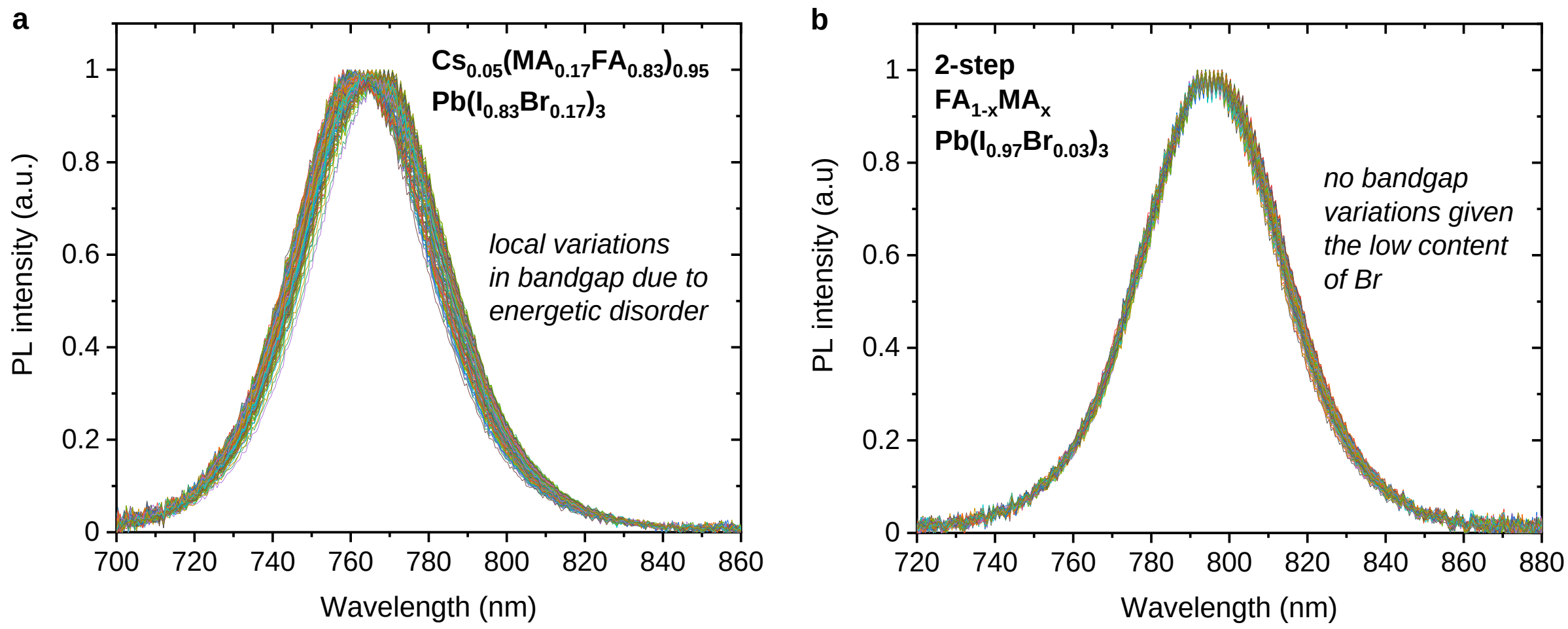

**Figure S24.** Representative confocal PL spectra measured at ~200 different spots on one sample of **a**, a ~400 nm-thick triple-cation (TC) perovskite film and **b**, a ~600 nm-thick double-cation perovskite film. The TC perovskite film exhibits variations in the PL peak position on the microscale due to energetic disorder induced by iodide-rich and bromide-rich regions, as recently discussed by Feldmann *et al.*[78] Therefore, the determination of the absorption coefficient spectrum with Equation (S9) based upon an averaged confocal PL spectra over many spots on the sample (compare Figure S5) can only serve as an approximation and the absolute results for the TC perovskite film derived with our curve fitting fitting (see Figure S25 and Table S5) must be taken with care. Studying the exact impact of these microscale spatial inhomogeneities on $\bar{p}_e$ is beyond the scope of the current work and requires further research.

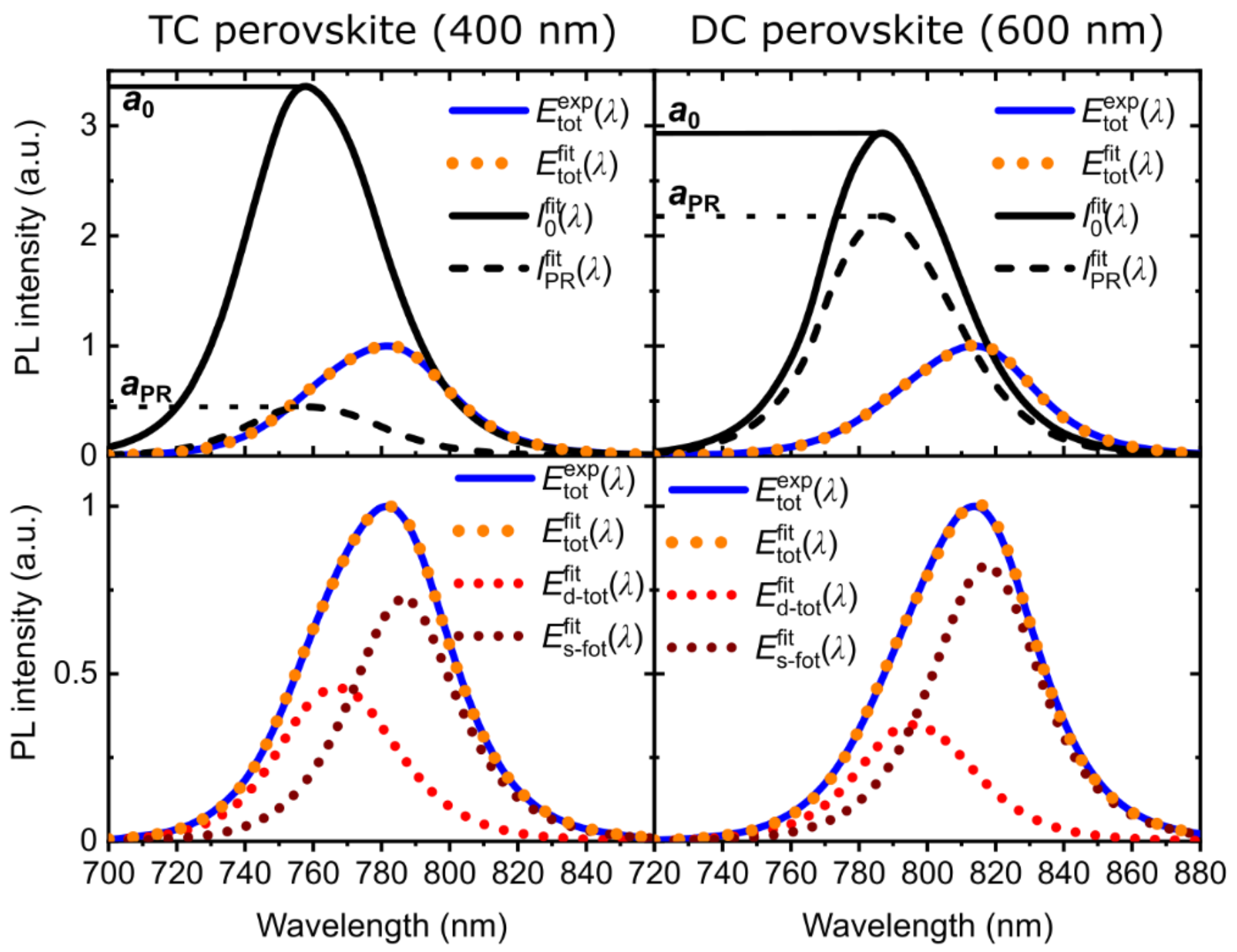

**Figure S25. Fits with Equation (12) to experimental PL spectra for two commonly employed perovskite compositions.** Left column, 400 nm-thick triple-cation (TC) ($Cs_{0.05}(MA_{0.17}FA_{0.83})_{0.95}Pb(I_{0.83}Br_{0.17})_3$) perovskite films. Right column, 600 nm-thick double-cation (DC) ($FA_{1-x}MA_xPb(I_{0.97}Br_{0.03})_3$) perovskite film (see Experimental Procedures). The fit results are summarized in Table S5. The corresponding confocal PL spectra for both compositions are discussed in more detail in Figure S24.

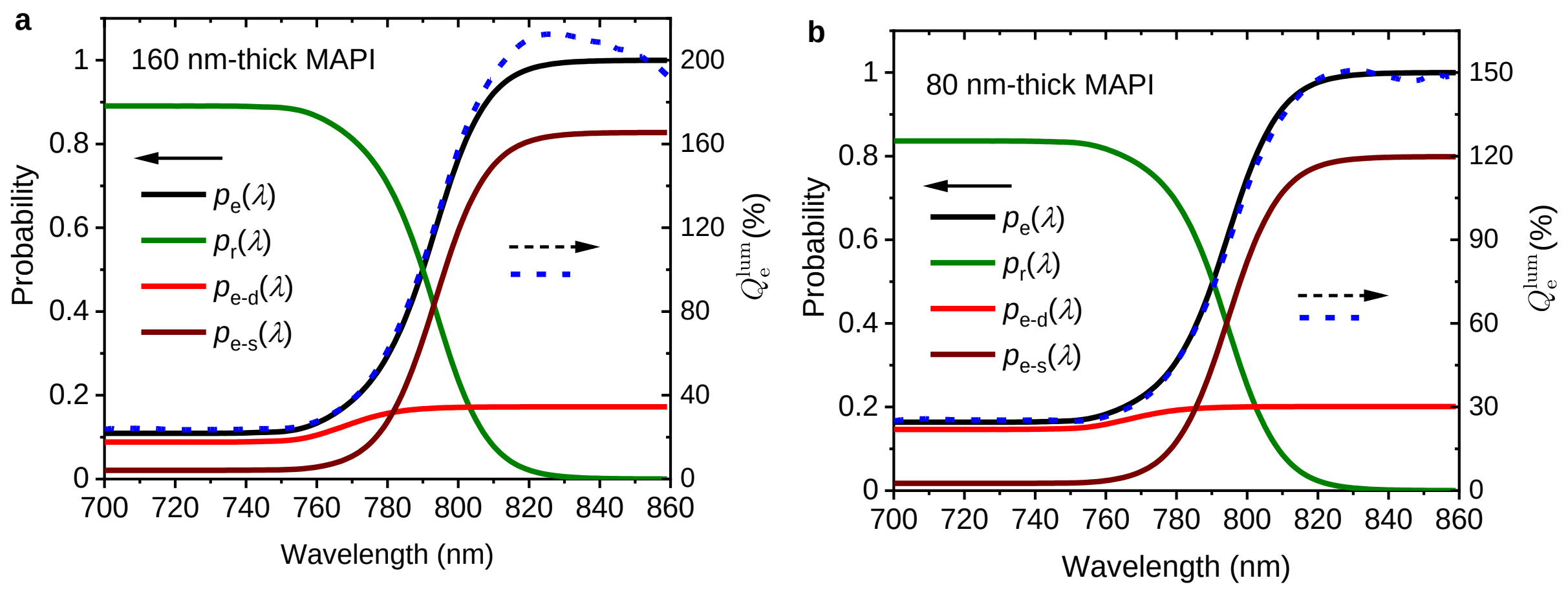

**Figure S26.** Wavelength-dependent escape and reabsorption probabilities (left y-axis; see Equations (9) and (10) in main text; $p_r(\lambda) = 1 - p_e(\lambda)$) and external luminescence quantum efficiency (right y-axis) for **a**, the 160 nm- and **b**, the 80 nm-thick MAPI film. Data for the 260 nm-thick MAPI film is shown in Figure 4A in the main text.

**Table S5. Results of curve fitting model for other perovskite compositions.**

| | TC perovskite | DC perovskite |
|---|---|---|
| Setup | IS | IS |
| Intensity [Sun] | ~1.00 | ~1.00 |
| $d$ [nm] | ~400 | ~600 |
| $R_q$ [nm] | ~16 | ~57 |
| $Q_{\text{e-max}}^{\text{lum}}$ [%] | 5.0% | 19.6% |
| $P_s$ | 0.005 | 0.005 |
| $d_{\text{eff}}$ [nm] | 372.4 | 355.4 |
| $z_{\text{av}}$ [nm] | 71.9 | 74.9 |
| $\bar{p}_{\text{e-d}}$ [%] | 11.0 | 6.3 |
| $\bar{p}_{\text{e-s}}$ [%] | 16.1 | 14.3 |
| $\bar{p}_{\text{e}}$ [%] | 27.1 | 20.7 |
| $a_{\text{PR}}/a_{\text{tot}}$ [%] | 11.8 | 42.6 |
| $M_s$ | 5.6 | 8.7 |
| $Q_i^{\text{lum}}$ [%] | 18.5 | 54.1 |
| $\Delta V_{\text{OC}}^{\text{PR}}$ [mV] | 3.75 | 14.7 |

Results of fits with Equation (12) (grey background) to experimental integrating sphere PL spectra of triple-cation (TC) ($Cs_{0.05}(MA_{0.17}FA_{0.83})_{0.95}Pb(I_{0.83}Br_{0.17})_3$) and double films (DC) ($FA_{1-x}MA_xPb(I_{0.97}Br_{0.03})_3$) perovskite films measured at intensities around ~1 Sun intensity. The measured champion external luminescence quantum efficiencies ($Q_{\text{e-max}}^{\text{lum}}$) and $P_s = 0.005$ are fixed input parameters (red). The internal luminescence quantum efficiency ($Q_i^{\text{lum}}$) is determined *via* Equation (3) using the fitted values of $\bar{p}_e$ and $Q_{\text{e-max}}^{\text{lum}}$ as input and the potential $\Delta V_{\text{OC}}^{\text{PR}}$ is calculated *via* Equation (4) neglecting parasitic absorption ($\bar{p}_r = 1 - \bar{p}_e$). $M_s = \frac{d}{z_{\text{av}}}$ serves to compare the magnitude of scattering, which is in line with the trend in surface roughness ($R_q$). We note that, as recently reported Feldmann *et al.*[78], TC perovskite films exhibit spatially varying energetic disorder causing local charge accumulation (due to bromide and iodide rich regions) and therefore the PL peak position does slightly vary on the microscale (more details in Figure S24a). Therefore, in case of TC perovskite the values for $\bar{p}_e$ should be taken with care as the effect of these bandgap variations on the overall $\bar{p}_e$ requires a more in-depth analysis, which is beyond the scope of the current work. Note that for the DC perovskite films no such variations in PL peak position are observed (Figure S24b).

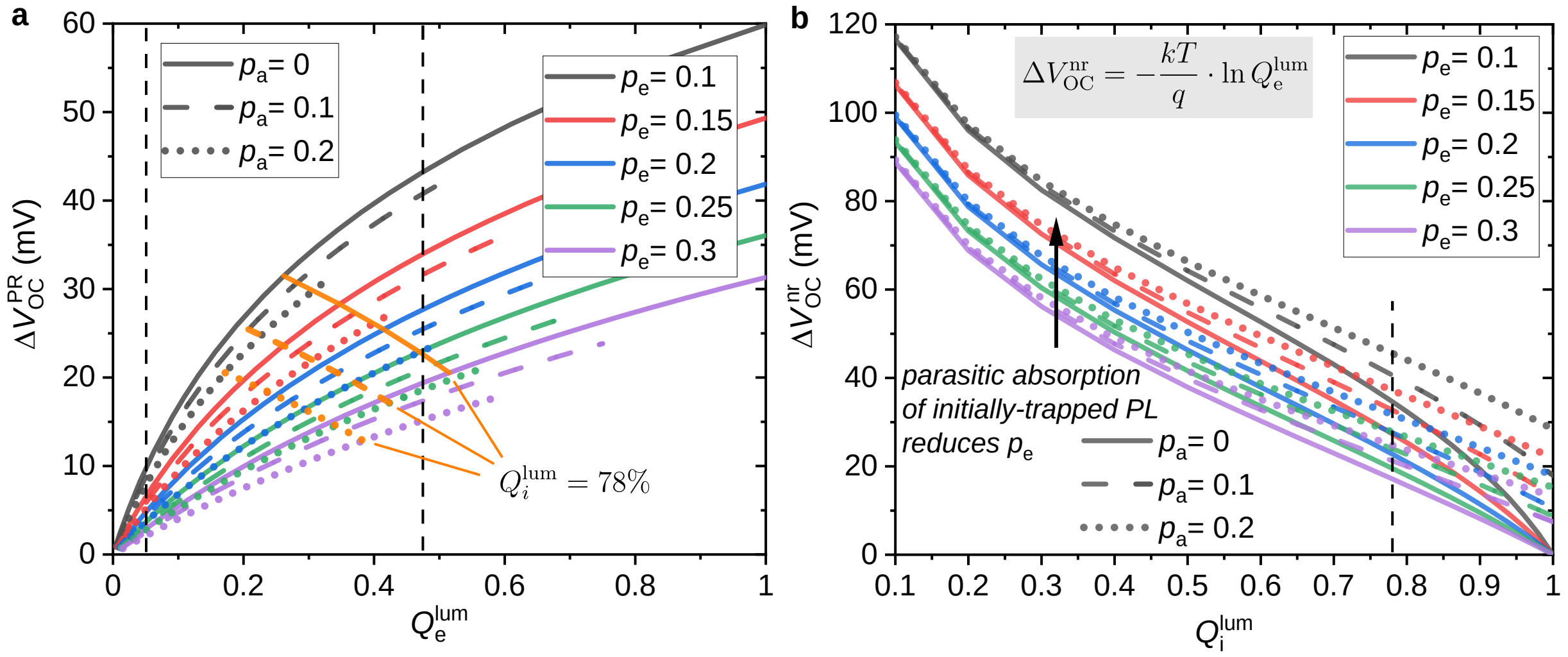

**Figure S27. Effects of photon recycling and parasitic absorption on the open-circuit voltage of PSCs. a,** Dependence of $\Delta V_{OC}^{PR}$ on $Q_e^{lum}$ and **b,** of non-radiative losses ($\Delta V_{OC}^{nr} = -\frac{kT}{q}\ln Q_e^{lum}$) on $Q_i^{lum}$ (compare Equations (S1) - (S4)) for various realistic photon escape probabilities ($\bar{p}_e$) and parasitic absorption values ($\bar{p}_a$) of 0% (solid lines), 10% (dashed lines) and 20% (dotted lines). In **a,** the two vertical dashed black lines visualize values of $Q_e^{lum}$ = 5% (the highest reported value for a complete perovskite solar cell to date[33]) and $Q_e^{lum}$ = 47.4% (the champion value measured in this work), respectively. The end of the lines correlate to $Q_i^{lum}$ = 100% (that is, $\Delta V_{OC}^{PR}$ in the radiative limit) and the orange lines visualize values of $Q_i^{lum}$ = 78% (the new benchmark determined in this work) for the respective value of $\bar{p}_a$. As an example, assuming that the interfaces in a complete PSC can be passivated such that $Q_i^{lum}$ = 78% and the values for $\bar{p}_e$ and $\bar{p}_a$ are both 10%, the measured $Q_e^{lum}$ would be 20.7% with $\Delta V_{OC}^{PR}$ = 25.4 mV. In **b,** it becomes apparent that in order to minimize $\Delta V_{OC}^{nr}$, minimizing $\bar{p}_a$ becomes crucial for high $Q_i^{lum}$ in excess of ~40%. However, in this work we reveal that parasitically absorbing layers as used in state-of-the-art PSCs can efficiently suppress the outcoupling of initially-trapped PL and thereby reduce $\bar{p}_e$ (see discussion in main text; Figure 4B). As examples related to the best 260 nm-thick MAPI film, assuming values of $Q_i^{lum}$ = 78% (see vertical dashed line), $\bar{p}_e$ = 25% and $\bar{p}_a$ = 0% in a complete PSC, $\Delta V_{OC}^{nr}$ accounts for a very low ~19.5 mV. In contrast, in case 15% of initially-trapped PL would be parasitically absorbed before coupling out upon scattering, that is, $\bar{p}_e$ = 10% and $\bar{p}_a$ = 15%, $\Delta V_{OC}^{nr}$ increases to ~43.2 mV. Critically, assuming the same values and changes for $\bar{p}_e$ and $\bar{p}_a$ in case $Q_i^{lum}$ = 20% (a realistic value for the best state-of-the-art PSCs), $\Delta V_{OC}^{nr}$ would increase from 73.2 mV to 96.9 mV, clearly showing that this kind of parasitic absorption affects PSCs independent of the value of $Q_i^{lum}$ and thus reduces the PCE of state-of-the-art PSCs.

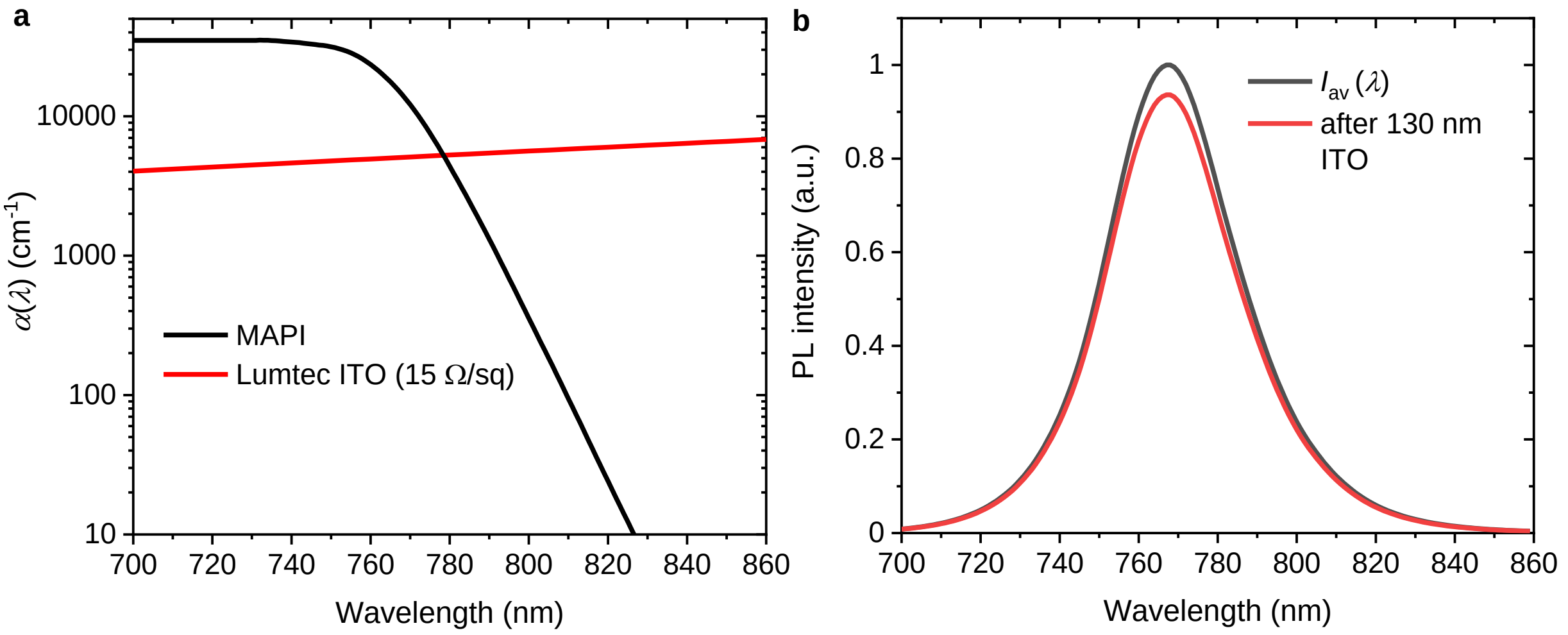

**Figure S28. a,** Absorption coefficient ($\alpha(\lambda)$) of MAPI (black; Figure S10) and commercial indium-tin-oxide (ITO) from Lumtec (red). Above ~780 nm, $\alpha(\lambda)$ of ITO is considerably larger given the exponential drop for MAPI, resulting in parasitic absorption of initially-trapped PL (see discussion in main text; Figure 4B). **b,** Expected effect on the internal PL spectra of MAPI ($I_{\text{av}}(\lambda)$; Figure S10) propagating once through 130 nm-thick ITO considering the Lambert-Beer law. The relative absorption accounts for ~6.4% (that is, ~3.2% for directly emitted PL considering only PL emitted within the bottom escape cone propagates through ITO; Figure 4B) and no spectral shift is observed.

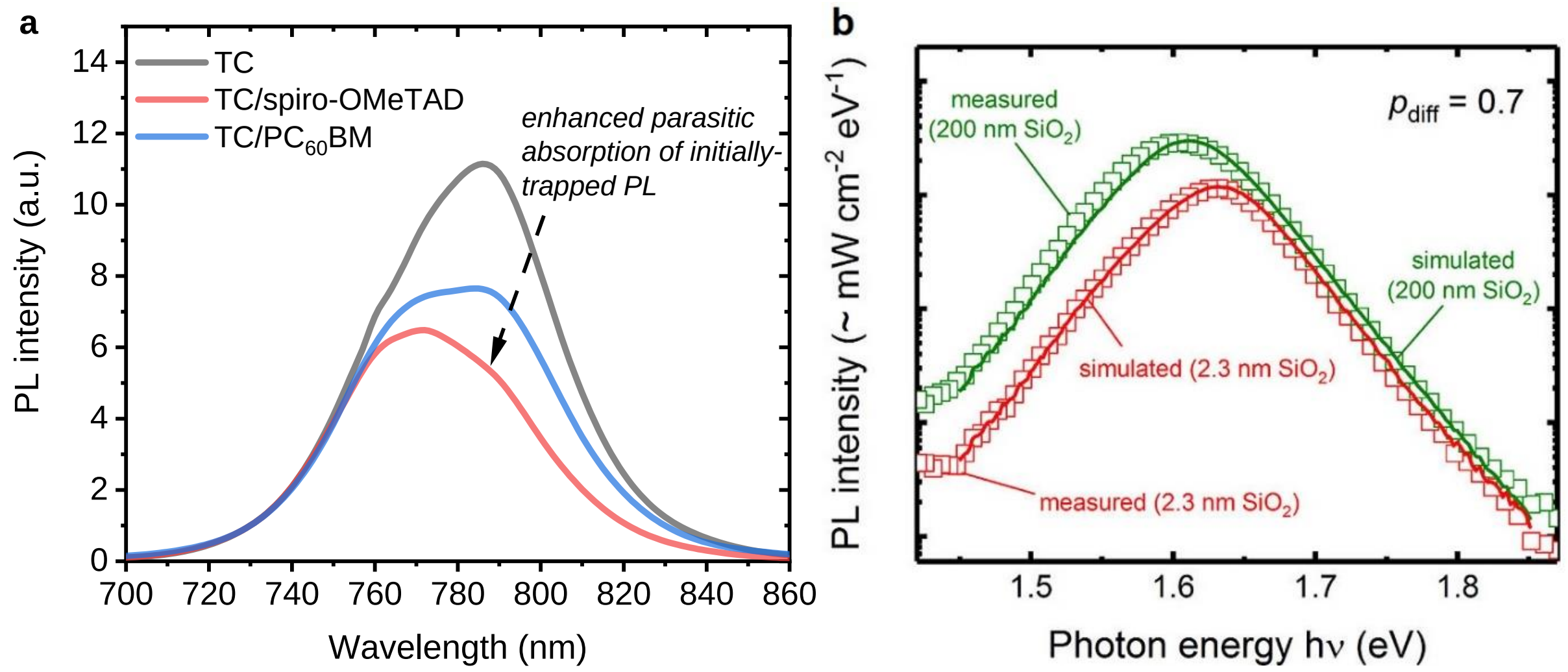

**Figure S29. PL peaks reported in literature that are affected by parasitic absorption similar to the results presented in Figure 4B. a**, PL peaks measured in an integrating sphere as reported by Stolterfoht *et al*.[69] (data extracted from the SI using WebPlotDigitizer, interpolated and normalized to the value at 730 nm), either of a bare triple-cation (TC) perovskite film on glass (black) or of a TC film covered with common charge transport layers (spiro-OMeTAD, red; $PC_{60}BM$, blue), which exhibit low yet non-negligible absorption within the spectral range of the TC PL emission. **b**, PL peaks of MAPI films reported by Kirchartz *et al*.[79] (Reprinted with permission from (F. Staub, T. Kirchartz, K. Bittkau, U. Rau, J. Phys. Chem. Lett. 2017, 8, 5084.). Copyright (2017) American Chemical Society.), either spin-coated on 2.3 nm- or 200 nm-thick $SiO_2$ and with a parasitically absorbing Si layer below the $SiO_2$. The observed PL spectral shapes were tried to be simulated by assuming a very large share of diffuse absorption ($p_{diff}$ = 70%). We argue that the (red-shifted and broad) PL peak in case of 200 nm-thick $SiO_2$ is explained by initially-trapped PL that couples out of the film towards the front detector (see discussion in main text; Figure 1C), while the measured fraction of scattered PL is strongly suppressed in case of 2.3 nm-thick $SiO_2$, given initially-trapped PL evanescently couples to the bottom Si layer where it is parasitically absorbed and, therefore, the PL blue-shifts (see discussion in main text; Figure 4B).

## Note S7 – Estimation of 1 Sun equivalent absorbed photon flux

To make the measurements of $Q_{\mathrm{e}}^{\mathrm{lum}}$ for a perovskite film relevant to photovoltaic operation, it should be measured at an illumination intensity (with monochromatic light) that generates a steady-state charge carrier density similar to that for a typical opaque perovskite solar cell (PSC) at AM1.5G illumination. First, we scale the absorption coefficient spectra derived in this work to spectra reported by Löper *et al.*[36] (from spectroscopic ellipsometry) to estimate the optical density for a $CH_3NH_3PbI_3$ film with thickness $d$ over the spectral region absorbed by a typical PSC (~350 nm < $\lambda$ < 860 nm):

$$\mathrm{OD}(\lambda) = -\log e^{-\alpha(\lambda)\cdot d} \quad \text{(S12)}$$

Next, we calculated the absorbed photon flux at AM1.5G illumination of a typical opaque PSC, for simplicity assuming zero front reflectivity and 100% back reflectivity (that is, the light passes twice through the photovoltaic film). We set it equal to the absorbed photon flux under monochromatic light, where $A$ is the absorptance of the films determined in the integrating sphere measurements[80]:

$$\frac{\sum_{\lambda=350}^{860} \phi_{\mathrm{sun}}(\lambda) \cdot \lambda \cdot 10^{-9}(1 - 10^{-2\cdot \mathrm{OD}(\lambda)})}{hc} = \frac{P_{\mathrm{laser}}(\lambda) \cdot \lambda \cdot 10^{-9} \cdot A}{hc} \quad \text{(S13)}$$

$$\rightarrow P_{\mathrm{laser}}(\lambda) = \frac{\sum_{\lambda=350}^{860} \phi_{\mathrm{sun}}(\lambda) \cdot \lambda \cdot (1 - 10^{-2\cdot \mathrm{OD}(\lambda)})}{\lambda \cdot A} \quad \text{(S14)}$$

Here, $P_{\mathrm{laser}}(\lambda)$ is the 1 Sun equivalent intensity for the respective laser wavelength, which is used to estimate the intensity in Sun employed in the IS measurements as shown in Table S6.

**Table S6. Estimation of 1 Sun equivalent intensities for the laser intensities employed in this work as well as in the work by Braly *et al.*[2] and Brenes *et al.*[1]**

| | MAPI (this work) | MAPI (this work) | MAPI (this work) | MAPI (Braly *et al.*) | MAPI (Brenes *et al.*) |
|---|---|---|---|---|---|
| Thickness [nm] | 80 | 160 | 260 | 250 | 250 |
| Laser intensity [mW/cm$^2$] | ~64.9 | ~65.4 | ~65.1 | ~60 | ~150 |
| $A$ [%] | 50 (measured) | 74.5 (measured) | 79.9 (measured) | 82 (estimated) | 79.9 (estimated) |
| Wavelength [nm] | 525 | 525 | 525 | 532 | 532 |
| 1 Sun equivalent [mW/cm$^2$] | 77.4 | 65.7 | 68.3 | 65.1 | 66.8 |
| **Intensity [Sun]** | **~0.84** | **~1.00** | **~0.95** | **~0.92** | **~2.24** |

The laser intensity shown into the integrating sphere was continuously tracked during the measurements (see Experimental Procedures); The absorptance $A$ is determined from the integrating sphere analysis procedure.[80]; 1 Sun equivalent intensity calculated with Equation (S14)

## Supplemental references